\documentclass[aip,cha, preprint,showpacs,showkeys]{revtex4-2}
\usepackage{amssymb, dcolumn,graphicx,latexsym,amsfonts,bm}

\begin{document}

\title{The nonmodal kinetic theory of the macroscale convective flows of magnetized plasma, generated by the 
inhomogeneous microturbulence}
\author{V. V. Mikhailenko}\email[E-mail:]{vladimir@pusan.ac.kr}
\affiliation{Plasma Research Center,  Pusan National University, Busan 46241, South Korea}
\author{V. S. Mikhailenko}\email[E-mail: ]{vsmikhailenko@pusan.ac.kr}
\affiliation{Plasma Research Center, Pusan National University, Busan 46241, South Korea}
\author{Hae June Lee}\email[E-mail: ]{haejune@pusan.ac.kr}
\affiliation{Department of Electrical Engineering, Pusan National University, Busan 46241, South Korea}

\date{\today}

\begin{abstract}
In this paper, we present the nonmodal kinetic theory of the macroscale two-dimensional compressed-sheared  non-diffusive 
convective flows of a magnetized plasma generated by the inhomogeneous microturbulence. 
This theory bases on the two-scales approach to the solution of the 
Vlasov-Poisson system of equations for magnetized plasma, in which the self-consistent  evolution of the plasma and of 
the electrostatic  field on the microscales, commensurable with the wavelength of the microscale instabilities and of the ion
 gyroradius, as well as on the macroscales of a bulk of plasma, is accounted for. It includes  the theory of the  formation of the 
 macroscale spatially inhomogeneous compressed-sheared convective flows by the inhomogeneous microturbulence,  
 the theory of the  back reaction of the macroscale convected flows on the microturbulence, and of the slow macroscale respond 
 of a bulk of plasma on the development of the  compressed-sheared convective flows. 
\end{abstract}

\pacs{52.35.Ra, 52.35.Kt}

\maketitle

\section{Introduction}\label{Sec1}

A common feature of the contemporary tokamaks is their operation in the regime of the enhanced confinement 
(known as H-mode)  of  a plasma, in which the microscale drift  turbulence, that cause anomalous loss 
of  heat and particles  in the edge region,  is suppressed inside the last closed flux surface (LCFS) by the 
“spontaneously” developed  poloidal sheared plasma flow.  The H-mode,  discovered in ASDEX 
tokamak\cite{Wagner_1, Wagner_2, Wagner_3} in 1982 at neutral beam heating experiment, is started as usual 
at the low confinement phase (L-mode). A critical condition for the L-H transition was found determined by a power 
threshold well above the ohmic power level of ASDEX. This condition was ion mass dependent. It met with deuterium 
target plasmas at lower power than hydrogen plasmas. The transition  from L- to H-mode  occurs with  
a dwell-time\cite{Wagner_1, Keilhacker},  estimated as $\sim 0,1 s$,  after the heating power has been increased from 
the ohmic level before the plasma transits into the H-mode. This transition occurs  without any interference from outside 
and at constant power.  During a short  time, which is estimated as  $\sim 100\, \mu s$, the tokamak edge plasma jumps
into H-mode regime. The development of the sheared poloidal  flow inside the separatrix that follows by the suppression 
of the drift turbulence and formation of a transport barrier  at the plasma edge  (2 to 4 cm from LCFS) with  steep edge 
temperature and density gradients  (commonly referred to as  the pedestal) is a generic feature of the H-mode.  
The pedestal development results in a  significant increase the core density and temperature that is beneficial for fusion reactors. 

One of the first nonlinear gyro-Landau fluid simulations\cite{Waltz, Waltz_1, Waltz_2} of the tokamaks plasmas
with sheared flows  has shown drift turbulence stabilization and suppression when the $E\times B$ velocity shearing rate $V'_{0}$ 
(the prime  denotes the derivative of the flow velocity $V_{0}$ with respect to the spatial coordinate) becomes comparable 
with or above the maximum linear growth rate $\gamma_{max}$ of  unstable drift modes of the suppressed drift turbulence.
Beginning from experiments on Doublet-III-D tokamak, this empiric "quench rule",  $V'_{0}\gtrsim \gamma_{max}$ was confirmed 
experimentally in numerous experiments in tokamaks as a rough estimate for the magnitude of the velocity  shear above which 
the suppression of the turbulence and formation of the transport barriers occur \cite{Lao}. In spite of that experimental evidence,
the underlying physics that relates to the quench rule and the suppression of the instabilities and turbulence and creating 
of transport barriers remained incomprehensible  for a long time and was the subject of significant debates.

One of the most challenging problems  in the theory of plasma shear flow turbulence was the development of analytical methods of
the investigations of the long time evolution of the turbulence in the sheared flow. The analytical description of the stability and 
turbulence of the magnetized plasma in strong inhomogeneous electric field can suffer from inadequate application of the spectral 
transforms to sheared plasma flow having an inhomogeneous radial profile of velocity. In typical treatment of the gyrokinetic theory  
of magnetized plasma in the presence of the inhomogeneous electric field, a transformation to the reference frame in the velocity space, 
that moves with inhomogeneous flow velocity  $V_{0}\left(x\right)\mathbf{e}_{y}$, but with unchanging the spatial 
(laboratory) coordinates, was usually used in the gyrokinetic theory of plasma sheared flows\cite{Brizard}. The   potential $\varphi$ of the 
electrostatic perturbations of a plasma was considered in a such approach in the canonical modal form, 
$\varphi\sim \exp \left[ 
-i\Big(\omega+k_{y}\left(V_{0}\left(x_{0}\right)+V'_{0}\left(x_{0}\right)\left(x-x_{0}\right)\right)\Big)t
-ik_{x}x-ik_{y}y-ik_{z}z\right]$ with the space-dependent Doppler shift. This form of solution is modal only during 
a finite time at which $k_{y}V'_{0}t\ll k_{x}$ .  At time $t\sim \gamma^{-1}$, where $\gamma$ is the modal growth 
rate of the drift instability, the modal structure of the perturbation with $k_{y}\sim k_{x}$  holds only for a weak flow velocity shear, 
$|V'_{0}|\ll \gamma$ i. e. for the 
conditions at which the suppression of the drift turbulence by the sheared flow is absent.  It was proved in Refs.\cite {Mikhailenko_5, 
Mikhailenko_6, Mikhailenko_4} that the "quench rule"  $|V'_{0}|\gg \gamma$ is the condition of the non-applicability of the modal approach to 
the stability analysis of the plasma  sheared flow.  This condition displays that the suppression of the turbulence by the sheared flow is strictly 
nonmodal process.  The detailed description of the nonmodal evolution and suppression of the kinetic drift instability, hydrodynamic and 
kinetic ion temperature gradient instabilities are presented in Refs.\cite {Mikhailenko_5, Mikhailenko_6, Mikhailenko_4}.

The heating of plasma by the fast ion flow, produced  after ionization of the injected energetic beam of neutrals, provides  
appreciable gradient across the magnetic field of  the ion temperature and little or no density change\cite{Rhodes}.
Such a plasma is unstable against the development of the  microscale turbulence\cite{Rhodes}, which is responsible  for the 
anomalous loss of a tokamak plasma heat and particles.  At L-mode phase, the microscale turbulence involves two disparate 
spatial scales: the microscale, commensurable with the wavelength of the most unstable microscale perturbation,
 and much larger macroscales of the  radial spatial inhomogeneity of the plasma density and of the ion temperature  
 and  of a spatial inhomogeneity of the spectral intensity of the microturbulence developed in the inhomogeneous plasma.  
 The L-H transition 
reveals in the  "spontaneous" realignment of the macroscale structure of the inhomogeneous plasma and of the spatially inhomogeneous 
microturbulence by development of the  sheared poloidal  flow inside the separatrix  and development of the transport barrier, 
resulted from the  suppression of the microturbulence outside the transport barrier. The formation of the pedestal structure 
near LCFS in H-mode regime introduces in the edge region third radial spatial scale intermediate between the macroscale 
and the  microscale. This spatial scale, determined by the radial gradient scale lengths of the ion density,  the ion temperature  
and of the microturbulence  in the pedestal, is referred to as the mesoscale.  The  kinetic  theory of the weak microscale turbulence,  
as well as the quasilinear theory,  of an inhomogeneous  plasma are based on the local approximation and are applicable 
to the  treatment the processes  on the microscales  such as the excitation and saturation of the microinstabilities,  the anomalous 
diffusion and heating of plasma components. The macroscales in this theory are involved as the parameters.  It is obvious that 
the evolution processes in plasma turbulence, which occur on the macroscales or on the mesoscales during the evolution  time 
much larger than the inverse linear or nonlinear growth rates,  such as at the L-H transition, are missed in the local theory.  

The goal of this paper is the development of the  kinetic theory of the microturbulence  of the inhomogeneous plasma, 
which provides  the self-consistent  two-scale treatment of the fast and of the slow evolution of the microturbulence on 
the microscales and on the macroscales. In Refs.\cite{Mikhailenko_2, Mikhailenko_3}, the two-scales 
kinetic theory was developed for the  investigations of  the temporal evolution of the spatially inhomogeneous 
electrostatic ion cyclotron (IC) parametric microturbulence, driven by the fast wave in the inhomogeneous pedestal 
plasma with a sheared poloidal flow.  The main result of that theory, which was based on  the Vlasov - Poisson system 
of equations,  is discovery the generation  of the radially inhomogeneous  non-diffusive  convective flows in a pedestal 
region  which resulted from the interaction of ions with  parametric microturbulence, that was not suppressed  by the 
poloidal sheared flow radially inhomogeneous on the scales commensurable with the pedestal width.  That theory was made under 
assumption that the electric field of the microscale turbulence is not changed with time by the convective flows formed 
by the inhomogeneous microturbulence itself. However any perturbations in the flow with spatially inhomogeneous flow velocity 
experiences the continuous distortion with time and becomes the perturbation with time dependent nonmodal structure. 

In the present paper, we develop the two-scales  nonmodal approach to the kinetic theory of the  radially inhomogeneous  
microturbulence.  Compared to the generally accepted modal theory of the microturbulence, the developed two-scales nonmodal theory 
reveals the generation of the macroscale convective flows resulted from the interaction of  ions with microturbulence 
inhomogeneous on the macroscales.  The developed nonmodal approach provides the analytical treatment of the slow macroscale 
evolution of the microturbulence in plasma with spatially inhomogeneous convective flows, and the analytical treatment 
of the slow macroscale evolution of the  plasma resulted from the interaction of the plasma components with inhomogeneous 
microturbulence. That theory is necessary  for the understanding the processes which occur in the time interval corresponding to the 
initial stage of the L-H transition commensurable with a  dwell-time\cite{Wagner_1, Keilhacker}.

In Sec. \ref{sec2}, we present for the first time the  theory of the generation of the ion convective flows by the microturbulence 
inhomogeneous on the macroscales. It is proved that the generation of the convestive flows is the inherent property of the temporal 
evolution of the inhomogeneous microturbulence. In this section, we derive the nonmodal ion and electron kinetic  equations,  
that determines the temporal evolution  of the microscale perturbations of  the ion and electron distribution functions in the macroscale 
convected flows  driven by the inhomogeneous ion temperature.  Also,  the kinetic equations for the ion and electron distribution 
functions which account for the slow macroscale respond of a bulk of plasma on the development of the  convective flows, 
are derived in that Section. The derived  microscale and macroscale kinetic equations are employed in Sections \ref{sec3} 
and  \ref{sec4}.  

In Sec. \ref{sec3}, we present the detailed derivation of  the integral equation for the potential of the microturbulence 
that determines the temporal evolution of the microscale turbulence potential in the inhomogeneous convective flows. 
It is the theory of the transformation of the initially modal microscale turbulence with a plane wave structure to the 
nonmodal turbulence the elementary perturbations of which experience the continuous distortion in the convective flow 
and become the sheared-compressed modes with time dependent structures. 

The macroscale evolution of bulk of ions in the convective flow is considered in  Sec. \ref{sec4}.  That section contains 

1) the nonmodal quasilinear theory, which governs the slow temporal evolution on the macroscales the ion and electron 
distribution functions, resulted from the interactions of ions  and electrons with ensemble of sheared-compressed microscale 
waves with random phases,  
2) the self-consistent theory of the slow temporal evolution of the macroscale electrostatic potential of the plasma respond 
on the developed macroscale convective flow. It is the theory of the generation of the secondary nonmodal macroscale/mesoscale 
instabilities  of a plasma with compressed-sheared convective flows. The integral equation for that macroscale potential  is the basic 
equation of 
the stability theory of the convective flows against the development of the secondary mesoscale instabilities of a plasma with 
inhomogeneous macroscale convective flows. 
Conclusions are presented in Sec. \ref{sec5}.

\section{The nonlocal two-scale approach to the kinetic theory of the macroscale convective plasma flows  generated by 
the spatially inhomogeneous microturbulence}\label{sec2} 

Our theory is based on the  Vlasov-Poisson system of equations in a slab geometry approximation, in which the coordinates 
$x$, $y$, $z$ for the microscale fast variations are viewed as corresponding to the radial, poloidal and toroidal directions, respectively, 
of the toroidal co-ordinate system.   Within this approximation,  the Vlasov equation for the velocity distribution function 
$F_{\alpha}$ of $\alpha$ plasma species ($\alpha=i$ for ions and $\alpha=e$ for electrons), in coordinates 
$\mathbf{r}=\left (x, y, z\right)$,
\begin{eqnarray}
&\displaystyle \frac{\partial F_{\alpha}\left(\mathbf{v}, \mathbf{r}, t\right)}{\partial t} 
+\mathbf{v}\frac{\partial F_{\alpha}\left(\mathbf{v}, \mathbf{r}, t\right)}
{\partial\mathbf{r}}
\nonumber 
\\ 
&\displaystyle
+\frac{e_{\alpha}}{m_{\alpha}}\left(\mathbf{E}_{0}\left(x, t\right) 
+\mathbf{E}\left(\mathbf{r}, t \right) 
+\frac{1}{c}\left[\mathbf{v}\times\mathbf{B}_{0}\right]\right)
\frac{\partial F_{\alpha}\left(\mathbf{v}, \mathbf{r}, t\right)}{\partial\mathbf{v}}=0,	\label{1}
\end{eqnarray}
contains the inhomogeneous along the coordinate $x$ electric field $\mathbf{E}_{0}\left(x, t\right) $
of the applied RF electromagnetic wave for the plasma heating and current drive, or the radially inhomogeneous  
electric field formed in plasma after ionization of the injected energetic beam of neutrals\cite{Xu}.  The electrostatic  
electric field   $\mathbf{E}\left(\mathbf{r}, t\right)$  is determined by the  Poisson equation 
\begin{eqnarray}
&\displaystyle 
\nabla_{\mathbf{r}}\cdot \mathbf{E}\left(\mathbf{r},  t\right) =
4\pi\sum_{\alpha=i,e} e_{\alpha}\int f_{\alpha}\left(\mathbf{v}, \mathbf{r},  t \right)d{\bf v}, 
\label{2}
\end{eqnarray}
in which $f_{\alpha}$ is the fluctuating part of the distribution function 
$F_{\alpha}$, $f_{\alpha}=F_{\alpha}-F_{0\alpha}$, where $F_{0\alpha}$ is the equilibrium 
distribution function.   $\mathbf{B}_{0}$ is the uniform confined magnetic field,   directed along $z$ axes.  
The radial extent of the magnetic shear does not seem to play a role in the L-H transition\cite{Wagner_2}.

Equation (\ref{1}) contains two disparate spatial inhomogeneity lengths. The macroscale inhomogeneity is introduced 
by the $\mathbf{E}_{0}\left(x, t\right) $ electric field.  Electric field $\mathbf{E}\left(\mathbf{r}, t\right)$ 
being the microscale respond of the inhomogeneous plasma on the inhomogeneous $\mathbf{E}_{0}\left(x, t\right) $ 
electric field, contains micro and macro spatial scales. For the deriving  solution for $\mathbf{E}\left(\mathbf{r}, t\right)$ field, 
which determines  the  temporal evolution on the both spatial scales,  we use in our theory jointly with the  variables 
$\mathbf{r}=\left (x, y, z\right)$ and time $t$ for the microscale fast variations,  the spatial variables $X=\varepsilon x$, 
$Y=\varepsilon y$ and time $T=\varepsilon t$ for the macroscale slow variations.

It was found in Ref.\cite{Mikhailenko_2} that with the velocity  $\mathbf{v}_{\alpha}$  and  position coordinates 
$\mathbf{r}_{\alpha}=\left(x_{\alpha},  y_{\alpha}\right)$, determined in the reference flow,  which 
moves relative to the laboratory frame with velocity $\mathbf{V}_{\alpha}\left(X, t\right)$ of  $\alpha$ species 
particle in  $\mathbf{E}_{0}\left(X, t \right)$ and in confined $\mathbf{B}_{0}$ fields, the spatially inhomogeneous field 
$\mathbf{E}_{0}\left(X, t\right)$ is presented in the Vlasov equation (\ref{1}) for $F_{\alpha}\left(\mathbf{v}_{\alpha}, 
\mathbf{r}_{\alpha}, t\right)$ only in terms of the order of  $\left| R_{\alpha x}/L_{E}\right| \ll 1$,
where $R_{\alpha x}$ is the $\alpha$ species particle displacement in the $\mathbf{E}_{0}\left(X, t\right)$ 
and $\mathbf{B}_{0}$ fields,  and $L_{E}$ is a spatial scale of the $\mathbf{E}_{0}\left(X, t\right)$ 
field inhomogeneity.  Without these terms, the  Vlasov equation for $F_{i}\left(\mathbf{v}_{i}, \mathbf{r}_{i},  t\right)$ 
with great accuracy has a form as for a steady plasma in the uniform magnetic field $\mathbf{B}_{0}$ without 
$\mathbf{E}_{0}\left(X, t\right)$  field, i. e.
\begin{eqnarray}
&\displaystyle 
\frac{\partial F_{i}\left(\mathbf{v}_{i}, \mathbf{r}_{i},  t, X_{i} \right)}
{\partial t}+ \mathbf{v}_{i}\frac{\partial F_{i}} {\partial
\mathbf{r}_{i}}+\frac{e_{i}}{m_{i}c}\left[\mathbf{v}_{i}\times\mathbf{B}_{0}\right]
\frac{\partial F_{i}}{\partial\mathbf{v}_{i}}
\nonumber 
\\ 
&\displaystyle
+\frac{e_{i}}{m_{i}}\mathbf{E}_{i}\left(\mathbf{r}_{i}, t,  X_{i}\right)
\frac{\partial F_{i}\left(\mathbf{v}_{i}, \mathbf{r}_{i}, t , X_{i}\right)}{\partial \mathbf{v}_{i}} =0.
\label{3}
\end{eqnarray}
The macroscale variable $X_{i}$ is presented in Eq.  (\ref{3}) as the parameter in  the electric 
field $\mathbf{E}_{i}\left(\mathbf{r}_{i}, t, X_{i}\right)$  of the electrostatic microturbulence, determined in the ion reference flow.  
Equation  (\ref{3}) for ions, the  Vlasov equation for electrons, 
determined in the electron reference flow by Eq. (\ref{3}) with  ion species subscript $i$ changed on the electron subscript $e$, 
and the Poisson equation for the electric field $\mathbf{E}_{i}\left(\mathbf{r}_{i}, t, X_{i}\right)$  form the set of equations, 
which determines in the local approximation the development and the temporal evolution of the microscale instabilities of the 
inhomogeneous plasma. The saturation of the microscale instability with the frequency $\omega\left(\mathbf{k}\right)$ 
and the growth rate $\gamma\left(\mathbf{k}\right)$,  which occurs at time $t\gtrsim\gamma^{-1}
\left(\mathbf{k}\right) \gtrsim 2\pi |\omega^{-1}\left(\mathbf{k}\right)|$, is followed by the formation at the fast time $t$ 
the steady level of the microturbulence spatially inhomogeneous along $x_{i}$. At that stage, the electric field $\mathbf{E}_{i}$ 
of the electrostatic microturbulence, directed almost across the magnetic field $\mathbf{B}_{0}$, may be presented in the ion 
reference flow in the form, that includes the microscale $\mathbf{r}_{i}$ and the large scale $X_{i}$ variables, i. e. 
\begin{eqnarray}
&\displaystyle 
\mathbf{E}_{i}\left( X_{i}, \psi\right)= \frac{1}{2\left(2\pi\right)^{3}}
\int d\mathbf{k}\left[\mathbf{E}_{i}\left(\mathbf{k}, X_{i}\right)
e^{i\psi}+\mathbf{E}_{i}^{\ast}\left(\mathbf{k}, X_{i}\right)e^{-i\psi} \right], 
\label{4}
\end{eqnarray}
with phase
\begin{eqnarray}
&\displaystyle
\psi=\psi\left(\mathbf{r}_{i}, t, X_{i}\right)= -\omega\left(\mathbf{k}, X_{i}\right)t+ \mathbf{k}
\mathbf{r}_{i}+\theta\left(\mathbf{k}\right) 
\label{5}
\end{eqnarray}
changed on the microscale $\mathbf{r}_{i}$ and fast time $t$, i. e. as a linear superposition of the electric fields 
of perturbations which has a modal form of the plane waves with frequencies $\omega\left(\mathbf{k}, X_{i}\right)$, 
and with the wave vectors $\mathbf{k}$ directed almost across the magnetic field. In this paper, we consider the case 
of the microturbulence with  frequency  $\omega\left(\mathbf{k}\right)$ which does not depend on the 
macroscales variables $X_{i}$ or $X_{e}$, but with amplitudes $\mathbf{E}_{i}\left(\mathbf{k}, X_{i}\right)$ 
dependent on $X_{i}$ in the steady state. In a plasma with turbulent electric field $\mathbf{E}_{i}\left( X_{i}, \psi\right)$,
the velocity $\mathbf{v}_{i}$ of ions is the total velocity of the ion thermal motion and of 
the ion motion in the turbulent electric field (\ref{5}) and in the magnetic field $\mathbf{B}_{0}$. 
It was found in Refs.\cite{Mikhailenko_2, Mikhailenko_3}, that some two-dimensional spatially 
inhomogeneous microturbulence-associated  ion reference flow, which moves with velocity
$\tilde{\mathbf{U}}_{i}\left(\tilde{\mathbf{r}}_{i},  \tilde{X}_{i}, t\right)$  relative to the laboratory frame, 
may be determined, in which  the ion velocity $\mathbf{v}_{i}$ and position vector $\mathbf{r}_{i}$  at time $t$ 
are determined by the relations
\begin{eqnarray}
&\displaystyle
\mathbf{v}_{i}=\tilde{\mathbf{v}}_{i}+\tilde{\mathbf{U}}_{i}\left(\tilde{\mathbf{r}}_{i},  \tilde{X}_{i}, t\right),
\label{6}
\\ 
&\displaystyle
\mathbf{r}_{i}=\tilde{\mathbf{r}}_{i}+\tilde{\mathbf{R}}_{i}\left(\tilde{\mathbf{r}}_{i}, \tilde{X}
_{i}, t\right)=\tilde{\mathbf{r}}_{i}+\int\limits^{t}_{t_{0}}
\tilde{\mathbf{U}}_{i}\left(\tilde{\mathbf{r}}_{i}, \tilde{X}_{i}, t_{1}\right)dt_{1},
\label{7}
\end{eqnarray}
or by the inverse transformations  $(\tilde{\mathbf{v}}_{i}, \tilde{\mathbf{r}}_{i},  \tilde{X}_{i}, t)
\rightarrow (\mathbf{v}_{i}, \mathbf{r}_{i},  X_{i}, t)$,
\begin{eqnarray}
&\displaystyle
\tilde{\mathbf{v}}_{i}=\mathbf{v}_{i}-\tilde{\mathbf{V}}_{i}\left(\mathbf{r}_{i}, 	X_{i}, t\right), 
\label{8}
\\ 
&\displaystyle
\tilde{\mathbf{r}}_{i}=\mathbf{r}_{i}-\int\limits^{t}_{t_{0}}\tilde{\mathbf{V}}_{i}
\left(\mathbf{r}_{i}, X_{i}, t_{1}\right)dt_{1}=\mathbf{r}_{i}-\mathbf{R}_{i}\left(\mathbf{r}_{i}, X_{i}, t\right),
\label{9}
\end{eqnarray}
where $\tilde{\mathbf{v}}_{i}$ is the  velocity  of  an ion in the microturbulence-associated reference flow,  in which 
$\tilde{\mathbf{r}}_{i}$  and $\tilde{X}_{i}$  are the microscale and the large scale  coordinates, respectively, 
of the ion position, determined in this reference flow.  
\begin{figure}[!htbp]
\includegraphics[width=0.7\textwidth]{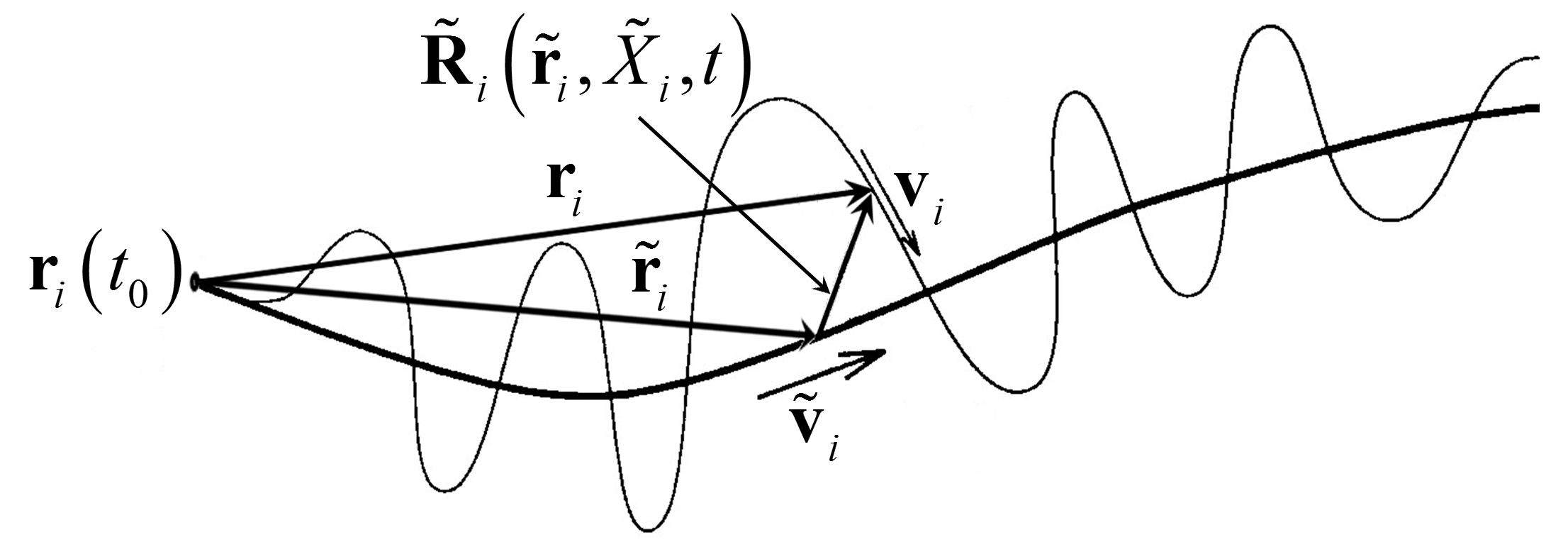}
\caption{\label{fig1} The actual ion position $\mathbf{r}_{i}$ in the ion reference frame at time $t$, 
$\tilde{\mathbf{r}}_{i}$ is the ion position at time $t$ in the mean microturbulence-associated ion reference flow, 
$\tilde{\mathbf{r}}_{i}+\tilde{\mathbf{R}}_{i}\left(\tilde{\mathbf{r}}_{i}, t, \tilde{X}_{i}\right)=\mathbf{r}_{i}$ is the ion 
position at the same time $t$ whose mean position was $\tilde{\mathbf{r}}_{i}$ and $\tilde{\mathbf{R}}_{i}\left(\tilde{\mathbf{r}}_{i}, t, 
\tilde{X}_{i}\right)$ is the ion displacement in the electric field $\mathbf{E}_{i}\left(\mathbf{r}_{i},  t, X_{i}\right)$
of the microturbulence.  $\left\langle \tilde{\mathbf{R}}_{i}\left(\tilde{\mathbf{r}}_{i}, \tilde{X}_{i}, t\right)\right\rangle =0$,
$\left\langle \mathbf{r}_{i}\right\rangle = \tilde{\mathbf{r}}_{i}$;  $\mathbf{v}_{i}$
is the actual ion velocity in the ion reference frame, $\tilde{\mathbf{v}}_{i} $ is velocity of  ion in the microturbulence-associated 
ion mean flow.}
\end{figure}

For the large scale co-ordinates  $X_{i}$, $Y_{i}$  and  $\tilde{X}_{i}$,  $\tilde{Y}_{i}$,
  Eqs. (\ref{7}), (\ref{9}) give the relations
\begin{eqnarray}
&\displaystyle
X_{i}=\tilde{X}_{i}+\int\limits^{T}_{T_{0}}\tilde{U}_{ix}\left(\tilde{X}_{i}, 
\tilde{\mathbf{r}}_{i}, T_{1}\right)dT_{1}=\tilde{X}_{i}+\tilde{R}_{ix}\left(\tilde{X}_{i}, 
\tilde{\mathbf{r}}_{i}, T\right), 
\label{10}
\\  
&\displaystyle 
Y_{i}=\tilde{Y}_{i}+\int\limits^{T}_{T_{0}}\tilde{U}_{iy}\left(\tilde{X}_{i}, 
\tilde{\mathbf{r}}_{i},T_{1}\right)dT_{1}=\tilde{Y}_{i}+\tilde{R}_{iy}\left(\tilde{X}_{i}, 
\tilde{\mathbf{r}}_{i}, T\right), 
\label{11}
\end{eqnarray}
and
\begin{eqnarray}
&\displaystyle
\tilde{X}_{i}=X_{i}-\int\limits^{T}_{T_{0}}
\tilde{V}_{ix}\left(X_{i}, \mathbf{r}_{i}, T_{1}\right)dT_{1}=X_{i}-R_{ix}\left(X_{i},\mathbf{r}_{i}, T\right), 
\label{12}
\\  
&\displaystyle 
\tilde{Y}_{i}=Y_{i}-\int\limits^{T}_{T_{0}}\tilde{V}_{iy}\left(X_{i},  \mathbf{r}_{i},T_{1}\right)dT_{1}
=Y_{i}-R_{iy}\left(X_{i},  \mathbf{r}_{i}, T\right),
\label{13}
\end{eqnarray}
where the slow time $T$ determined by $t=\varepsilon^{-1}T$ and the relations $x_{i}=\varepsilon^{-1}X_{i} $, 
$y_{i}=\varepsilon^{-1}Y_{i}$ for  $ \mathbf{r}_{i}$ were used.  The electrostatic potential 
$\varphi\left(\mathbf{r}_{i},  t, X_{i}\right)$, which determines the electrostatic field   
$\mathbf{E}_{i}\left(\mathbf{r}_{i},  t, X_{i}\right)= -\nabla_{\mathbf{r}_{i}}\varphi\left(\mathbf{r}_{i}, t, X_{i}\right)$, 
is presented in the form
\begin{eqnarray}
&\displaystyle
\varphi\left(\mathbf{r}_{i},  X_{i}, Y_{i}, t\right)= \tilde{\varphi}_{i}
\left(\tilde{\mathbf{r}}_{i}+\tilde{\mathbf{R}}_{i}\left(t\right),  \tilde{X}_{i}+\tilde{R}_{ix}\left(T\right), t\right)
+\Phi_{i}\left(\tilde{X}_{i}, \tilde{Y}_{i}, T\right),
\label{14}
\end{eqnarray}
where $\tilde{\varphi}_{i}$ is the electrostatic potential of the microscale turbulence,
\begin{eqnarray}
&\displaystyle 
\tilde{\mathbf{E}}_{i}\left(\tilde{\mathbf{r}}_{i}+\tilde{\mathbf{R}}_{i}\left(t\right), 
\tilde{X}_{i}+\tilde{R}_{ix}\left(T\right), t\right)
=-\nabla_{\tilde\mathbf{r}_{i}}\tilde{\varphi}_{i}\left(\tilde{\mathbf{r}}_{i}+\tilde{\mathbf{R}}_{i}\left(t\right),  
\tilde{X}_{i}+\tilde{R}_{ix}\left(T\right), t\right);
\label{15}
\end{eqnarray} 
 $\Phi_{i}\left(\tilde{X}_{i}, \tilde{Y}_{i}, T\right)$ is the potential of the large scale plasma response on 
the formation and slow evolution of the large scale plasma inhomogeneities, observed in the reference flow.
\begin{eqnarray}
&\displaystyle
\bar{\mathbf{E}}_{i}\left(\tilde{X}_{i}, 
\tilde{Y}_{i}, T\right)=-\nabla\Phi_{i}\left(\tilde{X}_{i}, \tilde{Y}_{i}, T\right). 
\label{16}
\end{eqnarray}

For the treatment of the slow evolution of a plasma on the macroscales, we present the Vlasov equation (\ref{3}) 
in variables $X_{i}$, $Y_{i}$, $T$,
\begin{eqnarray}
&\displaystyle 
\frac{\partial F_{i}\left(\mathbf{v}_{i}, X_{i}, Y_{i}, T\right)}
{\partial T}+ v_{ix}\frac{\partial F_{i}} {\partial
X_{i}}+v_{iy}\frac{\partial F_{i}} {\partial
Y_{i}}+\frac{\omega_{ci}}{\varepsilon}v_{iy}
\frac{\partial F_{i}}{\partial v_{ix}}-\frac{\omega_{ci}}{\varepsilon}v_{ix}
\frac{\partial F_{i}}{\partial v_{iy}}
\nonumber 	
\\ 
&\displaystyle
+\frac{e_{i}}{\varepsilon m_{i}}\mathbf{E}_{i}\left(X_{i}, \psi\right)\frac{\partial F_{i}}{\partial \mathbf{v}_{i}}=0,
\label{17}
\end{eqnarray}
where
\begin{eqnarray}
&\displaystyle 
\psi=\frac{1}{\varepsilon}\Psi\left(X_{i}, Y_{i}, z, T\right)=\frac{1}{\varepsilon}\left(-\omega\left(\mathbf{k}\right)T
+ k_{x}X_{i}+k_{y}Y_{i}+k_{z}Z+\varepsilon\theta\left(\mathbf{k}\right) \right).	
\label{18}
\end{eqnarray}	
The Vlasov equation (\ref{17}) for the ion distribution function $F_{i}\left(\tilde{\mathbf{v}}_{i}, \tilde{X}_{i}, 
\tilde{Y}_{i}, T\right)$  has a form
\begin{eqnarray}
&\displaystyle 
\frac{\partial F_{i}\left(\tilde{\mathbf{v}}_{i}, \tilde{X}_{i}, \tilde{Y}_{i}, T\right)}
{\partial T}+\tilde{v}_{ix}\frac{\partial F_{i}} {\partial \tilde{X}_{i}}+\tilde{v}_{iy}
\frac{\partial F_{i}}{\partial \tilde{Y}_{i}}
\nonumber 
\\ 
&\displaystyle
-\left(\tilde{v}_{ix}+\tilde{U}_{ix}\left(\tilde{X}_{i}, \tilde{Y}_{i}, 
T, \varepsilon\right)\right)\int\limits_{T_{0}}^{T}\frac{\partial \tilde{V}_{ix}} 
{\partial X_{i}}dT_{1}\frac{\partial F_{i}}{\partial\tilde{X}_{i}}-\left(\tilde{v}_{ix}
+\tilde{U}_{ix}\left(\tilde{X}_{i}, \tilde{Y}_{i}, T, \varepsilon\right)\right)\int\limits_{T_{0}}^{T}
\frac{\partial \tilde{V}_{iy}} {\partial
X_{i}}dT_{1}\frac{\partial F_{i}}{\partial\tilde{Y}_{i}}
\nonumber 
\\ 
&\displaystyle
+\left(\frac{\omega_{ci}}{\varepsilon}\tilde{v}_{iy}-\tilde{v}_{ix}\frac{\partial \tilde{V}_{ix}} 
{\partial X_{i}}\right)
\frac{\partial F_{i}}{\partial \tilde{v}_{ix}}-\left(\frac{\omega_{ci}}{\varepsilon}
+\frac{\partial \tilde{V}_{iy}} {\partial X_{i}}\right)\tilde{v}_{ix}
\frac{\partial F_{i}}{\partial \tilde{v}_{iy}}-\frac{e_{i}}{m_{i}}\tilde{\nabla}\Phi\left(\tilde{X}_{i}, 
\tilde{Y}_{i}, T\right)\frac{\partial F_{i}}{\partial \tilde{\mathbf{v}}_{i}}
\nonumber 
\\ 
&\displaystyle
-\left[\frac{\partial\tilde{U}_{ix}}{\partial T}-\frac{\omega_{ci}}{\varepsilon}\tilde{U}_{iy}\right.
\nonumber 
\\ 
&\displaystyle
\left.-\frac{e_{i}}{\varepsilon  m_{i}}\tilde{E}_{ix}\left(\tilde{X}_{i}+\tilde{R}_{ix}
\left(\tilde{X}_{i}, \varepsilon^{-1}\tilde{X}_{i}, \varepsilon^{-1}\tilde{Y}_{i}, 
\varepsilon^{-1}T\right), \varepsilon^{-1}X_{i}, \varepsilon^{-1}Y_{i}, 
\varepsilon^{-1}T\right)\right]\frac{\partial F_{i}}{\partial \tilde{v}_{ix}}
\nonumber 
\\ 
&\displaystyle
-\left[\frac{\partial\tilde{U}_{iy}}{\partial T}+\frac{\omega_{ci}}{\varepsilon}\tilde{U}_{ix}\right.
\nonumber 
\\ 
&\displaystyle
\left.-\frac{e_{i}}{\varepsilon m_{i}}\tilde{E}_{iy}\left(\tilde{X}_{i}+\tilde{R}_{ix}
\left(\tilde{X}_{i}, \varepsilon^{-1}\tilde{X}_{i}, 
\varepsilon^{-1}\tilde{Y}_{i}, \varepsilon^{-1}T\right), \varepsilon^{-1}X_{i}, \varepsilon^{-1}Y_{i}, 
\varepsilon^{-1}T\right)\right]\frac{\partial F_{i}}{\partial \tilde{v}_{iy}}=0,
\label{19}
\end{eqnarray}
where  $\tilde{U}_{ix,iy}=\tilde{U}_{ix,iy}\left(\tilde{X}_{i}, \varepsilon^{-1}X_{i}, \varepsilon^{-1}Y_{i}, 
\varepsilon^{-1}T\right)$. In Eq. (\ref{19}), the identities $\tilde{\mathbf{V}}_{i}\left(\mathbf{r}_{i}, X_{i}, t\right)
=\mathbf{\tilde{U}}_{i}\left(\tilde{\mathbf{r}}_{i}, \tilde{X}_{i}, t\right)$ and 
\begin{eqnarray}
&\displaystyle
\frac{\partial\tilde{\mathbf{V}}_{i}\left(X_{i}, Y_{i}, T\right)}{\partial T}+\tilde{V}_{ix}\left(X_{i}, Y_{i}, T\right)
\frac{\partial\tilde{\mathbf{V}}_{i}\left(X_{i}, Y_{i}, T\right)}{\partial X_{i}}=
\frac{\partial\tilde{\mathbf{U}}_{i}\left(\tilde{X}_{i}, \tilde{Y}_{i}, T, \varepsilon\right)}{\partial T},
\label{20}
\end{eqnarray}
which follow from Eqs. (\ref{6})-(\ref{13}), were used. 

 The first task in the solution of Eq.  (\ref{19}) is determining the velocity $\tilde{\mathbf{U}}_{i}\left(\tilde{X}_{i}, 
 \tilde{Y}_{i}, T, \varepsilon\right)$  of the reference flow.   The proper selection of $\tilde{\mathbf{U}}_{i}\left(\tilde{X}_{i}, 
 \tilde{Y}_{i}, T, \varepsilon\right)$ will provide the development of the kinetic theory
in which the coupled self-consistent  evolution of the plasma and of  the electrostatic  field on the microscales, 
as well as on the macroscales of a bulk of plasma, will be involved. 
In what follows we use the expansion
\begin{eqnarray}
&\displaystyle 
\tilde{\mathbf{E}_{i}}\left(\tilde{X}_{i}+\tilde{R}_{ix}\left(\tilde{X}_{i}, \varepsilon^{-1}\tilde{X}_{i}, 
\varepsilon^{-1}\tilde{Y}_{i}, \varepsilon^{-1}T\right), \varepsilon^{-1}X_{i}, \varepsilon^{-1}Y_{i},  \varepsilon^{-1}T\right)
\nonumber
\\ 
&\displaystyle
=\tilde{\mathbf{E}}_{i0}\left(\tilde{X}_{i}, \varepsilon^{-1}X_{i}, \varepsilon^{-1}Y_{i}, 
\varepsilon^{-1}T\right)+\tilde{\mathbf{E}}_{i1}\left(\tilde{X}_{i}, \varepsilon^{-1}X_{i}, \varepsilon^{-1}Y_{i}, 
\varepsilon^{-1}T\right),
\label{21}
\end{eqnarray}
where
\begin{eqnarray}
&\displaystyle 
\tilde{\mathbf{E}}_{i1}\left(\tilde{X}_{i}, \varepsilon^{-1}X_{i}, \varepsilon^{-1}Y_{i}, \varepsilon^{-1}T\right)
=\frac{\partial \tilde{\mathbf{E}}_{i0}}{\partial \tilde{X}_{i}}\tilde{R}_{ix}\left(\tilde{X}_{i}, 
\varepsilon^{-1}\tilde{X}_{i}, 	\varepsilon^{-1}\tilde{Y}_{i}, \varepsilon^{-1}T\right).
\label{22}
\end{eqnarray}
assuming  the small  displacement,  $|\tilde{R}_{ix}|\ll L_{E}$, of an ion in the crossed magnetic field $\mathbf{B}_{0}$ 
and electric field $\tilde{\mathbf{E}}_{i}$ inhomogeneous  along $\tilde{X}_{i}$. 
With  expansion (\ref{21}), the Vlasov equation (\ref{19}) for the ion distribution function 
$F_{i}\left(\tilde{\mathbf{v}}_{i}, \tilde{X}_{i}, \tilde{Y}_{i}, T\right)$  becomes
\begin{eqnarray}
&\displaystyle 
\frac{\partial F_{i}}{\partial T}+\tilde{v}_{ix}\frac{\partial F_{i}} {\partial \tilde{X}_{i}}
+\tilde{v}_{iy}\frac{\partial F_{i}}{\partial \tilde{Y}_{i}}
\nonumber 
\\ 
&\displaystyle	-\left(\tilde{v}_{ix}+\tilde{U}_{ix}\left(\tilde{X}_{i}, \tilde{Y}_{i}, 
T, \varepsilon\right)\right)\int\limits_{T_{0}}^{T}\frac{\partial \tilde{V}_{ix}} {\partial X_{i}}dT_{1}
\frac{\partial F_{i}}{\partial\tilde{X}_{i}}-\left(\tilde{v}_{ix}+\tilde{U}_{ix}\left(\tilde{X}_{i}, \tilde{Y}_{i}, 
T, \varepsilon\right)\right)\int\limits_{T_{0}}^{T}\frac{\partial \tilde{V}_{iy}} {\partial	X_{i}}dT_{1}
\frac{\partial F_{i}}{\partial\tilde{Y}_{i}}
\nonumber 
\\ 
&\displaystyle
+\left(\frac{\omega_{ci}}{\varepsilon}\tilde{v}_{iy}-\tilde{v}_{ix}\frac{\partial \tilde{V}_{ix}} {\partial X_{i}}\right)
\frac{\partial F_{i}}{\partial \tilde{v}_{ix}}-\left(\frac{\omega_{ci}}{\varepsilon}
+\frac{\partial \tilde{V}_{iy}} {\partial X_{i}}\right)\tilde{v}_{ix}
\frac{\partial F_{i}}{\partial \tilde{v}_{iy}}
\nonumber 	
\\ 
&\displaystyle
+\frac{e_{i}}{m_{i}}\left(\frac{1}{\varepsilon}\tilde{\mathbf{E}}_{i0}\left(\tilde{X}_{i}, 
\varepsilon^{-1}\tilde{X}_{i}, 
\varepsilon^{-1}\tilde{Y}_{i}, \varepsilon^{-1}T\right)-\tilde{\nabla}\Phi\left(\tilde{X}_{i}, \tilde{Y}_{i}, T\right)\right)
\frac{\partial F_{i}}{\partial \tilde{\mathbf{v}}_{i}}
\nonumber 	
\\ 
&\displaystyle
-\left[\frac{\partial\tilde{U}_{ix}\left(\tilde{X}_{i}, T, \varepsilon\right)}{\partial T}
-\frac{\omega_{ci}}	{\varepsilon}\tilde{U}_{iy}-\frac{e_{i}}{\varepsilon m_{i}}\tilde{E}_{i1x}\left(\tilde{X}_{i}, 
\varepsilon^{-1}X_{i}, \varepsilon^{-1}Y_{i}, 	\varepsilon^{-1}T\right)\right]
\frac{\partial F_{i}}{\partial \tilde{v}_{ix}}
\nonumber 
\\ 
&\displaystyle
-\left[\frac{\partial\tilde{U}_{iy}\left(\tilde{X}_{i}, T, \varepsilon\right)}{\partial T}
+\frac{\omega_{ci}}	{\varepsilon}\tilde{U}_{ix}-\frac{e_{i}}{\varepsilon m_{i}}\tilde{E}_{i1y}\left(\tilde{X}_{i}, 
\varepsilon^{-1}X_{i},  \varepsilon^{-1}Y_{i}, \varepsilon^{-1}T\right)\right]\frac{\partial F_{i}}{\partial 
\tilde{v}_{iy}}=0.
\label{23}
\end{eqnarray}
At time $T\gtrsim \gamma^{-1}$, electric field $\tilde{\mathbf{E}}_{i0}$ becomes the random 
function of the initial phase $\theta\left(\mathbf{k}\right)$ with the averaged 
$\left\langle \tilde{E}_{i0x}\left(\tilde{X}_{i}, \tilde{x}_{i}, \tilde{y}_{i}, t\right)\right\rangle=\left\langle 
\tilde{E}_{i0y}\left(\tilde{X}_{i}, \tilde{x}_{i}, \tilde{y}_{i}, t\right)\right\rangle=0$, zero mean value.
The velocities $\bar{U}_{ix}\left(\tilde{X}_{i}, t\right)=\left\langle \tilde{U}_{ix}\left(\tilde{X}_{i}, \tilde{x}_{i}, 
\tilde{y}_{i}, t\right)\right\rangle$ and $\bar{U}_{iy}\left(\tilde{X}_{i}, t\right)=
\left\langle \tilde{U}_{iy}\left(\tilde{X}_{i}, \tilde{x}_{i}, \tilde{y}_{i}, t\right)\right\rangle$ are determined 
by the equations
\begin{eqnarray}
	&\displaystyle
	\frac{\partial\bar{U}_{ix}}{\partial t}
	-\omega_{ci}\bar{U}_{iy}
	=\frac{e_{i}}{m_{i}}\left\langle  \tilde{E}_{i1x}\left(\tilde{X}_{i}, \tilde{x}_{i}, \tilde{y}_{i}, t\right)\right\rangle  ,
	\label{24}
	\\ 
	&\displaystyle
	\frac{\partial\bar{U}_{iy}}{\partial t}
	+\omega_{ci}\bar{U}_{ix}
	=\frac{e_{i}}{m_{i}} \left\langle \tilde{E}_{i1y}\left(\tilde{X}_{i}, \tilde{x}_{i}, \tilde{y}_{i}, t\right)\right\rangle .
	\label{25}
\end{eqnarray}
The averaged over the fast time $t\gg \omega_{ci}^{-1}$ solutions to Eqs. (\ref{24}) and (\ref{25}) for the velocities 
of  the reference flow, $\langle\langle \bar{U}_{ix}\left(\tilde{X}_{i}, t\right)\rangle\rangle=\bar{U}_{ix}\left(\tilde{X}_{i}\right)$ 
and  $\langle\langle \bar{U}_{iy}\left(\tilde{X}_{i}, t\right)\rangle\rangle=\bar{U}_{iy}\left(\tilde{X}_{i}\right)$, where 
the double angle brackets $ \langle\langle ...\rangle\rangle $ designates the averaging of the expression in it over the fast time 
$t=\frac{T}{\varepsilon}$ and over the initial phases $\theta\left(\mathbf{k}\right) $ of the microscale perturbations, are 
derived  in Appendix A.  With allowing for that  for a tokamak plasma  $\omega_{ci}\gg \varepsilon |\frac{\partial 
\bar{U}_{iy}}{\partial\tilde{X}_{i}}|, \varepsilon |\frac{\partial \bar{U}_{iy}}{\partial\tilde{X}_{i}}|$, Eq. (\ref{23}) 
obtains a simple form:
\begin{eqnarray}
	&\displaystyle 
	\frac{\partial F_{i}}{\partial T}+\tilde{v}_{ix}\left(1-\int\limits_{T_{0}}^{T}
	\frac{\partial \bar{U}_{ix}(\tilde{X}_{i})}{\partial \tilde{X}_{i}}dT_{1}\right)\frac{\partial F_{i}}{\partial 
		\tilde{X}_{i}}+\left(\tilde{v}_{iy}-\tilde{v}_{ix}\int\limits _{T_{0}}^{T}\frac{\partial \bar{U}_{iy}(\tilde{X}_{i})}
	{\partial \tilde{X}_{i}}dT_{1}\right)\frac{\partial F_{i}}{\partial \tilde{Y}_{i}}
	\nonumber 
	\\ 
	&\displaystyle
	+\tilde{v}_{iy}\frac{\omega_{ci}}{\varepsilon}\frac{\partial F_{i}}{\partial \tilde{v}_{ix}}
	-\tilde{v}_{ix}\frac{\omega_{ci}}{\varepsilon}\frac{\partial F_{i}}{\partial \tilde{v}_{iy}}
	\nonumber 
	\\ 
	&\displaystyle
	+\left(\frac{e_{i}}{\varepsilon m_{i}}\tilde{\mathbf{E}}_{i0}\left(\tilde{X}_{i}, \tilde{x}_{i}, \tilde{y}_{i}, t\right)
	-\frac{e_{i}}{m_{i}}\tilde{\nabla} \Phi\left(\tilde{X}_{i}, \tilde{Y}_{i}, Z, T \right)\right)
	\frac{\partial F_{i}}{\partial \tilde{\mathbf{v}}_{i}}=0,
	\label{26}
\end{eqnarray}f
in which the terms on the order of  $O\left(\left|\tilde{E}_{i0}\right|^{4}\right)$ are neglected.

In Eq. (\ref{26}), we will present the ion distribution function $F_{i}$ in the form 
\begin{eqnarray}
&\displaystyle 
F_{i}=\bar{F}_{i}\left(\tilde{\mathbf{v}}_{i}, \tilde{X}_{i}, \tilde{Y}_{i}, T\right)
+f_{i}\left(\tilde{\mathbf{v}}_{i}, \tilde{X}_{i}, \varepsilon^{-1}X_{i}, \varepsilon^{-1}Y_{i}, \varepsilon^{-1}Z_{i},
\varepsilon^{-1}T\right),
\label{27}
\end{eqnarray}
 where $\bar{F}_{i}=\left\langle F_{i}\right\rangle $ is the averaged $F_{i}$ over the ensemble of the initial phases, and  
 $f_{i}$ is the microscale perturbation of $F_{i}$ with $\left\langle f_{i}\right\rangle=0$.  For deriving the simplest solution 
 to Eq. (\ref{26}) for $\bar{F}_{i}\left(\tilde{\mathbf{v}}_{i}, \tilde{X}_{i}, 
\tilde{Y}_{i}, T\right)$ and   for $f_{i}\left(\tilde{\mathbf{v}}_{i}, \tilde{X}_{i}, \varepsilon^{-1}X_{i}, 
\varepsilon^{-1}Y_{i}, \varepsilon^{-1}Z_{i}, \varepsilon^{-1}T\right)$ we use the expansions for the velocities
\begin{eqnarray}
&\displaystyle
\bar{U}_{ix}\left(\tilde{X}_{i}\right)=\bar{U}_{ix}^{(0)}\left(\tilde{X}_{i}^{(0)}\right)
+ \bar{U}_{ix}'\left(\tilde{X}_{i}^{(0)}\right)\left(\tilde{X}_{i}-\tilde{X}_{i}^{(0)}\right),
\label{28}
\\
&\displaystyle
\bar{U}_{iy}\left(\tilde{X}_{i}\right)=\bar{U}_{iy}^{(0)}\left(\tilde{X}_{i}^{(0)}\right)
+ \bar{U}_{iy}'\left(\tilde{X}_{i}^{(0)}\right)\left(\tilde{X}_{i}-\tilde{X}_{i}^{(0)}\right).
\label{29}
\end{eqnarray}
In what follows, we consider the case of the uniform velocity compressing rate,  
$\bar{U}_{ix}'\left(\tilde{X}_{i}^{(0)}\right)=const$, and of the uniform velocity shearing rate,
$\bar{U}_{iy}'\left(\tilde{X}_{i}^{(0)}\right)=const$, and put  $\tilde{X}_{i}^{(0)}=0$.
With expansions (\ref{28}) and (\ref{29}), the equation for  
$\bar{F}_{i}\left(\tilde{\mathbf{v}}_{i}, \tilde{X}_{i}, \tilde{Y}_{i}, T\right)$, which determines the slow macroscale 
evolution of  $F_{i}$, is derived by averaging of Eq. (\ref{26}) over the ensemble of the initial phases,
\begin{eqnarray}
&\displaystyle 
\frac{\partial \bar{F}_{i}}{\partial T}+\tilde{v}_{ix}\left(1-\bar{U}'_{ix}T\right)\frac{\partial \bar{F}_{i}}
{\partial \tilde{X}_{i}}
+\left(\tilde{v}_{iy}-\tilde{v}_{ix}\bar{U}'_{iy}T\right)\frac{\partial \bar{F}_{i}}{\partial \tilde{Y}_{i}}
\nonumber 
\\ 
&\displaystyle
+\tilde{v}_{iy}\frac{\omega_{ci}}{\varepsilon}\frac{\partial \bar{F}_{i}}{\partial \tilde{v}_{ix}}
-\tilde{v}_{ix}\frac{\omega_{ci}}{\varepsilon}\frac{\partial \bar{F}_{i}}{\partial \tilde{v}_{iy}}
\nonumber 
\\ 
&\displaystyle
-\frac{e_{i}}{m_{i}}\tilde{\nabla} \Phi\left(\tilde{X}_{i}, \tilde{Y}_{i}, Z, T \right)
\frac{\partial \bar{F}_{i}}{\partial \tilde{\mathbf{v}}_{i}}=-\frac{e_{i}}{\epsilon m_{i}}\left\langle 
\tilde{\mathbf{E}}_{i0}\left(\tilde{\mathbf{r}}_{i}, t, \tilde{X}_{i}, T\right)
\frac{\partial f_{i}}{\partial \tilde{\mathbf{v}}_{i}}\right\rangle .
\label{30}
\end{eqnarray}
This equation involves the well known quasilinear effect of the microscale turbulence on the resonant ions, which is 
responsible for the local  processes of the anomalous diffusion and anomalous heating of the resonant ions. 
Also, Eq.   (\ref{30}) involves the macroscale response of  ions on the  spatially inhomogeneous 
sheared-compressed flows, resulted from the average  motion  of ions in the electric field of the microturbulence 
inhomogeneous on the macroscale.   Solution of this equation is presented in Sec. \ref{sec4}.

The fast microscale evolution of $F_{i}$ is determined by the equation for  
\begin{eqnarray}
	&\displaystyle f_{i}\left(\tilde{\mathbf{v}}_{i}, \tilde{X}_{i}, 
\varepsilon^{-1}X_{i}, \varepsilon^{-1}Y_{i}, \varepsilon^{-1}Z_{i}, \varepsilon^{-1}T\right)
=f_{i}\left(\tilde{\mathbf{v}}_{i}, \tilde{\mathbf{r}}_{i},  t, \tilde{X}_{i} \right),
\nonumber
\\
&\displaystyle
\frac{\partial f_{i}}{\partial t}+\tilde{v}_{ix}\left(1-\bar{u}'_{ix}t\right)\frac{\partial f_{i}}{\partial 
\tilde{x}_{i}}+\left(\tilde{v}_{iy}-\tilde{v}_{ix}\bar{u}'_{iy}t\right)\frac{\partial f_{i}}{\partial 
\tilde{y}_{i}}+\omega_{ci}\tilde{v}_{iy}\frac{\partial f_{i}}{\partial \tilde{v}_{ix}}
-\omega_{ci}\tilde{v}_{ix}\frac{\partial f_{i}}{\partial \tilde{v}_{iy}}
\nonumber
\\  
&\displaystyle
-\frac{e_{i}}{m_{i}}\nabla_{\tilde{\mathbf{r}}_{i}}\tilde{\varphi}_{i0}\left(\tilde{\mathbf{r}}_{i},  \tilde{X}_{i}, 
t\right)\frac{\partial} {\partial \tilde{\mathbf{v}}_{i}}\left(\bar{F}_{i}\left(\tilde{\mathbf{v}}_{i}, \tilde{X}_{i}, 
\tilde{Y}_{i}, T\right)+f_{i}\left(\tilde{\mathbf{v}}_{i}, \tilde{\mathbf{r}}_{i},  t, \tilde{X}_{i} \right)\right)
=0,
\label{31} 
\end{eqnarray}
where $\tilde{\varphi}_{i0}$ is the electrostatic potential of the microscale turbulence,
\begin{eqnarray} 
&\displaystyle
\tilde{\mathbf{E}}_{i0}\left( \tilde{\mathbf{r}}_{i}, \tilde{X}_{i}, t\right)=-\nabla_{\tilde{\mathbf{r}}_{i}} 
\tilde{\varphi}_{i0}\left( \tilde{\mathbf{r}}_{i}, \tilde{X}_{i},  t\right),
\label{32} 
\end{eqnarray}
and  $\bar{u}'_{ix}$,  $\bar{u}'_{iy}$  in Eq.   (\ref{31})  denotes   the derivatives of   $\bar{U}_{ix}$,  
$\bar{U}_{iy}$ over the microscale co-ordinate $\tilde{x}_{i}=\varepsilon^{-1}\tilde{X}_{i}$ and the identity 
$\bar{u}'_{ix}t=\bar{U}'_{ix}T$ is used. In Eq.  (\ref{31}),
the variables  $\tilde{X}_{i}$, $\tilde{Y}_{i}$, $Z$, $T$ enter as the parameters.  Equations  (\ref{30})  and (\ref{31}) 
presents the two-scale expansion of the ion Vlasov equation in the  frame of references co-moving with the 
ion convective  flow with flow velocities inhomogeneous along the coordinate  $\tilde{X}_{i}$. 

As  it follows from Eqs. (\ref{122})  and (\ref{123}),  the electron convective velocities $\bar{U}_{ex}$ and 
$\bar{U}_{ey}$ are negligible small and are assumed here to be equal to zero.  Therefore, the  equations for  
$\bar{F}_{e}$ and for $f_{e}$ are determined in the laboratory frame in a form
\begin{eqnarray}
&\displaystyle 
\frac{\partial \bar{F}_{e}}{\partial T}+\tilde{v}_{ex}\frac{\partial \bar{F}_{e}}
{\partial \tilde{X}_{e}}+\tilde{v}_{ey}\frac{\partial \bar{F}_{e}}{\partial \tilde{Y}_{e}}
-\frac{e_{e}}{m_{e}}\tilde{\nabla} \Phi\left(\tilde{X}_{e}, \tilde{Y}_{e}, Z, T \right)
\frac{\partial \bar{F}_{e}}{\partial \tilde{\mathbf{v}}_{e}}
\nonumber 
\\ 
&\displaystyle
= -\frac{e}{\varepsilon m_{e}}\left\langle \tilde{\mathbf{E}}_{e0}\left(\tilde{\mathbf{r}}_{e}, t, \tilde{X}_{e}, T\right)
\frac{\partial f_{e}}{\partial \tilde{\mathbf{v}}_{e}}\right\rangle ,
\label{33}
\end{eqnarray}
\begin{eqnarray}
&\displaystyle
\frac{\partial f_{e}}{\partial t}+\tilde{v}_{ex}\frac{\partial f_{e}}{\partial 	\tilde{x}_{e}}+\tilde{v}_{ey}\frac{\partial 
f_{e}}{\partial \tilde{y}_{ie}}+\omega_{ce}\tilde{v}_{ey}
\frac{\partial f_{e}}{\partial \tilde{v}_{ex}}	-\omega_{ce}\tilde{v}_{ex}\frac{\partial f_{e}}{\partial \tilde{v}_{ey}}
\nonumber
\\  
&\displaystyle
-\frac{e}{m_{e}}\nabla_{\tilde{\mathbf{r}}_{e}}\tilde{\varphi}_{e0}\left(\tilde{\mathbf{r}}_{e},  \tilde{X}_{e}, t\right)
\frac{\partial }{\partial \tilde{\mathbf{v}}_{e}}\left(\bar{F}_{e}\left(\tilde{\mathbf{v}}_{e}, \tilde{X}_{e}, \tilde{Y}_{e}, 
T\right)+f_{e}\left(\tilde{\mathbf{v}}_{e}, \tilde{\mathbf{r}}_{e},  t, \tilde{X}_{e} \right)\right)=0,
\label{34} 
\end{eqnarray}
where $\tilde{\mathbf{E}}_{e0}\left( \tilde{\mathbf{r}}_{e}, \tilde{X}_{e}, t\right)$ and $\tilde{\varphi}_{e0}$  are 
the electric field and the electrostatic potential of the microscale turbulence determined in the electron frame,
\begin{eqnarray} 
&\displaystyle
\tilde{\mathbf{E}}_{e0}\left( \tilde{\mathbf{r}}_{e}, \tilde{X}_{e}, t\right)=-\nabla_{\tilde{\mathbf{r}}_{e}} 
\tilde{\varphi}_{e0}\left( \tilde{\mathbf{r}}_{e}, \tilde{X}_{e},  t\right).
\label{35} 
\end{eqnarray}

The system of Eqs. (\ref{30}),  (\ref{31}), (\ref{33}),  (\ref{34}), and the Poisson equations 
for the macroscale potential $\Phi$ and for the potential of the microscale turbulence  $\varphi$ compose 
the two-scale Vlasov-Poisson system, which describe the back-reaction effects of the convective flows on 
the microturbulence (Eqs.  (\ref{31}), (\ref{34}) ), and the  
macroscale plasma respond on the development convective flows in plasma (Eqs.  (\ref{30}),  (\ref{33})).

\section{The evolution of the microscale turbulence in the macroscale convective flows generated by 
the microturbulence}\label{sec3}
In the guiding center coordinates $\hat{x}_{i}$, $\hat{y}_{i}$, determined by the relations
\begin{eqnarray}
&\displaystyle
\tilde{x}_{i}=\hat{x}_{i}-\frac{\hat{v}_{i\bot}}{\omega_{ci}}\left(1-\bar{u}_{ix}'t\right)\sin\left(\phi_{1}
-\omega_{ci}t\right)+O\left(\frac{\bar{U}_{ix}}{\omega_{ci}}\right),
\label{36}
\\  
&\displaystyle 
\tilde{y}_{i}=\hat{y}_{i}+\frac{\hat{v}_{i\bot}}{\omega_{ci}}\cos\left(\phi_{1}-\omega_{ci}t\right)
+\frac{\hat{v}_{i\bot}}{\omega_{ci}}\sin\left(\phi_{1}-\omega_{ci}t\right)\bar{u}_{iy}'t
+O\left(\frac{\bar{U}_{iy}}{\omega_{ci}}\right),
\label{37} 
\end{eqnarray}
the linearized Vlasov equation (\ref{31}) for 
$f_{i}\left(\hat{v}_{i\bot}, \phi_{1}, v_{z}, \hat{x}_{i}, \hat{y}_{i}, z,
\hat{X}_{i}, t\right)$ has a form
\begin{eqnarray} 
&\displaystyle
\frac{\partial  f_{i}}{\partial t}= \frac{e_{i}}{m_{i}} \left[ -\frac{\omega_{ci}}{\hat{v}_{i\bot}}
\frac{\partial \tilde{\varphi}_{i0}}{\partial \phi_{1}}\frac{\partial \bar{F}_{i0}}{\partial  \hat{v}_{i\bot}}+
\frac{1}{\omega_{ci}}\left(1-\bar{u}'_{ix}t \right) 
\frac{\partial \tilde{\varphi}_{i0}}{\partial \hat{y}_{i}}\frac{\partial \bar{F}_{i0}}{\partial  \hat{X}_{i}}
+ \frac{\partial \tilde{\varphi}_{i0}}{\partial z_{i}}\frac{\partial \bar{F}_{i0}}{\partial  v_{iz}}\right] ,
\label{38} 
\end{eqnarray}
where the potential $\tilde{\varphi}_{i0}$ is equal to
\begin{eqnarray} 
&\displaystyle
\tilde{\varphi}_{i0}\left(\tilde{x}_{i},  \tilde{y}_{i}, z, \tilde{X}_{i }, t\right)=\frac{1}{\left(2\pi\right)^{3}}\int 
dk_{\tilde{x}_{i}}dk_{\tilde{y}_{i}}dk_{z}
\tilde{\varphi}_{i0}\left(\tilde{\mathbf{k}}_{i}, \tilde{X}_{i}, 
t\right)e^{ik_{\tilde{x}_{i}}\tilde{x}_{i}+ik_{\tilde{x}_{i}}k_{\tilde{y}_{i}}+ik_{z}z_{i}}
\nonumber
\\  
&\displaystyle 
=\frac{1}{\left(2\pi\right)^{3}}\int dk_{\tilde{x}_{i}}dk_{\tilde{y}_{i}}dk_{z}
\tilde{\varphi}_{i0}\left(\tilde{\mathbf{k}}_{i}, \hat{X}_{i}, t\right)
\nonumber
\\  
&\displaystyle 
\times\sum_{n=-\infty}^{\infty}J_{n}\left(\frac{\hat{k}_{i\bot}\left(t\right)\hat{v}_{i\bot}}{\omega_{ci}}\right)
e^{ik_{\tilde{x}_{i}}\hat{x}_{i}+ik_{\tilde{y}_{i}}\hat{y}_{i}+ik_{z}z_{i}
-in\left(\phi_{1}-\omega_{ci}t-\chi_{i}\left(t\right)\right)},
\label{39} 
\end{eqnarray}
with  $\tilde{\mathbf{k}}_{i}=\left(k_{\tilde{x}_{i}}, k_{\tilde{y}_{i}}, k_{z}\right)$ and  
$\hat{k}_{i\bot}\left(t\right)$ and 
$\chi_{i}\left(t\right)$ determined by the relations
\begin{eqnarray}
&\displaystyle
\hat{k}^{2}_{i\bot}\left(t\right)=\left(k_{\tilde{x}_{i}}-\left(k_{\tilde{x}_{i}}\bar{u}'_{ix}
+k_{\tilde{y}_{i}}\bar{u}'_{iy}\right)t\right)^{2}+k^{2}_{\tilde{y}_{i}}, \qquad	
\sin \chi_{i}\left(t\right)=\frac{k_{\tilde{y}_{i}}}{k_{i\bot}\left(t\right)}.	
\label{40}
\end{eqnarray}
The solution to  Eq. (\ref{38})  with nonmodal microscale potential (\ref{39}),
\begin{eqnarray}
&\displaystyle
f_{i}\left(\hat{v}_{i\bot}, \phi_{1}, v_{z}, \hat{x}_{i}, \hat{y}_{i}, z,  \hat{X}_{i}, t\right)=i\frac{e_{i}}{ m_{i}}	
\frac{1}{\left(2\pi\right)^{3}}\int dk_{\tilde{x}_{i}}dk_{\tilde{y}_{i}}dk_{z}
\nonumber	
\\  
&\displaystyle
\times
\sum_{n_{1}=-\infty}^{\infty}\int\limits_{t_{0}}^{t} dt_{1}\tilde{\varphi}_{i0}\left(\tilde{\mathbf{k}}_{i},
\hat{X}_{i}, t_{1}\right) J_{n_{1}}\left(\frac{\hat{k}_{i\bot}\left(t_{1}\right)\hat{v}_{i\bot}}{\omega_{ci}}\right)	
\nonumber	
\\  
&\displaystyle
\times 
\left[\frac{n_{1}\omega_{ci}}{\hat{v}_{i\bot}}\frac{\partial \bar{F}_{i0}}{\partial  
\hat{v}_{i\bot}}+\frac{k_{\tilde{y}_{i}}}{\omega_{ci}}\left(1-\bar{u}'_{ix}t_{1}\right)
\frac{\partial \bar{F}_{i0}}{\partial \hat{X}_{i}}+k_{z}\frac{\partial \bar{F}_{i0}}{\partial v_{iz}}\right]
\nonumber	
\\  
&\displaystyle
\times 
e^{ik_{\tilde{x}_{i}}\hat{x}_{i}+ik_{\tilde{y}_{i}}\hat{y}_{i}+ik_{z}z_{i}+ik_{z}v_{iz}t_{1}
-in_{1}\left(\phi_{1}-\omega_{ci}t_{1}-\chi_{i}\left(t_{1}\right)\right)}.
\label{41} 
\end{eqnarray}
displays two time-dependent effects of the sheared-compressed flow on the  temporal evolution of the perturbation $f_{i}$ 
of the ion distribution function.  The first one is the effect of the time dependence of the argument 
$\hat{k}_{i\bot}\left(t_{1}\right)\hat{v}_{i\bot}/\omega_{ci}$ of the Bessel function $J_{n}$.  It was found in
Refs. \cite{Mikhailenko_6, Mikhailenko_4}, that the  static spatial structure $\sim \exp\left(ik_{x}x+ik_{y}y+ik_{z}z\right)$  
of the perturbation in the sheared  flow may be determined only in the frame convected with a  sheared flow. 
In the laboratory frame, this  perturbation is 
observed as the sheared mode with a time dependent structure, which stems from the continuous distortion with time the 
perturbation by the sheared flow. Therefore,  an ion, the Larmor orbit of which experiences 
negligible small distortion in a sheared flow across the magnetic field,  interacts with perturbation which 
has a time dependent structure caused by the sheared flow. Equation  (\ref{41}) extends this basic linear nondissipative 
nonmodal effect on the interaction of  ions with  wave in the two-dimensional sheared-compressed convective flow. 
The second  effect is a new nonmodal time dependent effect of the compressed flow along $X_{i}$ on the ion drift  along 
coordinate $Y_{i}$. 

 The Fourier transformed over coordinates $\hat{x}_{i}$, $\hat{y}_{i}$ the perturbation \\  
 $n_{i}\left(\tilde{\mathbf{k}}_{i}, \tilde{X}_{i}, t \right)
=\int d\hat{\mathbf{v}}_{i}f_{i}\left(\hat{v}_{i\bot}, v_{z}, \tilde{\mathbf{k}}_{i}, \tilde{X}_{i}, 
t \right)$  of ions with the Maxwellian distribution\\ $\bar{F}_{i0}\left(\mathbf{v}_{i}, \hat{X}_{i}, T\right)$  with 
inhomogeneous ions density and ion temperature,
\begin{eqnarray}
&\displaystyle 
\bar{F}_{i0}\left(\mathbf{v}_{e}, \hat{X}_{i}\right)=\frac{n_{i0}\left(\hat{X}_{i}\right)}{\left(2\pi 
v^{2}_{Ti}\left(\hat{X}_{i}\right)\right)^{3/2}}\exp\left(-\frac{v^{2}_{i\bot}+v^{2}_{z}}{2v^{2}_{Ti}
\left(\hat{X}_{i}\right)}\right),
\label{42} 
\end{eqnarray}
was found in the form
\begin{eqnarray}
&\displaystyle 
n_{i}\left(\tilde{\mathbf{k}}_{i}, \tilde{X}_{i}, t \right)=\frac{e_{i}n_{0i}\left(\hat{X}_{i}\right)}{T_{i}\left(\hat{X}_{i}\right)}
\sum_{n=-\infty}^{\infty}\int_{t_{0}}^{t}dt_{1}\varphi_{i}\left(\tilde{\mathbf{k}}_{i}, \tilde{X}_{i}, t_{1}\right)
\nonumber
\\  
&\displaystyle 
\times \exp\left[  -\frac{1}{2}k^{2}_{z}v^{2}_{Ti}\left(t-t_{1}\right)^{2}-in\omega_{ci}\left(t-t_{1}\right)
-in\left(\chi_{i}\left(t\right)-\chi_{i}\left(t_{1}\right)\right)\right] 
\nonumber
\\  
&\displaystyle 
\times\left\lbrace 
-\left(1+i\left(1-\bar{u}'_{ix}t_{1}\right)\frac{1}{2}\omega_{Ti}\left(t-t_{1}\right)\right)k_{z}^{2}v_{Ti}^{2}\left(t-t_{1}\right) 
A^{(0)}_{in}\left(t, t_{1}\right)\right.
\nonumber
\\  
&\displaystyle
\left. + \Big(\left(1-\bar{u}'_{ix}t_{1}\right)ik_{y}v_{di} -in\omega_{ci} \Big)	
A^{(0)}_{in}\left(t, t_{1}\right)+ \omega_{Ti}\left(1-\bar{u}'_{ix}t_{1}\right)A^{(1)}_{in}\left(t, t_{1}\right)
\right\rbrace,
\label{43} 
\end{eqnarray}
where
\begin{eqnarray} 
	&\displaystyle
	A^{(0)}_{in}\left(t, t_{1}\right)= I_{n}\left(\hat{k}_{i\bot}\left(t\right)\hat{k}_{i\bot}\left(t_{1}\right)\rho^{2}_{i} 
	\right)e^{-\frac{1}{2}\rho^{2}_{i}\left(\hat{k}^{2}_{i\bot}\left(t\right)+\hat{k}^{2}_{i\bot}\left(t_{1}\right)\right)},
	\label{44} 
\end{eqnarray} 
and  
\begin{eqnarray} 
	&\displaystyle
	A^{(1)}_{in}\left(t, t_{1}\right)=e^{-\frac{1}{2}\rho^{2}_{i}\left(\hat{k}^{2}_{i\bot}\left(t\right)
		+\hat{k}^{2}_{i\bot}\left(t_{1}\right)\right)} \left[-\frac{1}{2}\rho^{2}_{i}\left(\hat{k}^{2}_{i\bot}\left(t\right)
	+\hat{k}^{2}_{i\bot}\left(t_{1}\right)\right)I_{n}\left(\hat{k}_{i\bot}
	\left(t\right)\hat{k}_{i\bot}\left(t_{1}\right)\rho^{2}_{i}\right)\right.
	\nonumber
	\\  
	&\displaystyle 
	\left.+\frac{1}{2}\rho^{2}_{i}\hat{k}_{i\bot}\left(t\right)\hat{k}_{i\bot}\left(t_{1}\right)\left(
	I_{n-1}\left(\hat{k}_{i\bot}\left(t\right)\hat{k}_{i\bot}\left(t_{1}\right)\rho^{2}_{i}\right)
	-I_{n+1}\left(\hat{k}_{i\bot}\left(t\right)\hat{k}_{i\bot}\left(t_{1}\right)\rho^{2}_{i}\right)\right)\right].
	\label{45} 
\end{eqnarray} 
The integration by parts of the first term in the braces on the right-hand of Eq.  (\ref{43}) gives the expression for 
$n_{i}\left(\tilde{\mathbf{k}}_{i}, \tilde{X}_{i}, t \right)$ which appears to be  more convenient and transparent 
for the solution of  the integral equations for the potential $\varphi_{i}\left(\tilde{\mathbf{k}}_{i}, \tilde{X}_{i}, t_{1}\right)$
or $\varphi_{e}\left(\tilde{\mathbf{k}}_{e}, \tilde{X}_{e}, t_{1}\right) $, derived from the Poisson equation, 
as the initial value problem\cite{Mikhailenko_5, Mikhailenko_4},
\begin{eqnarray}
&\displaystyle 
n_{i}\left(\tilde{\mathbf{k}}_{i}, \tilde{X}_{i}, t \right)=-\frac{e_{i}n_{0i}\left(\hat{X}_{i}\right)}{T_{i}\left(\hat{X}_{i}\right)}
\int_{t_{0}}^{t}dt_{1}\frac{d}{dt_{1}}\varphi_{i}\left(\tilde{\mathbf{k}}_{i}, \tilde{X}_{i}, t_{1}\right)
\nonumber
\\  
&\displaystyle 
+\frac{e_{i}n_{0i}\left(\hat{X}_{i}\right)}{T_{i}\left(\hat{X}_{i}\right)}\sum_{n=-\infty}^{\infty}\int_{t_{0}}^{t}dt_{1}
e^{-\frac{1}{2}k^{2}_{z}v^{2}_{Ti}\left(t-t_{1}\right)^{2}}\frac{d}{dt_{1}}\left[\varphi_{i}\left(\tilde{\mathbf{k}}_{i}, 
\tilde{X}_{i}, t_{1}\right)	A^{(0)}_{in}\left(t, t_{1}\right)\right.
\nonumber
\\  
&\displaystyle 
\left. \times \left(1+i\frac{\omega_{Ti}}{2}\left(1-\bar{u}'_{ix}t_{1}\right)\left(t-t_{1}\right)\right)e^{
-in\omega_{ci}\left(t-t_{1}\right)-in\left(\chi_{i}\left(t\right)-\chi_{i}\left(t_{1}\right)\right)} \right]
\nonumber
\\  
&\displaystyle 
+i\frac{e_{i}n_{0i}\left(\hat{X}_{i}\right)}{T_{i}\left(\hat{X}_{i}\right)}\sum_{n=-\infty}^{\infty}\int_{t_{0}}^{t}dt_{1}
\varphi_{i}\left(\tilde{\mathbf{k}}_{i}, \tilde{X}_{i}, t_{1}\right) 
A_{in}^{(0)}\left(t, t_{1}\right)\left(k_{\tilde{y}_{i}}v_{di}\left(1-\bar{u}'_{ix}t\right)-n\omega_{ci}\right)
\nonumber
\\  
&\displaystyle 
\times 
\exp\left[
 -\frac{1}{2}k^{2}_{z}v^{2}_{Ti}\left(t-t_{1}\right)^{2}-in\omega_{ci}\left(t-t_{1}\right)
-in\left(\chi_{i}\left(t\right)-\chi_{i}\left(t_{1}\right)\right)
\right] 
\nonumber
\\  
&\displaystyle 
+i\frac{e_{i}n_{0i}\left(\hat{X}_{i}\right)}{T_{i}\left(\hat{X}_{i}\right)}\sum_{n=-\infty}^{\infty}
\int_{t_{0}}^{t}dt_{1}\varphi_{i}\left(\tilde{\mathbf{k}}_{i}, \tilde{X}_{i}, t_{1}\right) 
e^{-\frac{1}{2}k^{2}_{z}v^{2}_{Ti}\left(t-t_{1}\right)^{2}}
\nonumber
\\  
&\displaystyle 
\times 
\omega_{Ti}\left(1-\bar{u}'_{ix}t_{1}\right)A^{(1)}_{in}\left(t, t_{1}\right)-Q_{i}\left(\tilde{\mathbf{k}}_{i}, \tilde{X}_{i}, t,\ 
t_{0}\right) .
\label{46} 
\end{eqnarray}
where $Q_{i}\left(\tilde{\mathbf{k}}_{i}, \tilde{X}_{i}, t,\ t_{0}\right) $ is equal to
\begin{eqnarray}
&\displaystyle
Q_{i}\left(\tilde{\mathbf{k}}_{i}, \hat{X}_{i}, t,\ t_{0}\right) =\frac{e_{i}n_{0i}\left(\hat{X}_{i}\right)}{T_{i}
\left(\hat{X}_{i}\right)}\varphi_{i}\left(\tilde{\mathbf{k}}_{i}, \hat{X}_{i}, t_{0}\right)
\nonumber
\\  
&\displaystyle 
\times
\left[
1+\sum_{n=-\infty}^{\infty}I_{n}\left(\hat{k}_{i\bot}\left(t\right)\hat{k}_{i\bot}\left(t_{0}\right)\rho^{2}_{i}\right)
\left(1+\frac{i}{2}\omega_{Ti}
\left(1-\bar{u}'_{ix}t_{0}\right) 
\left(t-t_{0}\right)\right)
\right] 
\nonumber
\\  
&\displaystyle 
\times 
e^{
-\frac{\rho^{2}_{i}}{2}\left(\hat{k}^{2}_{i\bot}\left(t\right)+\hat{k}^{2}_{i\bot}\left(t_{0}\right)\right)
-\frac{k^{2}_{z}v^{2}_{Ti}}{2}\left(t-t_{0}\right)^{2}-in\omega_{ci}\left(t-t_{0}\right)
-in\left(\chi_{i}\left(t\right)-\chi_{i}\left(t_{0}\right)\right)
}.
\label{47} 
\end{eqnarray}
In Equation (\ref{43}) and in what follows,   $v_{d\alpha}\left(X_{\alpha}\right)=\left(cT_{\alpha}/eB\right) d\ln 
n_{0\alpha}\left(X_{\alpha}\right)/dX_{\alpha}$ is the ion(electron)  $\left( \alpha=i(e)\right)$ diamagnetic velocity, 
$\omega_{Ti}=k_{y}v_{di}\left(X_{i}\right)\eta_{i}$,  $\eta_{\alpha}=d\ln T_{\alpha}/d\ln n_{\alpha}$,  \\ 
$\rho_{i}=v_{Ti}/\omega_{ci}$ is the ion thermal 
Larmor radius respectively.  For the low frequency electrostatic perturbations, for which  $\frac{\partial \tilde{\varphi}_{i0}}{\partial 
 t}\ll\omega_{ci}\tilde{\varphi}_{i0}$, only the terms with $n=n_{1}=0$ should be retained in summations over
 $n$ and $n_{1}$ in Eqs. (\ref{39}) and (\ref{41}). 

In the electron guiding center coordinates $\hat{x}_{e},  \hat{y}_{e}$, determined by the relations
\begin{eqnarray}
&\displaystyle
\tilde{x}_{e}=\hat{x}_{e}-\frac{\hat{v}_{e\bot}}{\omega_{ce}}\sin\left(\phi_{1}-\omega_{ce}t\right), \qquad	
\tilde{y}_{e}=\hat{y}_{e}+\frac{\hat{v}_{e\bot}}{\omega_{ce}}\cos\left(\phi_{1}-\omega_{ce}t\right),	
\label{48}
\end{eqnarray}
the linearized Vlasov equation (\ref{34}) for $f_{e}\left(\hat{v}_{e\bot},  \phi_{1}, v_{z}, \hat{x}_{e}, \hat{y}_{e}, z,
\hat{X}_{e}, t\right)$ has a form
\begin{eqnarray} 
&\displaystyle
\frac{\partial  f_{e}}{\partial t}= \frac{e}{m_{e}} \left[ \frac{1}{\omega_{ce}}
\frac{\partial \varphi_{e0}}{\partial \hat{y}_{e}}\frac{\partial \bar{F}_{e0}}{\partial  \hat{X}_{e}}
+ \frac{\partial \varphi_{e0}}{\partial z_{e}}\frac{\partial \bar{F}_{e0}}{\partial  v_{ez}}\right] ,
\label{49} 
\end{eqnarray}
where $\varphi_{e0}=\varphi_{e0}\left( \mathbf{r}_{e}, \tilde{X}_{e}, t\right)$.  The solution 
$f_{e}\left(\hat{v}_{e\bot}, v_{z},  k_{\tilde{x}_{e}}, k_{\tilde{y}_{e}}, k_{z},  \tilde{X}_{e}, t\right)$ 
to  Eq. (\ref{49}),  Fourier transformed over $\tilde{x}_{e}$,  $\tilde{y}_{e}$ , 
\begin{eqnarray}
&\displaystyle
f_{e}\left(\hat{v}_{e\bot}, v_{z}, k_{\tilde{x}_{e}}, k_{\tilde{y}_{e}}, k_{z}, \tilde{X}_{e}, 
t\right)=i\frac{e}{m_{e}}	
\int_{t_{0}}^{t}dt_{1}\varphi_{e}\left(\mathbf{k}_{e}, \tilde{X}_{e}, t_{1}\right)	
\nonumber	
\\  
&\displaystyle
\times
\left[
\frac{k_{\tilde{y}_{e}}}{\omega_{ce}}
\frac{\partial \bar{F}_{e0}}{\partial \tilde{X}_{e}}+k_{z}\frac{\partial \bar{F}_{e0}}{\partial v_{ez}}
\right]
e^{-ik_{z}v_{ez}\left(t-t_{1}\right)},
\label{50} 
\end{eqnarray}
determines the temporal evolution of the separate spatial Fourier harmonic  of the perturbation  
$f_{e}\left(\hat{v}_{e\bot}, v_{z}, k_{\tilde{x}_{e}}, k_{\tilde{y}_{e}}, k_{z}, \tilde{X}_{e}, t\right)$ 
in the laboratory frame. 

The separate harmonic of the long wavelength, $k_{e\bot}\rho_{e}\ll 1$,  electron density perturbation 
$n_{e}\left(\tilde{\mathbf{k}}_{e}, \tilde{X}_{e}, t_{1}\right)=\int 
d\hat{\mathbf{v}}_{e}f_{e}\left(\hat{\mathbf{v}}_{e}, 
\tilde{\mathbf{k}}_{e}, \tilde{X}_{e}, t\right)$  for the Maxwellian distribution of electrons, with inhomogeneous 
density and with inhomogeneous temperature,
\begin{eqnarray}
&\displaystyle 
\bar{F}_{e0}\left(\mathbf{v}_{e}, \hat{X}_{e}\right)=\frac{n_{e0}\left(\hat{X}_{e}\right)}{\left(2\pi 
v^{2}_{Te}\left(X_{e}\right)\right)^{3/2}}\exp\left(-\frac{v^{2}_{i\bot}+v^{2}_{z}}{2v^{2}_{Te}\left(X_{e}\right)}\right),
\label{51} 
\end{eqnarray}
is given  by the relation derived for $k_{e\bot}\rho_{e}=0$,
\begin{eqnarray}
&\displaystyle 
n_{e}\left(\tilde{\mathbf{k}}_{e}, \hat{X}_{e}, t\right)=\frac{en_{0e}\left(\hat{X}_{e}\right)}{T_{e}\left(\hat{X}_{e}\right)}
\int_{t_{0}}^{t}dt_{1}\varphi_{e}\left(\tilde{\mathbf{k}}_{e}, 
\hat{X}_{e}, t_{1}\right)e^{-\frac{1}{2}k^{2}_{z}v^{2}_{Te}\left(t-t_{1}\right)^{2}}
\nonumber	
\\  
&\displaystyle
\times 
\left(-\left(1+i\frac{1}{2}\omega_{Te}\left(t-t_{1}\right)\right)k^{2}_{z}v^{2}_{Te}
\left(t-t_{1}\right)+ik_{y}v_{de}\right),
\label{52} 	
\end{eqnarray}
which after integration in Eq. (\ref{52}) by parts becomes
\begin{eqnarray}
&\displaystyle 
n_{e}\left(\tilde{\mathbf{k}}_{e}, \hat{X}_{e},t\right)=\frac{en_{0e}\left(\hat{X}_{e}\right)}{T_{e}\left(\hat{X}_{e}\right)}
\int_{t_{0}}^{t}dt_{1}\left\lbrace -\frac{d\varphi_{e}\left(\tilde{\mathbf{k}}_{e}, \hat{X}_{e},t_{1}\right)}{dt_{1}}\right.
\nonumber	
\\  
&\displaystyle
\left.+e^{-\frac{1}{2}k^{2}_{z}v^{2}_{Te}\left(t-t_{1}\right)^{2}}\left[\frac{d\varphi_{e}\left(\tilde{\mathbf{k}}_{e}, \hat{X}_{e}, 
t_{1}\right)}{dt_{1}}\left(1+\frac{i}{2}\omega_{Te}\left(t-t_{1}\right)\right)\right.\right.
\nonumber	
\\  
&\displaystyle
\left.\left.+ik_{\tilde{y}_{e}}v_{de}\left(1-\frac{1}{2}\eta_{e}\right)\varphi_{e}\left(\tilde{\mathbf{k}}_{e}, 
\hat{X}_{e}, t_{1}\right)\right]\right\rbrace 
-Q_{e}\left(\tilde{\mathbf{k}}_{e}, \tilde{X}_{e}, t,\ t_{0}\right) ,
\label{53} 
\end{eqnarray}
where	$Q_{e}\left(\tilde{\mathbf{k}}_{e}, \tilde{X}_{e}, t,\ t_{0}\right) $ is equal to
\begin{eqnarray}
&\displaystyle
Q_{e}\left(\tilde{\mathbf{k}}_{e}, \tilde{X}_{e}, t,\ t_{0}\right)  
= -\frac{en_{e0}\left(\hat{X}_{e}\right)}{T_{e}\left(\hat{X}_{e}\right)}\varphi_{e}\left(\tilde{\mathbf{k}}_{e}, 
\hat{X}_{e}, t_{0}\right)
\nonumber 
\\ 
&\displaystyle
\times\left[1+\left(1+\frac{i}{2}\omega_{Te}\left(t-t_{0}\right)\right)
e^{-\frac{1}{2}k^{2}_{z}v^{2}_{Te}\left(t-t_{0}\right)^{2}}\right].
\label{54} 
\end{eqnarray}
It is important to note that for electrons, which have thermal velocity much larger than the phase velocity  along the magnetic field  of the 
perturbations, $k^{2}_{z}v^{2}_{Te}\left(t-t_{1}\right)^{2}\gg 1$ occurs for the time $t-t_{1}$ much less than the period of  the
perturbations.  Therefore the approximation of the adiabatic electrons corresponds in Eqs. (\ref{53})  and (\ref{54})  to the 
exponentially small  values of  $e^{-\frac{1}{2}k^{2}_{z}v^{2}_{Te}\left(t-t_{1}\right)^{2}}$ and 
$e^{-\frac{1}{2}k^{2}_{z}v^{2}_{Te}\left(t-t_{0}\right)^{2}}$.

Because the perturbation of the ion density (\ref{43}) is determined as a function of the potential 
$\varphi_{i}\left(\tilde{x}_{i}, \tilde{y}_{i}, z, t \right)$, whereas the perturbation  of the electron density (\ref{51})  
is determined as a function of the potential $\varphi_{e}\left(\tilde{x}_{e}, \tilde{y}_{e}, z, t \right)$,  
the Poisson equation (\ref{2}) for the electric field 
of the electrostatic microscale plasma turbulence may be  derived as the equation  for the potential 
$\varphi_{e}\left(\tilde{x}_{e}, \tilde{y}_{e}, z, t \right)$,  determined in variables $\tilde{x}_{e}, \tilde{y}_{e}$ 
 of the laboratory frame, or as the equation  
for the potential $\varphi_{i}\left(\tilde{x}_{i}, \tilde{y}_{i}, z, t \right)$,  determined in variables $\tilde{x}_{i}, \tilde{y}_{i}$  
of the ion convective frame. In this section, we derive the Poisson equation for the microscale potential 
 $\varphi_{e}\left(\tilde{x}_{e}, \tilde{y}_{e}, z, t \right)$, 
\begin{eqnarray}
&\displaystyle 
\frac{\partial^{2}\varphi_{e}\left(\tilde{x}_{e}, \tilde{y}_{e}, z, \tilde{X}_{e}, t\right)}
{\partial^{2}\tilde{x}_{e}}+ \frac{\partial^{2}	\varphi_{e}\left(\tilde{x}_{e}, \tilde{y}_{e}, z,  \tilde{X}_{e}, t\right)}
{\partial^{2}\tilde{y}_{e}}+\frac{\partial^{2} \varphi_{e}\left(\tilde{x}_{e}, \tilde{y}_{e}, \tilde{X}_{e}, z, t\right)}
{\partial^{2}\tilde{z}_{e}}
\nonumber 
\\ 
&\displaystyle
= 
- 4\pi\left[e_{i}n_{i} \left(\tilde{x}_{i}, \tilde{y}_{i}, z, \tilde{X}_{i}, t\right)	
-|e|n_{e} \left(\tilde{x}_{e}, \tilde{y}_{e}, z,
\tilde{X}_{e},  t\right)\right].
\label{55} 
\end{eqnarray} 

The Fourier transform of Eq.  (\ref{55})  over  $\tilde{x}_{e}$,  $\tilde{y}_{e}$ and $z_{e}$,
\begin{eqnarray}
&\displaystyle
\left(k^{2}_{\tilde{x}_{e}}+k^{2}_{\tilde{y}_{e}} +k^{2}_{z_{e}}\right)\varphi_{e}\left(\mathbf{k}_{e},  
\tilde{X}_{e}, t \right)
=4\pi e_{i}n^{(e)}_{i}\left(\tilde{\mathbf{k}}_{e}, \tilde{X}_{e}, t\right)
+4\pi en_{e}\left(\tilde{\mathbf{k}}_{e}, \tilde{X}_{e}, 
t\right),
\label{56} 
\end{eqnarray} 
contains the Fourier transform $n^{(e)}_{i}\left(\tilde{\mathbf{k}}_{e}, \tilde{X}_{e}, t\right)$
 of the  ion density perturbation $n_{i} \left(\tilde{x}_{i}, \tilde{y}_{i}, z, \tilde{X}_{i}, t\right)$ 
 performed over $\tilde{x}_{e}$ and $\tilde{y}_{e}$, i. e
\begin{eqnarray}
&\displaystyle
n^{(e)}_{i}\left(\tilde{\mathbf{k}}_{e}, \tilde{X}_{e}, t\right)=\int d\tilde{x}_{e}\int d\tilde{y}_{e} 
n_{i} \left(\tilde{x}_{i}, \tilde{y}_{i}, k_{z}, \tilde{X}_{i}, t\right)e^{-ik_{\tilde{x}_{e}}\tilde{x}_{e}-ik_{\tilde{y}_{e}
}\tilde{y}_{e}}
\nonumber 
\\ 
&\displaystyle
=\int d\tilde{x}_{e}\int d\tilde{y}_{e} n_{i} \left(\tilde{x}_{i}, \tilde{y}_{i}, k_{z}, \tilde{X}_{i}, 
t\right)
e^{-ik_{\tilde{x}_{e}}\tilde{x}_{i}-ik_{\tilde{y}_{e}}\tilde{y}_{i}-ik_{\tilde{x}_{e}}\left(\tilde{x}_{e}
-\tilde{x}_{i}\right)-ik_{\tilde{y}_{e}}\left(\tilde{y}_{e}-\tilde{y}_{i}\right)}
\nonumber 
\\ 
&\displaystyle
=\int d\tilde{x}_{i}\int d\tilde{y}_{i} n_{i} \left(\tilde{x}_{i}, \tilde{y}_{i}, k_{z}, \tilde{X}_{i}, t\right) 
\left|\frac{\partial\left(\tilde{x}_{e},  
\tilde{y}_{e}\right)}{\partial\left(\tilde{x}_{i},  \tilde{y}_{i}\right)}\right|
e^{-ik_{\tilde{x}_{e}}\tilde{x}_{i}-ik_{\tilde{y}_{e}}\tilde{y}_{i}-ik_{\tilde{x}_{e}}\left(\tilde{x}_{e}-\tilde
{x}_{i}\right)-ik_{\tilde{y}_{e}}\left(\tilde{y}_{e}-\tilde{y}_{i}\right)}.
\label{57} 
\end{eqnarray} 
 It follows from Eqs. (\ref{7})  and (\ref{9}) that  
 \begin{eqnarray}
 &\displaystyle 
 \tilde{x}_{e}= \tilde{x}_{i}\left(1+\bar{u}'_{ix}t\right)+\bar{U}^{(0)}_{ix}{(0)}t,
 \label{58} 
\\
 &\displaystyle 
\tilde{x}_{i}=\frac{\tilde{x}_{e}-\bar{U}^{(0)}_{ix}t}{1+\bar{u}'_{ix}t},
\label{59} 
\end{eqnarray}
and
\begin{eqnarray}
&\displaystyle
\tilde{y}_{e}=\tilde{y}_{i}+\left(\bar{U}^{(0)}_{iy}+\bar{u}'_{iy}\tilde{x}_{i}\right)t,
\label{60} 
\\
&\displaystyle
\tilde{y}_{i}=\tilde{y}_{e}-\frac{\bar{u}'_{iy}t}{1+\bar{u}'_{ix}t}\tilde{x}_{e}       
-\bar{U}^{(0)}_{iy}t +\frac{\bar{u}'_{iy}\bar{U}^{(0)}_{iy}t^{2}}{1+\bar{u}'_{ix}t}.
\label{61} 
\end{eqnarray}
With  Eqs. (\ref{58}) and (\ref{60}), Eq. (\ref{57}) becomes
\begin{eqnarray}
&\displaystyle
n^{(e)}_{i}\left(\tilde{\mathbf{k}}_{e}, \tilde{X}_{e}, t\right)=\int d\tilde{x}_{i}\int d\tilde{y}_{i} n_{i} \left(\tilde{x}_{i}, 
\tilde{y}_{i}, k_{z}, \tilde{X}_{i}, t\right) \left|1+\bar{u}'_{ix}t\right|
\nonumber 
\\ 
&\displaystyle
\times \exp\left[-ik_{\tilde{x}_{e}}\left(\tilde{x}_{i}+\left(\bar{U}^{(0)}_{ix}+\bar{u}'_{ix}\tilde{x}_{i}\right)t\right)	
-ik_{\tilde{y}_{e}}\left(\tilde{y}_{i}+\left(\bar{U}^{(0)}_{iy}+\bar{u}'_{iy}\tilde{x}_{i}\right)t\right)\right]	
\nonumber 
\\ 
&\displaystyle
=e^{-ik_{\tilde{x}_{e}}\bar{U}^{(0)}_{ix}t-ik_{\tilde{y}_{e}}\bar{U}^{(0)}_{iy}t}\left|1+\bar{u}'_{ix}t\right|
n_{i}\left(k_{\tilde{x}_{e}}\left(1+\bar{u}'_{ix}t\right)
+k_{\tilde{y}_{e}}\bar{u}'_{iy}t, k_{\tilde{y}_{e}},  k_{z}, t\right).
\label{62} 
\end{eqnarray}

Equation (\ref{44}) for  $n_{i}\left(\tilde{\mathbf{k}}_{i}, \tilde{X}_{i}, t \right)$  contains the Fourier transform
$\varphi_{i}\left(\tilde{\mathbf{k}}_{i}, \tilde{X}_{i}, t\right)$ of the  potential $\varphi_{i}\left(\tilde{x}_{i}, \tilde{y}_{i}, 
\tilde{X}_{i}, t_{1}\right)$. The connection relation of $\varphi_{i}\left(\tilde{\mathbf{k}}_{i}, \tilde{X}_{i}, t_{1}\right)$ 
with $\varphi_{e}\left(\tilde{\mathbf{k}}_{e}, \tilde{X}_{e}, t_{1}\right)$ follows from the relation
\begin{eqnarray}
&\displaystyle
\varphi_{i}\left(\tilde{\mathbf{k}}_{i}, \tilde{X}_{i}, t_{1}\right)=\int d\tilde{x}_{i}\int 
d\tilde{y}_{i}\varphi_{i}\left(\tilde{x}_{i}, 
\tilde{y}_{i}, k_{z},\tilde{X}_{i}, t_{1}\right)e^{-ik_{\tilde{x}_{i}}\tilde{x}_{i}-ik_{\tilde{y}_{i}}\tilde{y}_{i}}
\nonumber 
\\ 
&\displaystyle
=\int d\tilde{x}_{e}\int d\tilde{y}_{e}\varphi_{e}\left(\tilde{x}_{e}, \tilde{y}_{e}, k_{z}, \tilde{X}_{e}, t_{1}\right)
\left|\frac{\partial\left(\tilde{x}_{i}\left(t_{1}\right), 	\tilde{y}_{i}\left(t_{1}\right)\right)}{\partial\left(\tilde{x}_{e},  
\tilde{y}_{e}\right)}\right|
\nonumber 
\\ 
&\displaystyle
\times 
e^{-ik_{\tilde{x}_{i}}\tilde{x}_{e}-ik_{\tilde{y}_{i}}\tilde{y}_{e}-ik_{\tilde{x}_{i}}
\left(\tilde{x}_{i}-\tilde{x}_{e}\right)-ik_{\tilde{y}_{i}}\left(\tilde{y}_{ie}-\tilde{y}_{e}\right)}.
\nonumber 
\\ 
&\displaystyle
=\frac{1}{4\pi^{2}}\frac{1}{\left|1+\bar{u}'_{ix}t_{1}\right|}\int dk_{\tilde{x}_{e}}\int dk_{\tilde{y}_{e}}
\varphi_{e}\left(k_{\tilde{x}_{e}}, k_{\tilde{y}_{e}}, k_{z}, \tilde{X}_{i}, t_{1}\right)
\nonumber 
\\ 
&\displaystyle
\times\int d\tilde{x}_{e}\int d\tilde{y}_{e}\exp \left [ i\left(k_{\tilde{x}_{e}}-k_{\tilde{x}_{i}}\right)\tilde{x}_{e} 
+ i\left(k_{\tilde{y}_{e}}-k_{\tilde{y}_{i}}\right)\tilde{y}_{e} \right.
\nonumber 
\\ 
&\displaystyle
\left.
-ik_{\tilde{x}_{i}}\left(\tilde{x}_{i}-\tilde{x}_{e}\right) 
-ik_{\tilde{y}_{i}}\left(\tilde{y}_{i}-\tilde{y}_{e}\right) \right].
\label{63} 
\end{eqnarray}
The integrating of  Eq. (\ref{63})  over  $\tilde{x}_{e}$,  $\tilde{y}_{e}$, in which the relations
\begin{eqnarray}
&\displaystyle
\tilde{x}_{i}\left(t_{1}\right)-\tilde{x}_{e}=-\frac{\bar{U}^{(0)}_{ix}t_{1}}{1+\bar{u}'_{ix}t_{1}}-
\frac{\bar{u}'_{ix}t_{1}}{1+\bar{u}'_{ix}t_{1}}\tilde{x}_{e}=b_{0x}\left(t_{1}\right)+b_{1x}\left(t_{1}\right)\tilde{x}_{e},
\label{64} 
\\
&\displaystyle
\tilde{y}_{i}\left(t_{1}\right)-\tilde{y}_{e}=-\bar{U}^{(0)}_{ix}t_{1}+\frac{\bar{u}'_{iy}
\bar{U}^{(0)}_{ix}t^{2}_{1}}{1+\bar{u}'_{ix}t_{1}}-\frac{\bar{u}'_{iy}t_{1}}{1+\bar{u}'_{ix}t_{1}}\tilde{x}_{e}
=b_{0y}\left(t_{1}\right)+b_{1y}\left(t_{1}\right)\tilde{x}_{e},	
\label{65} 
\end{eqnarray}	
are employed, gives the relation
\begin{eqnarray}
&\displaystyle	
\varphi_{i}\left(k_{\tilde{x}_{i}}, k_{ \tilde{y}_{i}}, k_{z}, \tilde{X}_{i}, t_{1}\right)=\frac{1}{\left 
|1+\bar{u}'_{ix}t_{1}\right |}
e^{-ik_{\tilde{x}_{i}}b_{0x}\left(t_{1}\right)-ik_{\tilde{y}_{i}}b_{0y}\left(t_{1}\right)}
\nonumber 
\\ 
&\displaystyle
\times
\varphi_{e}\left(k_{\tilde{x}_{i}}\left(1+b_{1x}\left(t_{1}\right)\right)
+k_{\tilde{y}_{i}}b_{1x}\left(t_{1}\right),  k_{\tilde{y}_{i}},  k_{z}, \tilde{X}_{i}, t_{1}\right).
\label{66}
\end{eqnarray}	
It follows from Eq.  (\ref{62}) that the wave numbers of  $n_{i}^{(e)}$, which are conjugate with co-ordinates 
$\tilde{x}_{i}$,  $\tilde{y}_{i}$, are $k _{\tilde{x}_{e}}+\left(k _{\tilde{x}_{e}}\bar{u}'_{ix}+k_{\tilde{y}_{e}}
\bar{u}'_{iy}\right)t$ 
and $k_{\tilde{y}_{e}}$. Applying this result to Eq.  (\ref{62}) gives the presentation of the Fourier transform 
$\varphi_{i}\left(k _{\tilde{x}_{i}}, k _{\tilde{y}_{i}}, k_{z}, \tilde{X}_{i}, t_{1}\right)$ in variables $k _{\tilde{x}_{e}}$,
 $k_{\tilde{y}_{e}}$, $k_{z}$ ,
\begin{eqnarray}
&\displaystyle	
\varphi_{i}\left(k_{\tilde{x}_{i}}, k_{\tilde{y}_{i}}, \tilde{X}_{i}, t_{1}\right)=\frac{1}{\left |1+\bar{u}'_{ix}t_{1}\right |}
\exp\left[ik_{\tilde{x}_{e}}\bar{U}^{(0)}_{ix}t_{1}\left(1+\frac{\bar{u}'_{ix}\left(t-t_{1}\right)}
{1+\bar{u}'_{ix}t_{1}}\right)\right.
\nonumber 
\\ 
&\displaystyle
\left.
+ik_{\tilde{y}_{e}}\bar{U}^{(0)}_{iy}t_{1}\left(1+\frac{\bar{u}'_{iy}\left(t-t_{1}\right)}
{1+\bar{u}'_{ix}t_{1}}\right)\right]
\nonumber 
\\ 
&\displaystyle
\times
\varphi_{e}\left(k_{\tilde{x}_{e}}\left(1+\frac{\bar{u}'_{ix}\left(t-t_{1}\right)}
{1+\bar{u}'_{ix}t_{1}}\right)+k _{\tilde{y}_{e}}\frac{\bar{u}'_{iy}\left(t-t_{1}\right)}
{1+\bar{u}'_{ix}t_{1}}, k _{\tilde{y}_{e}},  k_{z}, \tilde{X}_{i}, t_{1}\right)
\nonumber 
\\ 
&\displaystyle
=\varphi^{(e)}_{i}\left(k _{\tilde{x}_{e}}, k _{\tilde{y}_{e}}, k_{z}, \tilde{X}_{i}, t, t_{1}\right).
\label{67}
\end{eqnarray}	
Equation (\ref{67})  displays, that the separate spatial Fourier harmonic $\varphi_{i}\left(k _{\tilde{x}_{i}}, 
k _{\tilde{y}_{i}}, k_{z}, \tilde{X}_{i}, t_{1}\right)$ of the electrostatic potential, determined in the frame of references, 
which moves with velocities (\ref{28}), (\ref{29}), is  perceived in the electron (laboratory) frame as  the Doppler-shifted 
continuously sheared and compressed mode with time-dependent wave vectors.

For deriving  the  ion density perturbation $n_{i}^{(e)}$,  determined  by  Eq. (\ref{62}),  the potential 
$\varphi_{i}\left(\tilde{\mathbf{k}}_{i}, \tilde{X}_{i}, t_{1}\right) $ in Eq. (\ref{57}), which determines 
$n_{i}\left(\tilde{\mathbf{k}}_{i}, \tilde{X}_{i}, t \right)$, should be changed on $\varphi^{(e)}_{i}\left(k _{\tilde{x}_{e}}, k 
_{\tilde{y}_{e}}, k_{z}, \tilde{X}_{i}, t, t_{1}\right)$ given by Eq. (\ref{67}),
\begin{eqnarray}
&\displaystyle	
n^{(e)}_{i}\left(\tilde{\mathbf{k}}_{e}, \tilde{X}_{e}, t\right)
=-\frac{e_{i}n_{0i}\left(\tilde{X}_{i}\right)}{T_{i}\left(\tilde{X}_{i}\right)}\int\limits_{t_{0}}^{t}dt_{1}\frac{d}{dt_{1}}
\left\lbrace \left| \frac{1+\bar{u}'_{ix}t}{1+\bar{u}'_{ix}t_{1}} \right|
e^{-\frac	{1}{2}k^{2}_{z}v^{2}_{Ti}\left(t-t_{1}\right)^{2}}\right.
\nonumber 
\\ 
&\displaystyle
\left.
\times
 \exp\left[-ik _{\tilde{x}_{e}}\bar{U}^{(0)}_{ix}\left(t-t_{1}\right)\left(1-\frac{\bar{u}'_{ix}t_{1}}
{1+\bar{u}'_{ix}t_{1}}\right)-ik _{\tilde{y}_{e}}\bar{U}^{(0)}_{iy}\left(t-t_{1}\right)
 \left(1-\frac{\bar{u}'_{iy}t_{1}}{1+\bar{u}'_{ix}t_{1}}\right)\right]\right.
\nonumber 
\\ 
&\displaystyle
\left.
\times
\varphi_{e}\left( k _{\tilde{x}_{e}}\left(1+\frac{\bar{u}'_{ix}\left(t-t_{1}\right)}
{1+\bar{u}'_{ix}t_{1}}\right) +k _{\tilde{y}_{e}}\frac{\bar{u}'_{iy}\left(t-t_{1}\right)}
{1+\bar{u}'_{ix}t_{1}}, k_{\tilde{y}_{e}}, k_{z}, \tilde{X}_{i}, t_{1}\right)\right\rbrace 
\nonumber 
\\ 
&\displaystyle
+\frac{e_{i}n_{0i}\left(\tilde{X}_{i}\right)}{T_{i}\left(\tilde{X}_{i}\right)}\sum_{n=-\infty}^{\infty}
\int\limits_{t_{0}}^{t}dt_{1}e^{-\frac	{1}{2}k^{2}_{z}v^{2}_{Ti}\left(t-t_{1}\right)^{2}}
\frac{d}{dt_{1}}\left[ \left| \frac{1+\bar{u}'_{ix}t}{1+\bar{u}'_{ix}t_{1}}\right|
 \left(1+\frac{i}{2}\omega_{Ti}\left(1-u'_{ix}t_{1}\right)\left(t-t_{1}\right)\right)\right.
\nonumber 
\\ 
&\displaystyle
\times 
\exp\left[-ik _{\tilde{x}_{e}}\bar{U}^{(0)}_{ix}\left(t-t_{1}\right)\left(1-\frac{\bar{u}'_{ix}t_{1}}
{1+\bar{u}'_{ix}t_{1}}\right)-ik _{\tilde{y}_{e}}\bar{U}^{(0)}_{iy}\left(t-t_{1}\right)
\left(1-\frac{\bar{u}'_{iy}t_{1}}{1+\bar{u}'_{ix}t_{1}}\right)\right]
\nonumber 
\\ 
&\displaystyle
\times
\varphi_{e}\left( k _{\tilde{x}_{e}}\left(1+\frac{\bar{u}'_{ix}\left(t-t_{1}\right)}
{1+\bar{u}'_{ix}t_{1}}\right) +k _{\tilde{y}_{e}}\frac{\bar{u}'_{iy}\left(t-t_{1}\right)}
{1+\bar{u}'_{ix}t_{1}}, k_{\tilde{y}_{e}}, k_{z}, \tilde{X}_{i}, t_{1}\right)
\nonumber 
\\ 
&\displaystyle
\left.\times
A_{in}^{(0)}\left(t, t_{1}\right)e^{-in\omega_{ci}\left(t-t_{1}\right)
-in\left(\chi_{i}\left(t\right)-\chi_{i}\left(t_{1}\right)\right)} \right]  +
 \nonumber 
 \\ 
 &\displaystyle
+i\frac{e_{i}n_{0i}\left(\tilde{X}_{i}\right)}{T_{i}\left(\tilde{X}_{i}\right)}\sum_{n=-\infty}^{\infty}
 \int\limits_{t_{0}}^{t} dt_{1}\left| \frac{1+\bar{u}'_{ix}t}{1+\bar{u}'_{ix}t_{1}} \right|
 \nonumber 
 \\ 
 &\displaystyle
 \times 
 \exp\left[-ik _{\tilde{x}_{e}}\bar{U}^{(0)}_{ix}\left(t-t_{1}\right)\left(1-\frac{\bar{u}'_{ix}t_{1}}
 {1+\bar{u}'_{ix}t_{1}}\right)-ik _{\tilde{y}_{e}}\bar{U}^{(0)}_{iy}\left(t-t_{1}\right)
 \left(1-\frac{\bar{u}'_{iy}t_{1}}{1+\bar{u}'_{ix}t_{1}}\right)\right]
 \nonumber 
 \\ 
 &\displaystyle
 \times\varphi_{e}\left( k _{\tilde{x}_{e}}\left(1+\frac{\bar{u}'_{ix}\left(t-t_{1}\right)}
 {1+\bar{u}'_{ix}t_{1}}\right) +k _{\tilde{y}_{e}}\frac{\bar{u}'_{iy}\left(t-t_{1}\right)}
 {1+\bar{u}'_{ix}t_{1}}, k_{\tilde{y}_{e}}, k_{z}, \tilde{X}_{i}, t_{1}\right) 
 \nonumber 
 \\ 
 &\displaystyle
 \times 
 e^{ -\frac{1}{2}k_{z}^{2}v^{2}_{Ti}\left(t-t_{1}\right)^{2}-in\omega_{ci}\left(t-t_{1}\right)
 -in\left(\chi_{i}\left(t\right)-\chi_{i}\left(t_{1}\right) \right)}
 \nonumber 
 \\ 
 &\displaystyle
 \times 
 \left[\left(k_{\tilde{y}_{e}}v_{di}\left(1-\bar{u}'_{ix}t_{1} \right)- n\omega_{ci} \right)A^{(0)}_{in}\left(t, t_{1}\right) +\omega_{Ti}\left(1-\bar{u}'_{ix}t_{1}\right)A_{in}^{(1)}\left(t, t_{1}\right)  \right] 
  \nonumber 
\\ 
&\displaystyle
  -Q^{(e)}_{i}\left(\tilde{\mathbf{k}}_{i}, \tilde{X}_{i}, t,\ t_{0}\right) ,
\label{68}
\end{eqnarray}	
where
\begin{eqnarray}
&\displaystyle
k^{2}_{i\bot}\left(t\right)=\left(k_{\tilde{x}_{e}}-\left(k_{\tilde{x}_{e}}\bar{u}'_{ix}
+k_{\tilde{y}_{e}}\bar{u}'_{iy}\right)\bar{u}'_{iy}t^{2}\right)^{2}+k^{2}_{\tilde{y}_{e}},
\nonumber 
\\ 
&\displaystyle 
k^{2}_{i\bot}\left(t_{1}\right)=\left(k_{\tilde{x}_{e}}-\left(k_{\tilde{x}_{e}}\bar{u}'_{ix}
+k_{\tilde{y}_{e}}\bar{u}'_{iy}\right)\bar{u}'_{iy}t_{1}^{2}\right)^{2}+k^{2}_{\tilde{y}_{e}},
\label{69}
\end{eqnarray}	
and $Q^{(e)}_{i}\left(\hat{\mathbf{k}}_{i}, \tilde{X}_{i}, t,\ t_{0}\right) $ is equal to
\begin{eqnarray}
&\displaystyle
Q^{(e)}_{i}\left(\tilde{\mathbf{k}}_{i}, \tilde{X}_{i}, t, t_{0}\right) =\frac{e_{i}n_{0i}\left(\hat{X}_{i}\right)}
{T_{i}\left(\hat{X}_{i}\right)}
 \left|\frac{1+\bar{u}'_{ix}t}{ 1+\bar{u}'_{ix}t_{0}} \right|
\nonumber 
\\ 
&\displaystyle
\times 
\exp\left[-ik _{\tilde{x}_{e}}\bar{U}^{(0)}_{ix}\left(t-t_{0}\right)\left(1-\frac{\bar{u}'_{ix}t_{0}}
{1+\bar{u}'_{ix}t_{0}}\right)-ik _{\tilde{y}_{e}}\bar{U}^{(0)}_{iy}\left(t-t_{0}\right)
\left(1-\frac{\bar{u}'_{iy}t_{0}}{1+\bar{u}'_{ix}t_{0}}\right)\right]
\nonumber 
\\ 
&\displaystyle
\times
\varphi_{e}\left( k _{\tilde{x}_{e}}\left(1+\frac{\bar{u}'_{ix}\left(t-t_{0}\right)}
{1+\bar{u}'_{ix}t_{0}}\right) +k _{\tilde{y}_{e}}\frac{\bar{u}'_{iy}\left(t-t_{0}\right)}
{1+\bar{u}'_{ix}t_{0}}, k_{\tilde{y}_{e}}, k_{z}, \tilde{X}_{i}, t_{0}\right)
\nonumber 
\\ 
&\displaystyle
\times 
\left. \left[1- \sum_{n=-\infty}^{\infty} I_{n}\left(k_{i\perp}\left(t \right)k_{i\perp}\left(t_{0} \right)\rho_{i}^{2} \right) 
e^{-\frac{1}{2}\left(k_{i\perp}^{2}\left(t \right) +k_{i\perp}^{2}\left(t_{0} 
\right)\right)\rho_{i}^{2}-\frac{1}{2}k_{z}^{2}v^{2}_{Ti}\left(t-t_{0}\right)^{2} } \right]  \right\rbrace.
\label{70}
\end{eqnarray}
It follows from Eq. (\ref{53}) for $n_{e}\left(\tilde{\mathbf{k}}_{e}, \tilde{X}_{e}, t\right)$ 
and from Eq. (\ref{68}) for $n^{(e)}_{i}\left(\tilde{\mathbf{k}}_{e}, \tilde{X}_{e}, t\right)$, that 
the Poisson equation  (\ref{56}) is the integral equation for the potential $\varphi_{e}\left(k_{\tilde{x}_{e}}, 
k_{\tilde{y}_{e}},  k_{z}, \tilde{X}_{e}, t \right)$ for the plasma with compressed-sheared convective flows.  

Equations (\ref{67}) and (\ref{68}) display that for the spatially uniform flow, for which 
$\bar{u}'_{ix}=\bar{u}'_{iy}=0$, the spatial Fourier harmonics of the electrostatic potential 
$\varphi_{i}\left(\mathbf{k}_{i}, \tilde{X}_{i}, t\right)$ and of  the  ion density perturbation 
$n_{i}\left(\mathbf{k}_{i}, \tilde{X}_{i}, t\right)$  are perceived in the  electron frame as the 
Doppler-shifted modes
\begin{eqnarray}
&\displaystyle
\varphi_{i}\left(\mathbf{k}_{i},  t_{1} \right) 
 = e^{i(k_{x}\bar{U}_{ix}^{\left(0 \right)}+ k_{y}\bar{U}_{iy}^{\left(0 \right)})t_{1}} 
\varphi_{i}^{\left(e \right) }\left(\mathbf{k}_{e}, t_{1} \right),
\label{71} 
\\  
&\displaystyle 
n_{i}\left(\mathbf{k}_{e}, t \right) = e^{-i(k_{x}\bar{U}_{ix}^{\left(0 \right)}
+k_{y}\bar{U}_{iy}^{\left(0 \right)})t} n_{i}^{\left(e \right) }\left(\mathbf{k}_{i}, t \right).
\label{72}
\end{eqnarray}  
In that case, Eq. (\ref{56}) becomes the integral equations of the convolution type, 
which can be solved by using various kinds of integral transform. In the $t_{0}\rightarrow -\infty$ limit 
explored by the eigenmode analysis, Eq. (\ref{56}) has known solution of the form $\varphi \left(\mathbf{k}, 
\omega\right)\varepsilon \left(\mathbf{k}, \omega \right)=0$ for the Fourier transformed over time variable potential 
$\varphi\left(\mathbf{k}, \omega\right)$ ,  where \cite{Mikhailovskii}
\begin{eqnarray}
&\displaystyle \varepsilon \left(\mathbf{k}, \omega \right)=1+\frac{1}{k^{2}\lambda^{2}_{Di}}\left[ 1+i\sqrt{\frac{\pi}{2}}
\frac{\left(\omega - k_{y}v_{di}\left(1-\frac{\eta_{i}}{2} \right)  
\right)}{k_{z}v_{Ti}}\sum_{n=-\infty}^{\infty}W\left(z_{in}\right)I_{n}\left(k^{2}_{\perp}\rho^{2}_{i}\right)
e^{-\rho^{2}_{i}k^{2}_{\perp}}\right] 
\nonumber
\\ 
&\displaystyle
-\frac{1}{k^{2}\lambda^{2}_{Di}}\frac{k_{y}v_{di}\eta_{i}}{\sqrt{2}k_{z}v_{Ti}}\sum_{n=-\infty}^{\infty}z_{in}
\left(1+i\sqrt{\frac{\pi}{2}}z_{in} W\left(z_{in}\right)\right)	I_{n}\left(k^{2}_{\perp}\rho^{2}_{i}\right)	e^{-\rho^{2}_{i}k^{2}_{\perp}}
\nonumber
\\ 
&\displaystyle
+\frac{1}{k^{2}\lambda^{2}_{Di}}\frac{k_{y}v_{di}\eta_{i}}{k_{z}v_{Ti}}\sum_{n=-\infty}^{\infty}
i\sqrt{\frac{\pi}{2}}W\left(z_{i}\right) k^{2}_{\perp}\rho^{2}_{i}
e^{-k^{2}_{\perp}\rho^{2}_{i}}\left(I_{n}\left(k^{2}_{\perp}\rho^{2}_{i}\right)
-I'_{n}\left(k^{2}_{\perp}\rho^{2}_{i}\right) \right)	
	\nonumber
	\\ 
	&\displaystyle
+\frac{1}{k^{2}\lambda^{2}_{De}}\left( 1+i\sqrt{\frac{\pi}{2}}\frac{\left( \omega -k_{x}\bar{U}_{ix}^{\left(0 \right)}- 
	k_{y}\bar{U}_{iy}^{\left(0 \right)} -k_{y}v_{de}\right) }{k_{z}v_{Te}}
W\left(z_{e}\right)\right),
\label{73}
\end{eqnarray}
and $W\left(z_{i(e)}\right)=e^{ - z_{i(e)}^{2}}\left(1 +\left(2i / \sqrt {\pi }\right)\int\limits_{0}^{z_{i(e)}}
 e^{t^{2}}dt \right)$ is the complex error function with arguments $ z_{in}=\left(\omega-n\omega_{ci}\right)/\sqrt{2}k_{z}v_{Ti}$, 
and  $ z_{e}=\left( \omega -k_{x}\bar{U}_{ix}^{\left(0 \right)}- k_{y}\bar{U}_{iy}^{\left(0 \right)} \right)/\sqrt{2}k_{z}v_{Te}$.
 The solution $\omega\left( \mathbf{k}\right) $ of the dispersion equation $\varepsilon 
\left(\mathbf{k}, \omega \right)=0$ reveals for $n=0$ the low frequency ($\omega\ll \omega_{ci}$) kinetic ion temperature 
gradient (ITG) instabilities\cite{Horton} which are the important contributors to turbulent transport in the tokamak core\cite{Garbet}. 

The presence of the sheared-compressed convective flow introduces substantial complication into integral equation (\ref{54}).   
 It follows from Eq. (\ref{67}), that the modal time dependence $\sim e^{-i\omega\left(\mathbf{k}_{i}\right)t}$ 
 of the potential $\varphi_{e}$ exists  only at the initial stage of the potential 
evolution at which  the sheared-compressed effects of the convected flow are negligible small, i. e. when
$\bar{u}'_{ix}\left(t-t_{0}\right)\ll1$, $\bar{u}'_{iy}\left(t-t_{0}\right)\ll 1$. At a longer time, the time dependence  
of the potential $\varphi_{e}$  becomes very different from a canonical modal form. The exceptional advantage 
of the nonmodal approach, which uses the wavenumber-time variables, is the ability to perform the analysis of the solutions 
to integral equation  (\ref{56}) with the electron and the ion density perturbations  (\ref{53}) and  (\ref{68}) at finite time 
domain and including an arbitrary initial time $t_{0}$. For the approximate solution of Eq.  (\ref{56}) we distinguish  
the characteristic times during which  the  nonmodal effects  becomes important. For the long-wavelength perturbations 
with $k_{i\perp}\left(t _{0}\right)\rho_{i}\ll 1$ the nonmodal effects for the potential $\varphi_{e}$ in Eq. (\ref{64}) 
for the ion density perturbation becomes important at time $t$, for which  $t \gg \left(\bar{u}'_{ix}\bar{u}'_{iy}\right)^{-1/2}$. At time $t_{s}\gg t\gg \left(\bar{u}'_{ix}\bar{u}'_{iy}\right)^{-1/2}$, where  
\begin{eqnarray}
&\displaystyle
t_{s}=\left[\rho_{i} \left( k_{x}\bar{u}'_{ix}+k_{y}\bar{u}'_{iy}\right)\bar{u}'_{iy} \right] ^{-1/2},
\label{74}
\end{eqnarray}
the initially long-wavelength perturbations with $k_{i\perp}\left(t _{0}\right)\rho_{i}\ll 1$,
will be long-wavelength perturbation with $k_{i\perp}\left(t\right)\rho_{i}\ll 1$.  At time $t\gg t_{s}$ these perturbations
will  become  the short wavelength perturbations with $k_{i\perp}\left(t \right)\rho_{i}\gg 1$.
  
The approximate non-modal analysis of the solutions to Eq.  (\ref{56})  may be performed, as it was done for the case of 
the sheared flow in Refs. \cite{Mikhailenko_4, Mikhailenko_5, Mikhailenko_6}, separately for 
the long-wavelength  perturbations for $k_{i\perp}\left(t\right)\rho_{i}\ll 1$ by employing the 
long wave asymptotic 
 \begin{eqnarray}
 &\displaystyle
 I_{0}\left(k_{i\perp}\left(t \right)k_{i\perp}\left(t_{1} \right)\rho_{i}^{2} \right) 
 e^{-\frac{1}{2}\left(k_{i\perp}^{2}\left(t \right) +k_{i\perp}^{2}\left(t_{1} 
 	\right)\right)\rho_{i}^{2}} \approx  1- \frac{1}{2}\left(k_{i\perp}^{2}\left(t \right) +k_{i\perp}^{2}
\left(t_{1}\right)\right)\rho_{i}^{2},
 	\label{75}
 \end{eqnarray}
and for $k_{i\perp}\left(t\right)\rho_{i}\gg 1$   by employing the asymptotic
\begin{eqnarray}
&\displaystyle
I_{0}\left(k_{i\perp}\left(t \right)k_{i\perp}\left(t_{1} \right)\rho_{i}^{2} \right) 
e^{-\frac{1}{2}\left(k_{i\perp}^{2}\left(t \right) +k_{i\perp}^{2}\left(t_{1} 
\right)\right)\rho_{i}^{2}} \approx  \frac{t_{s}}{\sqrt{2\pi t t_{1}}}.
\label{76}
\end{eqnarray}
These solutions  for Eq. (\ref{56}) for the kinetic ion temperature gradient driven instability,  
considered in Ref.\cite{Mikhailenko_4} for the case of the sheared flow across the magnetic field, will be presented soon.
 
\subsection{The Poisson equation for the microturbulence potential 	$\tilde{\varphi}_{i0}
\left(\tilde{\mathbf{r}}_{i}, \tilde{X}_{i}, t\right)$  in the ion convective reference flow}\label{sec3.1} 

It was found in Ref.\cite{Mikhailenko_5} for the sheared flow, and was confirmed above  for the sheared-compressed flow, 
that the separate spatial Fourier harmonic of the  potential of the electrostatic perturbations in the sheared, or in  the 
sheared-compressed flow,  may be determined only in the frame of references convected with such a flow. Therefore, 
for the investigations of the temporal evolution 
of the perturbations in a plasma far from the plasma boundary, the ion component of which moves with spatially inhomogeneous 
flow velocities relative to the electron component, the selection of the ion convective reference  frame for the  Poisson 
equation may be preferable.  In this subsection, we will derive the Poisson equation (\ref{2}) as the equation  for the potential 
$\varphi_{i}\left(\tilde{x}_{i}, \tilde{y}_{i}, z, t \right)$,  
\begin{eqnarray}
&\displaystyle 
\frac{\partial^{2}\varphi_{i}\left(\tilde{x}_{i}, \tilde{y}_{i}, z, \tilde{X}_{i}, t\right)}
{\partial^{2}\tilde{x}_{i}}+ \frac{\partial^{2}	\varphi_{i}\left(\tilde{x}_{i}, \tilde{y}_{i}, z,  \tilde{X}_{i}, t\right)}
{\partial^{2}\tilde{y}_{i}}+\frac{\partial^{2} \varphi_{i}\left(\tilde{x}_{i}, \tilde{y}_{i}, \tilde{X}_{i}, z, t\right)}
{\partial^{2}\tilde{z}_{i}}
\nonumber 
\\ 
&\displaystyle
= - 4\pi\left[e_{i}n_{i} \left(\tilde{x}_{i}, \tilde{y}_{i}, z, \tilde{X}_{i}, t\right)	
-|e|n_{e} \left(\tilde{x}_{e}, \tilde{y}_{e}, z, 	\tilde{X}_{e},  t\right)\right],	
\label{77} 
\end{eqnarray} 
determined in variables $\tilde{x}_{i}, \tilde{y}_{i}$  of the ion convective frame, which moves 
with velocities $\bar{U}_{ix}\left(\tilde{X}_{i}\right)$ and $\bar{U}_{ix}\left(\tilde{X}_{i}\right)$, 
determined by Eqs. (\ref{28})  and  (\ref{29}), relative to the electron (laboratory) frame, and will apply the derived 
equation to the analysis of the temporal evolution of the hydrodynamic ion temperature gradient (HITG) instability 
in a plasma sheared-compressed  flow.  

The Fourier transform of Eq.  (\ref{77})  over  $\tilde{x}_{i}$,  $\tilde{y}_{i}$ and $z$,
\begin{eqnarray}
&\displaystyle
\left(k^{2}_{\tilde{x}_{i}}+k^{2}_{\tilde{y}_{i}} +k^{2}_{z}\right)\varphi_{i}\left(\mathbf{k}_{i},  
\tilde{X}_{i}, t \right) 	=4\pi e_{i}n_{i}\left(\tilde{\mathbf{k}}_{i}, \tilde{X}_{i}, t\right)
+4\pi en^{(i)}_{e}\left(\tilde{\mathbf{k}}_{i}, \tilde{X}_{e}, t\right),
\label{78} 
\end{eqnarray} 
contains  the Fourier transform $n_{i}\left(\tilde{\mathbf{k}}_{i}, \tilde{X}_{i}, t\right)$
 of the ion density perturbation, determined by  Eq.  (\ref{43}), and the  Fourier transform  
$n^{(i)}_{e}\left(\tilde{\mathbf{k}}_{i}, \tilde{X}_{e}, t\right)$ of the  
electron density perturbation $n_{e} \left(\tilde{x}_{e}, \tilde{y}_{e}, z, \tilde{X}_{e}, t\right)$ 
performed over $\tilde{x}_{i}$ and $\tilde{y}_{i}$, i. e
\begin{eqnarray} 
&\displaystyle
n^{(i)}_{e}\left(\tilde{\mathbf{k}}_{i}, \tilde{X}_{e}, t\right)=-\frac{en_{e0}\left(X_{e}\right)}
{T_{e}\left(X_{e}\right)}\int_{t_{0}}^{t}dt_{1} \frac{d\varphi_{i}\left(\tilde{\mathbf{k}}_{i}, 
\tilde{X}_{i}, t_{1}\right)}{dt_{1}}
\nonumber 
\\ 
&\displaystyle
+\frac{en_{e0}\left(X_{e}\right)}{T_{e}\left(X_{e}\right)}
\int_{t_{0}}^{t}dt_{1}e^{-\frac{1}{2}k^{2}_{z}v^{2}_{Te}\left(t-t_{1}\right)^{2}}\left\lbrace
\left(1+\frac{i\omega_{Te}}{2}\left(t-t_{1}\right)\right) 
\frac{d}{dt_{1}}\left[ \left |\frac{1+\bar{u}'_{ix}t_{1}}{1+\bar{u}'_{ix}t}\right |\right.\right.
\nonumber 
\\ 
&\displaystyle
\left.\left. \times  \varphi_{i}\left(k_{\tilde{x_{i}}}-\frac{\bar{u}'_{ix}\left(t-t_{1} \right) }{1+\bar{u}'_{ix}t}
\left(k_{\tilde{x_{i}}}-k_{\tilde{y_{i}}}\bar{u}'_{iy}t_{1} \right) , k_{\tilde{y_{i}}}, \tilde{X_{e}}, t_{1}
\right)\right.\right.
 \nonumber 
 \\ 
 &\displaystyle
\left. \left. \times\exp\left( ik_{\tilde{x_{i}}}\bar{U}^{\left(0 \right) }_{ix}\left(t-t_{1} \right) 
\left(1-\frac{\bar{u}'_{ix}\left(t-t_{1} \right) }{1+\bar{u}'_{ix}t} \right) 
+ ik_{\tilde{y_{i}}}\bar{U}^{\left(0 \right) }_{iy}\left(t-t_{1} \right) 
\left(1-\frac{\bar{u}'_{iy}\left(t-t_{1} \right) }{1+\bar{u}'_{ix}t} \right)\right) \right]\right.
 \nonumber 
 \\ 
 &\displaystyle
\left. + ik_{y}v_{de}\left(1-\frac{\eta_{e}}{2} \right)  \left |\frac{1+\bar{u}'_{ix}t_{1}}{1+\bar{u}'_{ix}t}\right |
 \varphi_{i}\left(  k_{\tilde{x_{i}}}-\frac{\bar{u}'_{ix}\left(t-t_{1} \right) }{1+\bar{u}'_{ix}t}
 \left(k_{\tilde{x_{i}}}-k_{\tilde{y_{i}}}\bar{u}'_{iy}t_{1} \right) , k_{\tilde{y_{i}}}, \tilde{X_{e}}, t_{1} \right)\right.
  \nonumber 
 \\ 
 &\displaystyle
\left. \times  
\exp\left(ik_{\tilde{x_{i}}}\bar{U}^{\left(0 \right) }_{ix}\left(t-t_{1} \right) 
 \left(1-\frac{\bar{u}'_{ix}\left(t-t_{1} \right) }{1+\bar{u}'_{ix}t} \right) 
  + ik_{\tilde{y_{i}}}\bar{U}^{\left(0 \right) }_{iy}\left(t-t_{1} \right) 
  \left(1-\frac{\bar{u}'_{iy}\left(t-t_{1} \right) }{1+\bar{u}'_{ix}t} \right) \right)  \right\rbrace 
    \nonumber 
  \\ 
  &\displaystyle
  -Q_{e}^{\left(i \right) }\left(\mathbf{k}_{i}, t, t_{0} \right),
 \label{79}
\end{eqnarray} 
where
\begin{eqnarray} 
&\displaystyle	
Q_{e}^{\left(i \right) }\left(\mathbf{k}_{i},\tilde{X_{i}},t,t_{0} \right) 
=\frac{en_{e0}\left(X_{e}\right)}{T_{e}\left(X_{e}\right)}\varphi_{i}\left(\mathbf{k}_{i},\tilde{X_{i}},t_{0} \right)
 -\frac{en_{e0}\left(X_{e}\right)}{T_{e}\left(X_{e}\right)}
 \left|1+\bar{u}'_{ix} t_{0}\right|e^{-\frac{1}{2}k^{2}_{z}v^{2}_{Te}\left(t-t_{0} \right) ^{2}}
\nonumber 
\\ 
&\displaystyle
\times\left(1+\frac{i\omega_{Te}}{2}\left( t-t_{0}\right)  \right) 
\exp\left[-\frac{ik_{\tilde{x_{i}}}\bar{U}^{\left(0 \right) }_{ix}t_{0}}{1+\bar{u}'_{ix}t}
-ik_{\tilde{y_{i}}}\bar{U}^{\left(0 \right) }_{iy}t_{0}	\left( 1-	\frac{\bar{u}'_{ix}t_{0}}{1+\bar{u}'_{ix}t}
\right)\right] 
\nonumber 
\\ 
&\displaystyle
\times \varphi_{i}\left( k_{\tilde{x_{i}}}-\frac{\bar{u}'_{ix}\left(t-t_{0} \right) }{1+\bar{u}'_{ix}t}
\left(k_{\tilde{x_{i}}}-k_{\tilde{y_{i}}}\bar{u}'_{iy}t_{0} \right) , k_{\tilde{y_{i}}}, \tilde{X_{e}}, t_{0}\right).
\label{80}
\end{eqnarray}
In this subsection, we apply  Eq. (\ref{78}) to  the analysis of the temporal evolution of the HITG instability in the 
sheared-compressed flow. The  nonmodal analysis of the temporal evolution of this instability in a sheared flow 
was performed earlier in Ref. \cite{Mikhailenko_6}. The simplest form of Eq. (43) for 
$n_{i}\left(\mathbf{k}_{i}\tilde{X}_{i}, t \right) $ for this 
instability, the  frequency of which is much below the ion cyclotron frequency,
is derived by integration by parts of the second term in the braces of Eq. (43) using the identity 
$in\omega_{ci}e^{-in\omega_{ci}\left(t-t_{1}\right)}=\frac{d}{dt}\left(e^{-in\omega_{ci}\left(t-t_{1}\right)}\right)$ 
and retaining  after that integration only one term with $n=0$ in summation over $n$ for the low frequency perturbations for which 
$d\varphi_{i}/dt \ll \omega_{ci}\varphi_{i}$. The HITG instability  develops\cite{Rudakov} in the frequency range 
$k_{z}v_{Ti}<\omega<k_{z}v_{Te}$, where the perturbations driven by this instability experience  weak ion Landau damping 
and strong electron Landau damping. 

In the quasineutrality approximation, for which 
$k^{2}\lambda^{2}_{Di}\ll1$, Eq.  (\ref{78}), corresponding to the approximations used in the linear theory of the 
HITG instability for the steady plasma, has a form
\begin{eqnarray} 
&\displaystyle
\int\limits_{t_{0}}^{t}dt_{1}\frac{d}{dt_{1}}\left\lbrace\left[-\left(1+\frac{T_{i}}{T_{e}}\right)
+A_{0i}\left(t, t_{1}\right)\right]\varphi_{i}\left(\mathbf{k}_{i},  \tilde{X}_{i}, t _{1}\right) \right\rbrace
\nonumber 
\\ 
&\displaystyle
+\int\limits_{t_{0}}^{t}dt_{1}\varphi_{i}\left(\mathbf{k}_{i},  \tilde{X}_{i}, t _{1}\right) 
A_{0i}\left(t, t_{1}\right)
\nonumber 
\\ 
&\displaystyle
\times
\left[i\left(1-\bar{u}'_{ix}t_{1}\right)\left(k_{y}v_{di}-\frac{1}{2}\omega_{Ti}k^{2}_{z}v^{2}_{Ti}
\left(t-t_{1}\right)^{2}\right)-k^{2}_{z}v^{2}_{Ti}\left(t-t_{1}\right)\right]
\nonumber 
\\ 
&\displaystyle
+i\omega_{Ti}\int\limits_{t_{0}}^{t}\varphi_{i}\left(\mathbf{k}_{i},  \tilde{X}_{i}, t _{1}\right) \left(1-\bar{u}'_{ix}t_{1}\right)
A_{i1}\left(t, t_{1}\right)
\nonumber 
\\ 
&\displaystyle
=Q_{e}^{(i)}\left(\mathbf{k}_{i},\tilde{X_{i}},t,t_{0} \right) +Q_{i}\left(\mathbf{k}_{i},\tilde{X_{i}},t,t_{0} \right) 
\label{81}.
\end{eqnarray} 
where
\begin{eqnarray}  
	&\displaystyle
	A_{i0}\left(t, t_{1}\right)= I_{0}\left(\hat{k}_{i\bot}\left(t\right)\hat{k}_{i\bot}\left(t_{1}\right)\rho^{2}_{i} 
	\right)e^{-\frac{1}{2}\rho^{2}_{i}\left(\hat{k}^{2}_{i\bot}\left(t\right)+\hat{k}^{2}_{i\bot}\left(t_{1}\right)\right)},
	\label{82} 
	\\
	&\displaystyle
	A_{i1}\left(t, t_{1}\right)=e^{-\frac{1}{2}\rho^{2}_{i}\left(\hat{k}^{2}_{i\bot}\left(t\right)
		+\hat{k}^{2}_{i\bot}\left(t_{1}\right)\right)}\left(-\frac{1}{2}\rho^{2}_{i}\left(\hat{k}^{2}_{i\bot}\left(t\right)
	+\hat{k}^{2}_{i\bot}\left(t_{1}\right)\right)I_{0}\left(\hat{k}_{i\bot}
	\left(t\right)\hat{k}_{i\bot}\left(t_{1}\right)\rho^{2}_{i}\right)\right.
	\nonumber
	\\  
	&\displaystyle 
	\left.+\rho^{2}_{i}\hat{k}_{i\bot}\left(t\right)\hat{k}_{i\bot}\left(t_{1}\right)
	I_{1}\left(\hat{k}_{i\bot}\left(t\right)\hat{k}_{i\bot}\left(t_{1}\right)\rho^{2}_{i}\right)
	\right),
	\label{83} 
\end{eqnarray} 
where $\hat{k}_{i\bot}\left(t\right)$ and $\hat{k}_{i\bot}\left(t_{1}\right)$ are determined by Eq.  (\ref{40}).
In  Eqs. (\ref{81}), Eqs. (\ref{79}) and (\ref{80})  for the Fourier transform $n^{(i)}_{e}\left(\tilde{\mathbf{k}}_{i}, 
\tilde{X}_{e}, t\right)$ of the electron density perturbation was used  in the asymptotic limit  
$e^{-\frac{1}{2}k^{2}_{z}v^{2}_{Te}\left(t-t_{1}\right)^{2}}\approx 0$, corresponding to the strong electron 
Landau damping. The approximation $e^{-\frac{1}{2}k^{2}_{z}v^{2}_{Te}\left(t-t_{1}\right)^{2}}\approx 1$,
corresponding to the weak ion Landau damping, was used in Eq.  (\ref{81}) for $n_{i}\left(\tilde{\mathbf{k}}_{i}, 
\tilde{X}_{i}, t\right)$.  

In a steady plasma, the perturbations driven by the HITG instability have  long wavelengths  
with  $k_{\bot}\rho_{i}<1$.  In this long-wavelength limit 
\begin{eqnarray} 
&\displaystyle
A_{i0}\left(t, t_{1}\right) \approx 1-\frac{\rho_{i}^{2}}{2}\left(\hat{k}_{i\bot}\left(t\right)+\hat{k}_{i\bot}\left(t_{1}\right)\right)
\approx b_{i}+\rho_{i}^{2}\left(k_{\tilde{x}_{i}}\left(k_{\tilde{x}_{i}}\bar{u}'_{ix}+k_{\tilde{y}_{i}}\bar{u}'_{iy}\right)
\left(t+t_{1}\right)\right.
\nonumber 
\\ 
&\displaystyle
\left.-\frac{1}{2}\left(k_{ix}\bar{u}'_{ix}+k_{iy}\bar{u}'_{ix}\right)^{2}\left(t^{2}+t^{2}_{1}\right)\right),
\label{84} 
\\  
&\displaystyle
A_{i1}\left(t, t_{1}\right) \approx 1-\rho_{i}^{2}\left(\hat{k}_{i\bot}\left(t\right)+\hat{k}_{i\bot}\left(t_{1}\right)\right)
\approx b_{i1}+\rho_{i}^{2}\left(2k_{\tilde{x}_{i}}\left(k_{\tilde{x}_{i}}\bar{u}'_{ix}+k_{\tilde{y}_{i}}\bar{u}'_{iy}\right)
\left(t+t_{1}\right)\right.
\nonumber 
\\ 
&\displaystyle
\left.-\left(k_{ix}\bar{u}'_{ix}+k_{iy}\bar{u}'_{ix}\right)^{2}\left(t^{2}+t^{2}_{1}\right)\right),
\label{85} 
\end{eqnarray} 
where $b_{i}=1-k^{2}_{i\bot}\rho_{i}^{2}$,  $b_{i1}=1-2k^{2}_{i\bot}\rho_{i}^{2}$, $k^{2}_{i\bot}
=k^{2}_{ix}+k^{2}_{iy}$. With expansions  (\ref{84}),  (\ref{85}),  Eq.  (\ref{81}) with new variable $\Psi$, determined by 
$d^{2}\Psi\left(\mathbf{k}_{i},  t \right) /dt^{2}=\varphi_{i}\left(\mathbf{k}_{i},  t \right) $, becomes
\begin{eqnarray} 
&\displaystyle
\left(\frac{T_{i}}{T{e}}+k^{2}_{\bot}\rho_{i}^{2}\right)\int\limits_{t_{0}}^{t}dt_{1}\frac{d^{3}}{dt^{3}_{1}}
\Psi_{i}\left(\mathbf{k}_{i},  \tilde{X}_{i}, t _{1}\right) 
\nonumber 
\\ 
&\displaystyle
-\int\limits_{t_{0}}^{t}dt_{1}\frac{d^{2}}{dt^{2}_{1}}
\Psi_{i}\left(\mathbf{k}_{i},  \tilde{X}_{i}, t _{1}\right) 
\nonumber 
\\ 
&\displaystyle
\times\left[b_{i}\left(i\left(1-\bar{u}'_{ix}t_{1}\right)\left(k_{y}v_{di}-\frac{1}{2}
\omega_{Ti}k^{2}_{z}v^{2}_{Ti}\left(t-t_{1}\right)^{2}\right)-k^{2}_{z}v^{2}_{Ti}\left(t-t_{1}\right)\right)\right.
\nonumber 
\\ 
&\displaystyle
\left.-i\omega_{Ti}k^{2}_{\bot}\rho^{2}_{i}\left(1-\bar{u}'_{ix}t_{1}\right)\right]
\nonumber 
\\ 
&\displaystyle
=\rho_{i}^{2}\int\limits_{t_{0}}^{t}dt_{1}\frac{d}{dt_{1}}\left[ \frac{d^{2}\Psi_{i}\left(\mathbf{k}_{i},  \tilde{X}_{i}, t 
_{1}\right)}{dt^{2}_{1}}\left(k_{\tilde{x}_{i}}\left(k_{\tilde{x}_{i}}\bar{u}'_{ix}+k_{\tilde{y}_{i}}\bar{u}'_{iy}\right)
\left(t+t_{1}\right)\right.\right.
\nonumber 
\\ 
&\displaystyle
\left.\left.-\frac{1}{2}\left(k_{ix}\bar{u}'_{ix}+k_{iy}\bar{u}'_{ix}\right)^{2}\left(t^{2}+t^{2}_{1}\right)
\right) \right]
\nonumber 
\\ 
&\displaystyle
+\rho_{i}^{2}\int\limits_{t_{0}}^{t}dt_{1}\frac{d^{2}\Psi_{i}\left(\mathbf{k}_{i},  \tilde{X}_{i}, t_{1}\right)}{dt^{2}_{1}}
\left(k_{\tilde{x}_{i}}\left(k_{\tilde{x}_{i}}\bar{u}'_{ix}+k_{\tilde{y}_{i}}\bar{u}'_{iy}\right)
\left(t+t_{1}\right)-\frac{1}{2}\left(k_{ix}\bar{u}'_{ix}+k_{iy}\bar{u}'_{ix}\right)^{2}\left(t^{2}+t^{2}_{1}\right)\right)
\nonumber 
\\ 
&\displaystyle
\left(i\left(1-\bar{u}'_{ix}t_{1}\right)\left(k_{y}v_{di}-\frac{1}{2}\omega_{Ti}k^{2}_{z}v^{2}_{Ti}
\left(t-t_{1}\right)^{2}\right)-k^{2}_{z}v^{2}_{Ti}\left(t-t_{1}\right)\right)
\nonumber 
\\ 
&\displaystyle
+Q_{e}^{(i)}\left(\mathbf{k}_{i},\tilde{X_{i}},t,t_{0} \right) +Q_{i}\left(\mathbf{k}_{i},\tilde{X_{i}},t,t_{0} \right) 	
\label{86}.
\end{eqnarray} 
The HITG instability develops in the inhomogeneous steady plasma in the limit of a large value of the parameter 
$\eta_{i}=d\ln T_{i}/d\ln n_{i}=\omega_{Ti}/k_{y}v_{di}\gg 1$. \cite{Rudakov}. The solution to Eq. (\ref{86})  
for the conditions of HITG instability development was presented 
in Ref. \cite{Mikhailenko_6} for the case of the sheared flow for which the compressing rate $\bar{u}'_{ix}=0$.
Integrating second term in the left-hand side of Eq. (\ref{86}) by parts   and neglecting by all terms contained the initial values 
$\Psi_{i}\left(\mathbf{k}_{i},  \tilde{X}_{i}, t_{0}\right)$ at $t=t_{0}$ with the assumption of the exponential growth of 
$\Psi_{i}\left(\mathbf{k}_{i},  \tilde{X}_{i}, t\right)\gg  
\Psi_{i}\left(\mathbf{k}_{i},  \tilde{X}_{i}, t_{0}\right)$ at time $t\gg t_{0}$, and neglecting the right-hand side, proportional to 
$\rho_{i}^{2}$ we obtain for time $\left(\bar{u}'_{ix}\right)^{-1}>t>t_{0}$ the equation
\begin{eqnarray}
&\displaystyle
\int\limits_{t_{0}}^{t}dt_{1}\left[\left(\frac{T_{i}}{T_{e}}+k^{2}_{\bot}\rho_{i}^{2}\right)\frac{d^{3}}{dt^{3}_{1}}
\Psi_{i}\left(\mathbf{k}_{i},  \tilde{X}_{i}, t _{1}\right) \right.
\nonumber 
\\ 
&\displaystyle
\left.+ib_{i}\left(1-\bar{u}'_{ix}t_{1}\right)\omega_{Ti}k^{2}_{z}v^{2}_{Ti}\Psi_{i}\left(\mathbf{k}_{i},  
\tilde{X}_{i}, t _{1}\right)\right]=0.
\label{87}
\end{eqnarray} 
In  the zeroth order approximation,  the solution to Eq. (\ref{87}) with $\bar{u}'_{ix}=0$  
is $\Psi_{i}\left(\mathbf{k}_{i}, \tilde{X}_{i}, t 
\right)=C\exp\left(-i\omega\left(\mathbf{k}\right)\right)$, 
where the frequency $\omega\left(\mathbf{k}\right)$ is determined by the known equation \cite{Rudakov}
\begin{eqnarray}
&\displaystyle
\omega^{3}\left(\mathbf{k}\right)=-\omega_{Ti}k^{2}_{z}v^{2}_{Ti}\frac{\left(1-k^{2}_{\bot}\rho^{2}_{i}\right)}
{\left(\frac{T_{i}}{T_{e}}+k^{2}_{\bot}\rho^{2}_{i}\right)},
\label{88}
\end{eqnarray} 
one  root of which,  $\omega\left(\mathbf{k}\right)=\text{Re}\omega\left(\mathbf{k}\right)+\gamma\left(\mathbf{k}\right)\\=
\frac{1}{\sqrt{3}}\left(1+2i\right)\left(\omega_{Ti}k^{2}_{z}v^{2}_{Ti}
 \left(1-k^{2}_{\bot}\rho^{2}_{i}\right)/\left(\frac{T_{i}}{T_{e}}+k^{2}_{\bot}\rho^{2}_{i}\right)\right)^{1/3}$
gives the frequency and the growth rate of the HITG instability. The approximate solution to Eq. (\ref{87}), which 
displays the effect of the compressed flow  at time $t< \left(\bar{u}'_{ix}\right)^{-1}$ , we seek in the form 
\begin{eqnarray}
&\displaystyle
\Psi_{i}\left(\mathbf{k}_{i},  \tilde{X}_{i}, t \right)=C\exp\left(-i\omega\left(\mathbf{k}\right)t
+\sigma\left(\mathbf{k}, t \right) \right),
\label{89}
\end{eqnarray} 
where the correction $\sigma\left(\mathbf{k}, t \right) $ is found by the  procedure of successive approximation and is equal to
\begin{eqnarray}
&\displaystyle
\sigma\left(\mathbf{k}, t 
\right)=\frac{1}{6}\left(-i\text{Re}\omega\left(\mathbf{k}\right)+\gamma\left(\mathbf{k}\right)\right)\bar{u}'_{ix}t^{2}.
\label{90}
\end{eqnarray} 
Equation (\ref{90}) displays the nonmodal effect of the  linear growth with time,  as $\sim\bar{u}'_{ix}t>0$, of the frequency 
and the growth rate  of the HITG instability in the compressed flow for the perturbations which propagate 
 in the direction  of the compressed flow velocity. 
 
The  right-hand side of Eq. (\ref{86}), is of the order of  $\rho_{i}^{2}k_{\tilde{x}_{i}}\left(k_{\tilde{x}_{i}}\bar{u}'_{ix}
+k_{\tilde{y}_{i}}\bar{u}'_{iy}\right)t \ll \bar{u}'_{iy}t$  for $\rho_{i}^{2}k^{2}_{\tilde{x}_{i}}\ll 1$. Therefore the modification 
of the solution (\ref{90})  by the accounting for the right-hand side, which was found\cite{Mikhailenko_6} to be the main nonmodal 
effect for the sheared flow, appears to be negligible small for the sheared-compressed flow.

\section{The macroscale evolution of the  compressed-sheared  convective flows}\label{sec4}

The macroscale evolution  of  bulk of  ions in the  compressed-sheared convective flow is determined 
by Eq. (\ref{31}) which  with  velocity  variables $\tilde{v}_{i\bot}$ and $\phi$,  where 
$\tilde{v}_{ix}=\tilde{v}_{i\bot}\cos \phi$ and $\tilde{v}_{iy}=\tilde{v}_{i\bot}\sin \phi$, has a form
\begin{eqnarray}
&\displaystyle
\frac{\partial \bar{F}_{i}}{\partial T} + \tilde{v}_{i\bot}\cos\phi\left(1-\bar{U}_{ix}'T\right)\frac{\partial 
\bar{F}_{i}}{\partial \tilde{X}_{i}}
+\left(\tilde{v}_{i\bot}\sin\phi-\tilde{v}_{i\bot}\cos\phi\,\bar{U}_{iy}'T\right)\frac{\partial \bar{F}_{i}}{\partial 
\tilde{Y}_{i}} -\frac{1}{\varepsilon}\omega_{ci}\frac{\partial \bar{F}_{i}}{\partial \phi} 
\nonumber
\\  
&\displaystyle 
+ \tilde{v}_{iz}\frac{\partial \bar{F}_{i}}{\partial Z} 
- \frac{e_{i}}{m_{i}}\tilde{\nabla} \Phi\left(\tilde{X}_{i}, \tilde{Y}_{i}, T\right)\frac{\partial \bar{F}_{i}}{\partial 
\tilde{\mathbf{v}_{i}}}-\frac{e_{i}}{\varepsilon m_{i}}\left\langle \nabla_{\tilde{\mathbf{r}}_{i}} 
\tilde{\varphi}_{i0}\left(\tilde{\mathbf{r}}_{i}, \tilde{X}_{i},  T, \varepsilon\right)\frac{\partial f_{i}}
{\partial \tilde{\mathbf{v}}_{i}} \right\rangle =0.
\label{91} 
\end{eqnarray}

In the guiding center co-ordinates $\hat{X}_{i}$ and $\hat{Y}_{i}$, determined by the relations,
\begin{eqnarray}
&\displaystyle
\tilde{X}_{i}=\hat{X}_{i}-\varepsilon\frac{v_{i\bot}}{\omega_{ci}}\sin\left(\phi_{1}-
\frac{1}{\varepsilon}\omega_{ci}T\right)\left(1-\bar{U}_{ix}'T\right)+O\left(\varepsilon^{2}\right),
\label{92}
\\  
&\displaystyle 
\tilde{Y}_{i}=\hat{Y}_{i}+\varepsilon\frac{v_{i\bot}}{\omega_{ci}}\cos\left(\phi_{1}
-\frac{1}{\varepsilon}\omega_{ci}T\right)
+\varepsilon\frac{v_{i\bot}}{\omega_{ci}}\sin\left(\phi_{1}-\frac{1}{\varepsilon}\omega_{ci}T\right)
\bar{U}_{iy}'T+O\left(\varepsilon^{2}\right),
\label{93}
\end{eqnarray}
with $v_{i\bot}=\hat{v}_{i\bot}$ and $\phi=\phi_{1}-\frac{1}{\varepsilon}\omega_{ci}T$,  Eq. (\ref{75}) for 
$\bar{F}_{i}\left(\hat{v}_{i\bot},\phi, \hat{X}_{i}, \hat{Y}_{i}, T,\varepsilon\right)$ becomes
\begin{eqnarray}
&\displaystyle
\frac{\partial \bar{F}_{i}}{\partial T}-\frac{e_{i}}{m_{i}}\left\lbrace 
\frac{1}{\varepsilon}\frac{\omega_{ci}}{\hat{v}_{i\bot}}\left(\frac{\partial \Phi_{i}}{\partial 
\hat{v}_{i\bot}}\frac{\partial \bar{F}_{i}}{\partial \phi}
-\frac{\partial \Phi_{i}}{\partial \phi}\frac{\partial \bar{F}_{i}}{\partial \hat{v}_{i\bot}}\right) 
+\frac{\partial \Phi}{\partial Z}\frac{\partial \bar{F}_{i}}{\partial v_{z}}\right.
\nonumber
\\  
&\displaystyle 
\left.+\frac{\varepsilon}{\omega_{ci}}\left(\left(1-\bar{U}_{ix}'T\right)\frac{\partial\Phi_{i}}{\partial \hat{Y}_{i}}
\frac{\partial \bar{F}_{i}}{\partial \hat{X}_{i}}-\frac{\partial\Phi_{i}}{\partial \hat{X}_{i}}
\frac{\partial \bar{F}_{i}}{\partial \hat{Y}_{i}}-\frac{1}{2}\bar{U}_{ix}'T\frac{\partial\Phi_{i}}{\partial \hat{Y}_{i}}
\frac{\partial \bar{F}_{i}}{\partial \hat{Y}_{i}}\right)\right\rbrace 
\nonumber
\\  
&\displaystyle 
-\frac{1}{\varepsilon}\frac{e_{i}}{m_{i}}\left\langle \nabla_{\tilde{\mathbf{r}}_{i}} 
\tilde{\varphi}_{i0}\left(\tilde{\mathbf{r}}_{i}, \tilde{X}_{i},  t_{1}\right)\frac{\partial f_{i}}
{\partial \tilde{\mathbf{v}}_{i}} \right\rangle=0. 
\label{94}
\end{eqnarray}
The solution to Eq. (\ref{94}) we find in the form
\begin{eqnarray}
&\displaystyle
\bar{F}_{i}\left(\hat{v}_{i\bot},\phi, v_{z}, \hat{X}_{i}, \hat{Y}_{i}, Z_{i}, T\right)= 
\bar{F}_{i0}\left(\hat{v}_{i\bot}, v_{z}, \hat{X}_{i},  Z_{i}, T\right)
\nonumber
\\  
&\displaystyle 
+\bar{F}_{i1}\left(\hat{v}_{i\bot},\phi, v_{z}, \hat{X}_{i}, \hat{Y}_{i}, z_{i}, T,\varepsilon\right),
\label{95}
\end{eqnarray}
where $\bar{F}_{i0}$ is the equilbrium ion distribution function inhomogeneous along co-ordinate $ \hat{X}_{i}$. 
It is determined  by the quasilinear equation
\begin{eqnarray}
&\displaystyle
\frac{\partial \bar{F}_{i0}}{\partial T}=\frac{1}{\varepsilon}\frac{e_{i}}{m_{i}}\left\langle 
\nabla_{\tilde{\mathbf{r}}_{i}} 
\tilde{\varphi}_{i0}\left(\tilde{\mathbf{r}}_{i}, \tilde{X}_{i}, t\right)\frac{\partial f_{i}}
{\partial \tilde{\mathbf{v}}_{i}} \right\rangle.
\label{96}
\end{eqnarray}	
Employing  Eq.  (\ref{41})  for $f_{i}\left(\hat{v}_{i\bot}, \phi_{1}, v_{z}, \hat{x}_{i}, \hat{y}_{i}, z, 
 \hat{X}_{i}, t\right)$ and Eq. (\ref{39})  for $\tilde{\varphi}_{i0}\left(\tilde{\mathbf{r}}_{i}, \tilde{X}_{i}, 
 t\right)$ in Eq. (\ref{94}), 
and averaging over  the fast time $t=T/\varepsilon$, we derived the quasilinear equation for $\bar{F}_{i0}
\left(\hat{v}_{i\bot}, v_{z}, \hat{X}_{i},  T\right)$, 
\begin{eqnarray}
&\displaystyle
\frac{\partial \bar{F}_{i0}}{\partial T}=\frac{e_{i}^{2}}{m^{2}_{i}}\int\limits_{T_{0}}^{T}dT_{1}\int 
d\mathbf{k}_{i}\left(\frac{\varepsilon 
k_{\tilde{y}_{i}}}{\omega_{ci}}\left(1-\bar{U}'_{ix}T\right)\frac{\partial}{\partial \hat{X}_{i}}
+k_{z}\frac{\partial }{\partial v_{z}}\right)
\nonumber
\\  
&\displaystyle 
\times 
J_{0}\left(\frac{\hat{k}_{i\bot}\left(T\right)\hat{v}_{i\bot}}{\omega_{ci}}\right) 
J_{0}\left(\frac{ \hat{k}_{i\bot}\left(T_{1}\right)\hat{v}_{i\bot}}{\omega_{ci}}\right)
\langle\langle \tilde{\varphi}_{i0}\left(\tilde{\mathbf{k}}_{i},
\hat{X}_{i}, t\right) \tilde{\varphi}_{i0}\left(\tilde{\mathbf{k}}_{i}, \hat{X}_{i}, t_{1}\right) \rangle\rangle 
\nonumber
\\  
&\displaystyle 
\times 
\left(\frac{\varepsilon k_{\tilde{y}_{i}}}{\omega_{ci}}\left(1-\bar{U}'_{ix}T_{1}\right)\frac{\partial 
\bar{F}_{i0}}{\partial \hat{X}_{i}}+k_{z}\frac{\partial \bar{F}_{i0}}{\partial v_{z}}\right),
\label{97}
\end{eqnarray}
with  $\tilde{\mathbf{k}}_{i}=\left(k_{\tilde{x}_{i}}, k_{\tilde{y}_{i}}, k_{z}\right)$ and  
$\hat{k}_{i\bot}\left(T\right)$ and $\chi_{i}\left(T\right)$ determined by the relations
\begin{eqnarray}
	&\displaystyle
	\hat{k}^{2}_{i\bot}\left(T\right)=\left(k_{\tilde{x}_{i}}-\left(k_{\tilde{x}_{i}}\bar{U}'_{ix}
	+k_{\tilde{y}_{i}}\bar{U}'_{iy}\right)T\right)^{2}+k^{2}_{\tilde{y}_{i}}, \qquad	
	\sin \chi_{i}\left(T\right)=\frac{k_{\tilde{y}_{i}}}{k_{i\bot}\left(T\right)}.	
	\label{98}
\end{eqnarray}
In Eq. (\ref{97}), potential $\tilde{\varphi}_{i0} $ for times $t$, $t_{1}>t_{0}$ is determined by Eq. (\ref{67}), where 
$\varphi_{e}$ is the solution to Eq. (\ref{56}) with changed arguments 
$k_{x_{e}}\rightarrow k_{\tilde{x}_{i}}\left(1+b_{1x}\left(t\right)\right)+k_{\tilde{y}_{i}}b_{1x}\left(t\right),  
 k_{\tilde{y}_{e}}\rightarrow  k_{\tilde{y}_{i}}$ (here the time $t$ is equal to $t$ for  
 $\tilde{\varphi}_{i0}\left(t\right) $, and it is equal to $t_{1}$ for $\tilde{\varphi}_{i0}\left(t_{1}\right) $).

 The function $\bar{F}_{i1}\left(\hat{v}_{i\bot},\phi, v_{z}, \hat{X}_{i}, \hat{Y}_{i}, Z_{i}, T, 
 \varepsilon\right)$ is the perturbation of  $\bar{F}_{i0}$, caused by the electrostatic potential 
 $\Phi_{i}\left(\tilde{X}_{i}, \tilde{Y}_{i}, Z_{i}, T\right)$ of the plasma respond on the development in a plasma  
 the  macroscale sheared-compressed convective flows. The equation for  $\bar{F}_{i1}$, 
\begin{eqnarray}
&\displaystyle
\frac{\partial} {\partial T}\bar{F}_{i1}\left(\hat{v}_{i\bot},\phi, v_{z}, \hat{X}_{i}, \hat{Y}_{i}, Z_{i}, 
T,\varepsilon\right)
\nonumber 
\\ 
&\displaystyle
=\frac{e_{i}}{m_{i}}\left\lbrace\frac{\varepsilon}{\omega_{ci}}\left( 
1-\bar{U}_{ix}'T\right)\frac{\partial\Phi_{i}}{\partial \hat{Y}_{i}}
\frac{\partial \bar{F}_{i0}}{\partial \hat{X}_{i}}
-\frac{\omega_{ci}}{\varepsilon}\frac{1}{\hat{v}_{i\bot}}\frac{\partial \Phi_{i}}{\partial \phi}
\frac{\partial 	\bar{F}_{i0}}{\partial \hat{v}_{i\bot}}
+\frac{\partial \Phi_{i}}{\partial Z}\frac{\partial \bar{F}_{i0}}{\partial v_{z}}\right\rbrace,
\label{99}
\end{eqnarray}
follows from Eqs.  (\ref{78}) and  (\ref{79}).  In solution to Eq. (\ref{99}), we consider the potential $\Phi_{i}$ 
in the form
\begin{eqnarray}
&\displaystyle
\Phi_{i}\left(\tilde{X}_{i}, \tilde{Y}_{i}, Z, T\right)=\frac{1}{8\pi^{3}}\int \Phi_{i}\left(\mathbf{K}_{i}, 
\tilde{X}_{i}, 	T\right) 
e^{i\left(K_{\tilde{X}_{i}}\frac{\tilde{X}_{i}}{\varepsilon_{1}}+
K_{\tilde{Y}_{i}}\frac{\tilde{Y}_{i}}{\varepsilon_{1}}+K_{z}Z_{i}
\right)}	dK_{\tilde{X}_{i}}dK_{\tilde{Y}_{i}}dK_{z}
\nonumber	
\\  
&\displaystyle 
=\frac{1}{8\pi^{3}}\int\Phi_{i}\left(\mathbf{K}_{i}, \tilde{X}_{i}, T\right)	
e^{i\left(K_{\tilde{X}_{i}}\frac{\hat{X}_{i}}{\varepsilon_{1}}+K_{\tilde{Y}_{i}}\frac{\hat{Y}_{i}}
{\varepsilon _{1}}+K_{z}Z_{i}\right )}
\nonumber
\\  
&\displaystyle 
\times
\sum_{n=-\infty}^{\infty}J_{n}\left(\frac{\varepsilon 
K_{i\bot}\left(T\right)\hat{v}_{i\bot}}{\varepsilon_{1}\omega_{ci}}\right)	
e^{-in\left(\phi-\frac{\varepsilon_{1}}{\varepsilon}\omega_{ci}T-\chi_{i}\left(T\right)\right)}
dK_{\tilde{X}_{i}}dK_{\tilde{Y}_{i}}dK_{z}.
\label{100}
\end{eqnarray}
In Equation (\ref{100}),  the small parameter  $\varepsilon_{1}$, $1 \gg \varepsilon_{1}>\varepsilon$, is introduced  to
distinguish the intermediate scale of  the slow evolution of the amplitude $\Phi_{i}\left(\mathbf{K}_{i}, \tilde{X}_{i}, T\right) $ on the 
mesoscale $\varepsilon^{-1}_{1}\tilde{X}_{i}, T$ and the fast changed phase on the wavelengths that are much smaller 
than the scale lengths $L_{n}$, $L_{T_{i}}$, $L_{\bar{U}_{ix}}$, $L_{\bar{U}_{iy}}$ of the spatial inhomogeneity of the plasma 
density, of the ion temperature, and of the  convective flows velocities, respectively,  but which are much larger  than 
the wavelengths of the microturbulence, i. e. $|k_{x_{i}}|\gg |K_{X_{i}}|$, $|k_{y_{i}}|\gg |K_{Y_{i}}|$. In Equation (\ref{84}),
$K_{i\bot}\left(T\right)$ and $\chi_{i}\left(T\right)$  are determined by the relations 
\begin{eqnarray}
&\displaystyle
K^{2}_{i\bot}\left(T\right)=\left(K_{\tilde{X}_{i}}-\left(K_{\tilde{X}_{i}}\bar{U}'_{ix}+K_{\tilde{Y}_{i}}
\bar{U}'_{iy}\right)T\right)^{2}+K^{2}_{\tilde{Y}_{i}}, 
\qquad	\sin \chi_{i}\left(T\right)=\frac{K_{\tilde{Y}_{i}}}{K_{i\bot}\left(T\right)}.	
\label{101}
\end{eqnarray}
The  solution  to  Eq. (\ref{99}) for $\bar{F}_{i1}\left(\hat{v}_{i\bot}, v_{z}, 
K_{\tilde{X}_{i}}, K_{\tilde{Y}_{i}}, K_{z}, T\right)$,  
averaged over the fast time $t=\varepsilon_{1}T/\varepsilon\gg \omega_{ci}^{-1}$ , has a form
\begin{eqnarray}
&\displaystyle
\bar{F}_{i1}\left(\hat{v}_{i\bot}, v_{z}, K_{\tilde{X}_{i}}, K_{\tilde{Y}_{i}}, K_{z}, \tilde{X}_{i}, T\right)
=i\frac{e_{i}}{m_{i}}	\int \limits_{T_{0}}^{T}dT_{1}\Phi_{i}\left(\mathbf{K}_{i}, \tilde{X}_{i}, T_{1}\right)
\nonumber
\\  
&\displaystyle
\times 
J_{0}\left(\frac{\varepsilon K_{i\bot}\left(T\right)\hat{v}_{i\bot}}{\varepsilon_{1}\omega_{ci}}\right)	
J_{0}\left(\frac{\varepsilon K_{i\bot}\left(T_{1}\right)\hat{v}_{i\bot}}{\varepsilon_{1}\omega_{ci}}\right)	
\nonumber
\\  
&\displaystyle
\times
\left[\frac{\varepsilon 
	K_{\tilde{Y}_{i}}}{\varepsilon_{1}\omega_{ci}}\left(1-\bar{U}'_{ix}T_{1}\right)
\frac{\partial \bar{F}_{i0}}{\partial \tilde{X}_{i}}+K_{z}\frac{\partial \bar{F}_{i0}}{\partial v_{iz}}\right]
e^{-iK_{z}v_{iz}\left(T-T_{1}\right)},
\label{102} 
\end{eqnarray}
The  slow ion density perturbation $n_{i}\left(\mathbf{K}_{i}, T\right)$ in the convective flow is determined by relation
\begin{eqnarray}
&\displaystyle
n_{i}\left(\mathbf{K}_{i},  \tilde{X}_{i}, T\right)=\int d\hat{\mathbf{v}}_{i}\bar{F}_{i1}\left(\hat{v}_{i\bot}, 
v_{z}, K_{\tilde{X}_{i}}, K_{\tilde{Y}_{i}}, K_{z}, \tilde{X}_{i}, T\right)
\nonumber
\\  
&\displaystyle
=i\frac{2\pi e_{i}}{m_{i}}\int\limits_{T_{0}}^{T}dT_{1}\Phi_{i}\left(\mathbf{K}_{i}, \tilde{X}_{i}, T_{1}\right)
\int\limits_{\infty}^{\infty}dv_{iz}\int\limits _{0}^{\infty}d\hat{v}_{i\bot}\hat{v}_{i\bot}
\nonumber
\\  
&\displaystyle
\times
J_{0}\left(\frac{\varepsilon 	K_{i\bot}\left(T\right)\hat{v}_{i\bot}}{\varepsilon_{1}\omega_{ci}}\right)	
J_{0}\left(\frac{\varepsilon K_{i\bot}\left(T_{1}\right)\hat{v}_{i\bot}}{\varepsilon_{1}\omega_{ci}}\right)	
\nonumber
\\  
&\displaystyle
\times
 e^{-iK_{z}v_{iz}\left(T-T_{1}\right)}\left [\frac{\varepsilon K_{\tilde{X}_{i}}}{\varepsilon_{1}\omega_{ci}}
\left(1-\bar{U}'_{ix}T_{1}\right)
\frac{\partial  \bar{F}_{i0}}{\partial \tilde{X}_{i}}+K_{z}\frac{\partial  \bar{F}_{i0}}{\partial v_{iz}}\right ].
\label{103} 
\end{eqnarray}
The electron Vlasov equation (\ref{33}) for the average  electron  distribution function \\ $\bar{F}_{e}
\left(\hat{v}_{e\bot},\phi, v_{z}, \hat{X}_{e}, \hat{Y}_{e}, z_{e}, T,\varepsilon\right)$ in the electron guiding 
center coordinates  $\hat{X}_{e}\approx \tilde{X}_{e}$,  $\hat{Y}_{e}\approx \tilde{Y}_{e}$ \\
for $\tilde{X}_{e}\gg \rho_{e}$ and $\tilde{Y}_{e}\gg \rho_{e}$ becomes
\begin{eqnarray}
&\displaystyle 
\frac{\partial \bar{F}_{e}}{\partial T}-\frac{e}{m_{e}}\left\lbrace 
\frac{1}{\varepsilon}\frac{\omega_{ce}}{\hat{v}_{e\bot}}\left(\frac{\partial \Phi_{e}}{\partial 
\hat{v}_{e\bot}}\frac{\partial 
\bar{F}_{e}}{\partial \phi}	-\frac{\partial \Phi_{e}}{\partial \phi}\frac{\partial \bar{F}_{e}}{\partial 
\hat{v}_{e\bot}}\right) 
+\frac{\partial \Phi_{e}}{\partial Z}\frac{\partial \bar{F}_{e}}{\partial v_{z}}\right.
\nonumber	
\\  
&\displaystyle 
\left.+\frac{\varepsilon}{\omega_{ci}}\left(\frac{\partial\Phi_{e}}{\partial \hat{Y}_{e}}
\frac{\partial \bar{F}_{e}}{\partial \hat{X}_{e}}-\frac{\partial\Phi_{e}}{\partial \hat{X}_{e}}
\frac{\partial \bar{F}_{e}}{\partial \hat{Y}_{e}}\right)\right\rbrace 
\nonumber	
\\  
&\displaystyle 
+\frac{1}{\varepsilon}\frac{e}{m_{e}}\left\langle \tilde{\mathbf{E}}_{e0}\left(X_{e}, \varepsilon^{-1}X_{e}, 
\varepsilon^{-1}Y_{e}, Z_{e}, \varepsilon^{-1}T\right)\frac{\partial f_{e}}{\partial 
\tilde{\mathbf{v}}_{e}}\right\rangle=0. 
\label{104} 
\end{eqnarray}
The solution to Eq. (\ref{104}) for the electron distribution function $\bar{F}_{e}$  we derive in the form  
(\ref{95})  applied for $\bar{F}_{i}$,
\begin{eqnarray}
&\displaystyle
\bar{F}_{e}\left(\hat{v}_{e\bot},\phi, \hat{X}_{e}, \hat{Y}_{e}, T,\varepsilon\right)
=\bar{F}_{e0}\left(\hat{v}_{e\bot}, v_{ez}, \hat{X}_{e}, T\right)
\nonumber
\\  
&\displaystyle 
+\bar{F}_{e1}	\left(\hat{v}_{e\bot},\phi, v_{ez}, \hat{X}_{e}, \hat{Y}_{e}, Z_{e}, T, \varepsilon\right),	
\label{105}
\end{eqnarray}
in which we distinguish the equilibrium electron distribution function \\ $\bar{F}_{e0}\left(\hat{v}_{e\bot}, v_{ez}, 
\hat{X}_{e}, T\right)$, determined by the quasilinear equation 
\begin{eqnarray}
&\displaystyle
\frac{\partial \bar{F}_{e0}}{\partial T}= -\frac{1}{\varepsilon}\frac{e}{m_{e}}\left\langle 
\tilde{\mathbf{E}}_{e0}\left(\hat{X}_{e}, \hat{x}_{e}, \hat{y}_{e}, z_{e}, \varepsilon^{-1}T\right)\frac{\partial f_{e}}
{\partial \tilde{\mathbf{v}}_{e}}\right\rangle.
\label{106}
\end{eqnarray}
Employing Eq.  (\ref{50})  for  $f_{e}\left(\hat{v}_{e\bot}, v_{z}, k_{\tilde{x}_{e}}, k_{\tilde{y}_{e}}, k_{z}, 
\tilde{X}_{e}, t\right)$  and 	$\tilde{\varphi}_{e0}\left(\tilde{\mathbf{k}}_{e}, \hat{X}_{e}, t\right) $ 
as the solution to Eq.  (\ref{56})  in Eq. (\ref{106}), we derive the quasilinear equation for the electron distribution 
function $\bar{F}_{e0}$,
\begin{eqnarray}
&\displaystyle
\frac{\partial \bar{F}_{e0}}{\partial T}=\frac{e^{2}}{m^{2}_{e}}\int\limits_{T_{0}}^{T}dT_{1}\int 
d\mathbf{k}\left(\frac{\varepsilon k_{y}}{\omega_{ce}}\frac{\partial}{\partial \hat{X}_{e}}
+k_{z}\frac{\partial }{\partial v_{z}}\right)
\nonumber
\\  
&\displaystyle 
\times 	
\langle\langle \tilde{\varphi}_{e0}\left(\tilde{\mathbf{k}}_{e},
\hat{X}_{e}, t\right) \tilde{\varphi}_{e0}\left(\tilde{\mathbf{k}}_{e}, \hat{X}_{e}, t_{1}\right) \rangle\rangle 
\left(\frac{\varepsilon k_{y}}{\omega_{ce}}\frac{\partial \bar{F}_{e0}}{\partial \hat{X}_{e}}
+k_{z}\frac{\partial \bar{F}_{e0}}{\partial v_{z}}\right).
\label{107}
\end{eqnarray}

The perturbation $\bar{F}_{e1}\left(\hat{v}_{e\bot},\phi, v_{z}, \hat{X}_{e}, \hat{Y}_{e}, z_{e}, T,\varepsilon\right)$ 
of $\bar{F}_{e0}$ is caused  by the self-consistent potential $\Phi_{e}$ 
of the plasma  response on the development of the convective flows. The equation for $\bar{F}_{e1}$ follows 
from Eqs. (\ref{104})--(\ref{106}) and for the perturbations, for which $\frac{\partial  \Phi_{e}}{\partial T}\ll 
\omega_{ce}\Phi_{e}$,
it has a form
\begin{eqnarray}
&\displaystyle
\frac{\partial} {\partial T}\bar{F}_{e1}\left(\hat{v}_{e\bot},\phi, v_{z}, \hat{X}_{e}, \hat{Y}_{e}, Z, T,\epsilon\right)
=\frac{e}{m_{e}}\left\lbrace\frac{\varepsilon}{\omega_{ce}}\frac{\partial\Phi_{e}}{\partial \hat{Y}_{e}}	\frac{\partial 
\bar{F}_{e0}}{\partial \hat{X}_{e}}	+\frac{\partial \Phi_{e}}{\partial Z}\frac{\partial \bar{F}_{e0}}{\partial 
v_{z}}\right\rbrace.
\label{108}
\end{eqnarray}
The solution to Eq.	(\ref{108}), which determines the evolution of the separate spatial long wavelength, 
$K_{e\bot}\rho_{e}\ll 1$, macroscale Fourier harmonic $\bar{F}_{e1}\left(\hat{\mathbf{v}}_{e}, 
K_{\tilde{X}_{e}}, K_{\tilde{Y}_{e}}, K_{z}, T\right)$, 
\begin{eqnarray}
&\displaystyle
\bar{F}_{e1}\left(\hat{v}_{e\bot}, v_{z}, K_{\tilde{X}_{e}}, K_{\tilde{Y}_{e}}, K_{z}, \tilde{X}_{i}. T\right)
=i\frac{e}{ m_{e}}\int dT_{1}\Phi_{e}\left(\mathbf{K}_{e}, \tilde{X}_{i}, T_{1}\right)
\nonumber
\\  
&\displaystyle 
\times 	
\left[\frac{\varepsilon K_{\tilde{Y}_{e}}}{\varepsilon_{1}\omega_{ce}}
\frac{\partial \bar{F}_{i0}}{\partial \hat{X}_{e}}+K_{z}\frac{\partial \bar{F}_{e0}}{\partial v_{ez}}\right]
e^{-iK_{z}v_{iz}\left(t-t_{1}\right)},	
\label{109} 
\end{eqnarray}
was derived by the Fourier transforming of Eq. (\ref{108}) over $\hat{X}_{e}$, $\hat{Y}_{e}$ and $Z$.  In  Eq. (\ref{109}),  
the Fourier transformation of the potential $\Phi_{e}\left(\tilde{X}_{e}, \tilde{Y}_{e}, Z, T\right)$  over  coordinates 
$\tilde{X}_{e}$, $\tilde{Y}_{e}$, $Z$, 
\begin{eqnarray}
&\displaystyle
\Phi_{e}\left(\tilde{X}_{e}, \tilde{Y}_{e}, Z, T\right)=\frac{1}{8\pi^{3}}\int \Phi_{e}\left(\mathbf{K}_{e}, 
\tilde{X}_{e}, T\right)
\nonumber
\\  
&\displaystyle 
\times
\exp\left(iK_{\tilde{X}_{e}}\frac{\tilde{X}_{e}}{\varepsilon_{1}}
+iK_{\tilde{Y}_{e}}\frac{\tilde{Y}_{e}}{\varepsilon_{1}}+iK_{z}Z\right)
dK_{\tilde{X}_{e}}dK_{\tilde{Y}_{e}}dK_{z},
\label{110}
\end{eqnarray}
was used.  The macroscale  slow electron density perturbation $n_{e}\left(\mathbf{K}_{e}, T\right)$  is determined 
in the electron (laboratory) frame  by the relation
\begin{eqnarray}
&\displaystyle
n_{e}\left(\mathbf{K}_{e}, \tilde{X}_{e}, T\right)=\int d\hat{\mathbf{v}}_{e}\bar{F}_{e1}\left(\hat{v}_{e\bot}, v_{z}, 
K_{\tilde{X}_{e}},  K_{\tilde{Y}_{e}}, K_{z}, \tilde{X}_{e}, T\right)
\nonumber
\\  
&\displaystyle
=i\frac{2\pi e}{m_{e}}\int\limits_{T_{0}}^{T}dT_{1}\Phi_{e}\left(\mathbf{K}_{e}, \tilde{X}_{e}, T_{1}\right)
\int\limits_{-\infty}^{\infty}	dv_{ez}\int\limits _{0}^{\infty}d\hat{v}_{e\bot}\hat{v}_{e\bot}
\nonumber
\\  
&\displaystyle
\times 
e^{-iK_{z}v_{ez}\left(T-T_{1}\right)}\left [\frac{\varepsilon K_{\tilde{X}_{e}}}
{\varepsilon_{1}\omega_{ce}}\frac{\partial  \bar{F}_{e0}}{\partial \tilde{X}_{e}}+K_{z}\frac{\partial  
\bar{F}_{e0}}{\partial v_{ez}}\right ].
\label{111} 
\end{eqnarray}

The Poisson equation for the macroscale potential $\Phi_{e}$
\begin{eqnarray}
&\displaystyle 
\frac{\partial^{2} \Phi_{e}\left(\tilde{X}_{e}, \tilde{Y}_{e}, \tilde{Z}_{e}, t\right)}
{\partial^{2} \tilde{X_{e}}}+ \frac{\partial^{2}	\Phi_{e}\left(\tilde{X}_{e}, \tilde{Y}_{e}, \tilde{Z}_{e}, t\right)}
{\partial^{2}\tilde{Y_{e}}}+\frac{\partial^{2} \Phi_{e}\left(\tilde{X}_{e}, \tilde{Y}_{e},  \tilde{Z}_{e}, t\right)}
{\partial^{2} \tilde{Z_{e}}}
\nonumber 
\\ 
&\displaystyle
= - 4\pi\left[e_{i}n_{i} \left(\tilde{X}_{i}, \tilde{Y}_{i}, \tilde{Z}_{i},T\right)	-|e|n_{e} \left(\tilde{X}_{e}, 
\tilde{Y}_{e}, \tilde{Z}_{e}, T\right)\right],
\label{112} 
\end{eqnarray}
Fourier transformed over coordinates $\tilde{X}_{e}, \tilde{Y}_{e}, \tilde{Z}_{e}$,
\begin{eqnarray}
&\displaystyle
\left(K^{2}_{\tilde{X}_{e}}+K^{2}_{\tilde{Y}_{e}}+K^{2}_{Z}\right) \Phi_{e}\left(\mathbf{K}_{e},  
\tilde{X}_{e}, T\right)
=
4\pi e_{i}n^{(e)}_{i}\left(\mathbf{K}_{e}, \tilde{X}_{e}, T\right)+4\pi en_{e}\left(\mathbf{K}_{e}, 
\tilde{X}_{e}, T\right),
\label{113} 
\end{eqnarray} 
governs the kinetic macroscale nonmodal evolution of a macroscale potential 
$\Phi_{e}\left(\mathbf{K}_{e}, T\right)$ in convective flows,  formed by the spatially inhomogeneous 
microturbulence.  In Eq. (\ref{113}), $n_{e}\left(\mathbf{K}_{e}, \tilde{X}_{e}, T\right)$ is given 
by Eq. (\ref{111}), $n^{(e)}_{i}\left(\mathbf{K}_{e}, T\right)$  denotes the Fourier transform of the 
macroscale ion density perturbation $n_{i}\left(\tilde{X}_{i}, \tilde{Y}_{i}, Z, T\right)$ performed in 
the electron frame over $\tilde{X}_{e}, \tilde{Y}_{e}, Z$. By transforming  Eqs. (\ref{56})--(\ref{68}) to 
macroscale coordinates $X_{i}, Y_{i}, X_{e}, Y_{e}$ with accounting for  the identities 
$\bar{u}'_{ix}t=\bar{U}'_{ix}T$, 
$\bar{u}'_{iy}t=\bar{U}'_{iy}T$ we derived from Eq. (\ref{113}) the equation for the potential $\Phi_{e}\left(\mathbf{K}_{e}, 
\tilde{X}_{i}, T\right)$
\begin{eqnarray}
&\displaystyle	
K^{2}_{e}{\lambda^{2}_{Di}}\Phi_{e}\left(\mathbf{K}_{e}, \tilde{X}_{i}, T\right)
=-\int\limits_{T_{0}}^{T}dT_{1}\frac{d}{dT_{1}}	\left\lbrace \left| \frac{1+\bar{U}'_{ix}T}{1+\bar{U}'_{ix}T_{1}} \right|
e^{-\frac	{1}{2}K^{2}_{z}v^{2}_{Ti}\left(T-T_{1}\right)^{2}}\right.
\nonumber 
\\ 
&\displaystyle
\left.\times 
\exp\left[-iK_{\tilde{X}_{e}}\bar{U}^{(0)}_{ix}\left(T-T_{1}\right)\left(1-\frac{\bar{U}'_{ix}T_{1}}
{1+\bar{U}'_{ix}T_{1}}\right)-iK_{\tilde{Y}_{e}}\bar{U}^{(0)}_{iy}\left(T-T_{1}\right)
\left(1-\frac{\bar{U}'_{iy}T_{1}}{1+\bar{U}'_{ix}T_{1}}\right)\right]\right.
\nonumber 
	\\ 
	&\displaystyle
	\left.\times
	\Phi_{e}\left(K_{\tilde{X}_{e}}\left(1+\frac{\bar{U}'_{ix}\left(T-T_{1}\right)}
	{1+\bar{U}'_{ix}T_{1}}\right) +K_{\tilde{Y}_{e}}\frac{\bar{U}'_{iy}\left(T-T_{1}\right)}
	{1+\bar{U}'_{ix}T_{1}}, K_{\tilde{Y}_{e}}, K_{z}, \tilde{X}_{i}, T_{1}\right)\right\rbrace 
	\nonumber 
	\\ 
	&\displaystyle
	+	\int\limits_{T_{0}}^{T}dT_{1}e^{-\frac	{1}{2}K^{2}_{z}v^{2}_{Ti}\left(T-T_{1}\right)^{2}}
	\frac{d}{dT_{1}}\left\lbrace  \left| \frac{1+\bar{U}'_{ix}T}{1+\bar{U}'_{ix}T_{1}}\right|
	\left(1+\frac{i}{2}\omega_{Ti}\left(1-U'_{ix}T_{1}\right)\left(T-T_{1}\right)\right)\right.
	\nonumber 
	\\ 
	&\displaystyle
	\left.\times \exp\left[-iK_{\tilde{X}_{e}}\bar{U}^{(0)}_{ix}\left(T-T_{1}\right)\left(1-\frac{\bar{U}'_{ix}T_{1}}
	{1+\bar{U}'_{ix}T_{1}}\right)-iK_{\tilde{Y}_{e}}\bar{U}^{(0)}_{iy}\left(T-T_{1}\right)
	\left(1-\frac{\bar{U}'_{iy}T_{1}}{1+\bar{U}'_{ix}T_{1}}\right)\right]\right.
	\nonumber 
	\\ 
	&\displaystyle
\left.\times
\Phi_{e}\left( K _{\tilde{X}_{e}}\left(1+\frac{\bar{U}'_{ix}\left(T-T_{1}\right)}
	{1+\bar{U}'_{ix}T_{1}}\right) +K _{\tilde{Y}_{e}}\frac{\bar{U}'_{iy}\left(T-T_{1}\right)}
	{1+\bar{U}'_{ix}T{1}}, K_{\tilde{Y}_{e}}, K_{z}, \tilde{X}_{i}, T_{1}\right)\right.
	\nonumber 
	\\ 
	&\displaystyle
\left.\times 
I_{0}\left(K_{i\perp}\left(T \right)K_{i\perp}\left(T_{1} \right)\rho_{i}^{2} \right) 
	e^{-\frac{1}{2}\left(K_{i\perp}^{2}\left(T\right) +K_{i\perp}^{2}\left(T_{1}\right) \right)\rho_{i}^{2}}\right\rbrace 
	\nonumber 
	\\ 
	&\displaystyle
	+i\int\limits_{T_{0}}^{T} dT_{1}\left| \frac{1+\bar{U}'_{ix}T}{1+\bar{U}'_{ix}T_{1}} \right| 
	e^{-\frac{1}{2}\left(K_{i\perp}^{2}\left(T \right) +K_{i\perp}^{2}\left(T_{1}\right)\right)\rho_{i}^{2}
		-\frac{1}{2}K_{z}^{2}v^{2}_{Ti}\left(T-T_{1}\right)^{2}}
	\nonumber 
	\\ 
	&\displaystyle
	\times 
	\exp\left[-iK _{\tilde{X}_{e}}\bar{U}^{(0)}_{ix}\left(T-T_{1}\right)\left(1-\frac{\bar{U}'_{ix}T_{1}}
	{1+\bar{U}'_{ix}T_{1}}\right)-iK _{\tilde{Y}_{e}}\bar{U}^{(0)}_{iy}\left(T-T_{1}\right)
	\left(1-\frac{\bar{U}'_{iy}T_{1}}{1+\bar{U}'_{ix}T_{1}}\right)\right]
	\nonumber 
	\\ 
	&\displaystyle
	\times
	\Phi_{e}\left( K_{\tilde{X}_{e}}\left(1+\frac{\bar{U}'_{ix}\left(T-T_{1}\right)}
	{1+\bar{U}'_{ix}T_{1}}\right) +K_{\tilde{Y}_{e}}\frac{\bar{U}'_{iy}\left(T-T_{1}\right)}
	{1+\bar{U}'_{ix}T_{1}}, K_{\tilde{Y}_{e}}, K_{z}, \tilde{X}_{i}, T_{1}\right) 
	\nonumber 
	\\ 
&\displaystyle
\times 
\left[K_{\tilde{Y}_{e}}v_{di}\left(1-\bar{U}'_{ix}T_{1} \right)I_{0}\left(K_{i\perp}\left(T
\right)K_{i\perp}\left(T_{1} \right)\rho_{i}^{2} \right)   \right. 
\nonumber 
\\ 
&\displaystyle
\left.+\omega_{Ti}\left(1-\bar{U}'_{ix}T_{1}\right) \left[ - \frac{\rho_{i}^{2}}{2} \left(K_{i\perp}^{2}\left(T \right) 
+K_{i\perp}^{2}\left(T_{1} 	\right)\right)  I_{0}\left(K_{i\perp}\left(T\right)k_{i\perp}\left(T_{1} \right)\rho_{i}^{2} \right) 
 \right. \right.	\nonumber 
\\ 
&\displaystyle
\left.\left.  +\frac{\rho_{i}^{2}}{2} K_{i\perp}\left(T \right)K_{i\perp}\left(T_{1} \right)
 I_{1}\left(K_{i\perp}\left(T\right)K_{i\perp}\left(T_{1} \right)\rho_{i}^{2} \right)  \right] \right]
\nonumber 
\\ 
&\displaystyle
-Q^{(e)}_{i}\left(\mathbf{K}_{e},   \tilde{X}_{e}, T, T_{0}\right) 
- Q_{e}\left(\mathbf{K}_{e},   \tilde{X}_{e}, T,\ T_{0}\right),
\label{114}
\end{eqnarray}	
where
\begin{eqnarray}
&\displaystyle
K^{2}_{i\bot}\left(T\right)=\left(K_{\tilde{X}_{e}}-\left(K_{\tilde{X}_{e}}\bar{U}'_{ix}
+K_{\tilde{Y}_{e}}\bar{U}'_{iy}\right)\bar{U}'_{iy}T^{2}\right)^{2}+k^{2}_{\tilde{Y}_{e}},
\nonumber 
\\ 
&\displaystyle 
K^{2}_{i\bot}\left(T_{1}\right)=\left(K_{\tilde{X}_{e}}-\left(K_{\tilde{X}_{e}}\bar{U}'_{ix}
+K_{\tilde{Y}_{e}}\bar{U}'_{iy}\right)\bar{U}'_{iy}T_{1}^{2}\right)^{2}+K^{2}_{\tilde{Y}_{e}},
\label{115}
\end{eqnarray}	
and
\begin{eqnarray}
&\displaystyle
Q^{(e)}_{i}\left(\mathbf{K}_{e},   \tilde{X}_{i}, T,\ T_{0}\right) =\frac{e_{i}}{T_{i}}n_{0i}\left(\hat{X}_{i}\right)
\frac{1}{\left| 1+\bar{U}'_{ix}T_{0}\right|} 
\nonumber 
\\ 
&\displaystyle
\times 
\exp\left[-iK _{\tilde{X}_{e}}\bar{U}^{(0)}_{ix}\left(T-T_{0}\right)\left(1-\frac{\bar{U}'_{ix}T_{0}}
{1+\bar{U}'_{ix}T_{0}}\right)-iK _{\tilde{Y}_{e}}\bar{U}^{(0)}_{iy}\left(T-T_{0}\right)
\left(1-\frac{\bar{U}'_{iy}T_{0}}{1+\bar{U}'_{ix}T_{0}}\right)\right]
\nonumber 
\\ 
&\displaystyle
\times
\Phi_{e}\left( K _{\tilde{X}_{e}}\left(1+\frac{\bar{U}'_{ix}\left(T-T_{0}\right)}
{1+\bar{U}'_{ix}T_{0}}\right) +K _{\tilde{Y}_{e}}\frac{\bar{U}'_{iy}\left(T-T_{0}\right)}
{1+\bar{U}'_{ix}T_{0}}, K_{\tilde{Y}_{e}}, K_{Z},  T_{0}\right)
\nonumber 
\\ 
&\displaystyle
\times 
\left[1-I_{0}\left(\frac{\varepsilon^{2}}{\varepsilon_{1}^{2}}K_{i\perp}\left(T 
\right)K_{i\perp}\left(T_{0} \right)\rho_{i}^{2} \right) 
e^{-\frac{\varepsilon^{2}}{2\varepsilon^{2}_{1}}\left(K_{i\perp}^{2}\left(T \right) +K_{i\perp}^{2}\left(T_{0} 
\right)\right)\rho_{i}^{2}-\frac{1}{2}K_{Z}^{2}v^{2}_{Ti}\left(T-T_{0}\right)^{2} } \right].  
\label{116}
\end{eqnarray}	
Equation (\ref{114})  is the basic linear equation of the two-scale non-modal kinetic theory to investigate the temporal 
evolution of the potential $\Phi_{e}$ of the meso/macroscale perturbations in the compressed-sheared convective flow 
formed by the inhomogeneous microturbulence. 

\section{Conclusions}\label{sec5}
In this paper, we present the two-scale non-modal approach to the kinetic theory of the
microscale turbulence of a plasma, inhomogeneous on the macroscales across the confined magnetic
field. This approach reveals the effect of the macroscale convective flows formation in such a
plasma caused by the interaction of ions with microscale turbulence. The flow velocities are found
as the average velocities of the motion of ions and electrons in the spatially inhomogeneous
microturbulence and are proportional to the gradient of the spectral intensity of the electric field of
the microturbulence. It follows from Eqs. (\ref{122}) and (\ref{123}) that the velocities of the 
ion convective flow are ion mass dependent.  The velocities of the electron convective
flows are negligible small relative to the ion flow velocities and, therefore, the convective motion of
electron may be neglected. This result predicts that the macroscale convective flow transports mostly
the ions. For the ion flow, generated by the low frequency microturbulence with radially decreasing
spectral intensity, Eq. (\ref{122}) predicts that the radial velocity of the ion flow is directed outward of the
plasma core to the edge of the tokamak plasma.  This result displays that the non-diffusive convective 
ion heat flux to edge will play a key role in the determination of the edge radially inhomogeneous electric field,  
responsible for the formation of the poloidal sheared flow. It is interesting to note that this   result was obtained 
in the experiments carried out in the ASDEX Upgrade tokamak that the ion heat flux at the plasma edge plays a 
key role in the L-H transition physics, while the electron heat flux does not seem to play any role\cite{Ryter}. 
This result reveals the necessity in the 
investigations of the temporal evolution of the macroscale convective flow of ions in the edge region of the
tokamak plasma, investigation of the loss of ions  and formation of the localized radial electric field
and of the mesoscale poloidal sheared flow.

Any microscale perturbation in the radially inhomogeneous ion flows, which before the
development of the sheared-compressed flow had a plane wave structure, experiences the
continuous distortion in the flow and become the sheared-compressed mode with time dependent
structure. This distortion grows with time and forms a time-dependent nonmodal process, which
affects the microturbulence and the average ion distribution function. The derived quasilinear
equation  (\ref{97}), which determines the nonmodal evolution of the average ion distribution function,
resulted from the interactions of ions with ensemble of the microscale sheared-compressed waves,
and the integral equation  (\ref{56}) for the electrostatic potential $\varphi_{e}\left(\mathbf{k}_{e}, 
\tilde{X}_{e},t \right) $ of the microturbulence, which determines the macroscale nonlinear 
back-reaction of the sheared-compressed convective flows on the microscale perturbations, and 
the integral equation  (\ref{114}) for the macroscale potential $\Phi_{e}\left(\mathbf{K}_{e},  \tilde{X}_{e}, T\right)$  
of the nonmodal plasma respond on the development in plasma  the compressed - sheared convective flows, are the basic 
equations which determine the  macroscale evolution of the plasma with inhomogeneous microturbulence at time 
corresponding to the L--H transition before the formation of the poloidal sheared flow and the pedestal.

\begin{acknowledgments}
This work was supported by National R\&D Program through the National Research Foundation of 
Korea (NRF) funded by the Ministry of Education, Science and Technology (Grant No. NRF-2022R1A2C1012808126)
\end{acknowledgments}

\bigskip
{\bf DATA AVAILABILITY}

\bigskip
The data that supports the findings of this study are available within the article.

\appendix 
\section{{Solutions to Eqs. (\ref{24}) and (\ref{25}) for $\bar{U}_{ix}\left(X_{i}, t\right)$ and 
$\bar{U}_{iy}\left(X_{i}, t\right)$}}
 Direct integration of Eqs. (\ref{24}), (\ref{25}) gives 
\begin{eqnarray}
	&\displaystyle
	\bar{U}_{ix}\left(\tilde{X}_{i}, t\right)=\frac{e_{i}}{m_{i}}\int\limits_{0}^{t}dt_{1}\left[\left\langle  
	\tilde{E}_{i1x}\left(\tilde{X}_{i}, \tilde{x}_{i}, \tilde{y}_{i}, 
	t_{1}\right)\right\rangle\cos\omega_{ci}\left(t-t_{1}\right)\right.
	\nonumber
	\\  
	&\displaystyle 
	\left.
	+\left\langle \tilde{E}_{i1y}\left(\tilde{X}_{i}, 
	\tilde{x}_{i}, \tilde{y}_{i}, t_{1}\right)\right\rangle\sin\omega_{ci}\left(t-t_{1}\right)\right],
	\label{117}
	\\  
	&\displaystyle 
	\bar{U}_{iy}\left(\tilde{X}_{i}, t\right)=\frac{e_{i}}{m_{i}}\int\limits_{0}^{t}dt_{1}
	\left[-\left\langle \tilde{E}_{i1x}\left(\tilde{X}_{i}, \tilde{x}_{i}, \tilde{y}_{i}, 
	t_{1}\right)\right\rangle\sin\omega_{ci}\left(t-t_{1}\right)\right.
	\nonumber
	\\  
	&\displaystyle 
	\left.
	+\left\langle \tilde{E}_{i1y}\left(\tilde{X}_{i}, 
	\tilde{x}_{i}, \tilde{y}_{i}, t_{1}\right)\right\rangle\cos\omega_{ci}\left(t-t_{1}\right)\right],
	\label{118}
\end{eqnarray}
where
\begin{eqnarray}
&\displaystyle
\tilde{E}_{i1x}\left(\tilde{X}_{i}, \tilde{x}_{i}, \tilde{y}_{i}, t\right)=\frac{\partial}{\partial 
\tilde{X}_{i}}\left(\tilde{E}_{i0x}\left(\tilde{X}_{i}, \tilde{x}_{i}, 
\tilde{y}_{i},t\right)\right)\cdot\tilde{R}_{ix}\left(\tilde{X}_{i}, \tilde{x}_{i}, \tilde{y}_{i}, t\right),
	\label{119}
	\\ 
	&\displaystyle
	\tilde{E}_{i1y}\left(\tilde{X}_{i}, \tilde{x}_{i}, \tilde{y}_{i}, t\right)=\frac{\partial}{\partial
	\tilde{X}_{i}}\left(\tilde{E}_{i0y}\left(\tilde{X}_{i}, \tilde{x}_{i}, \tilde{y}_{i}, t\right)\right)\cdot\tilde{R}_{ix}
	\left(\tilde{X}_{i}, \tilde{x}_{i}, \tilde{y}_{i}, t\right),
	\label{120}
	\\ 
	&\displaystyle
	\tilde{R}_{ix}\left(\tilde{X}_{i}, \tilde{x}_{i}, \tilde{y}_{i}, t\right)=\int\limits_{0}^{t}dt_{1}\tilde{U}_{ix}^{(0)}
	\left(\tilde{X}_{i},\tilde{x}_{i}, \tilde{y}_{i}, t_{1}\right)
	\nonumber
	\\ 
	&\displaystyle
	=\frac{e_{i}}{m_{i}}\int\limits_{0}^{t}dt_{1}\int\limits_{0}^{t_{1}}dt_{2}
	\left[\tilde{E}_{i0x}\left(\tilde{X}_{i}, \tilde{x}_{i}, \tilde{y}_{i}, 
	t_{2}\right)\cos\omega_{ci}\left(t_{1}-t_{2}\right)\right.
	\nonumber
	\\  
	&\displaystyle 
	\left.
	+ \tilde{E}_{i0y}\left(\tilde{X}_{i}, \tilde{x}_{i}, \tilde{y}_{i}, 
	t_{2}\right)\sin\omega_{ci}\left(t_{1}-t_{2}\right)\right],
	\label{121}
\end{eqnarray}
where $\tilde{\mathbf{E}}_{i0}\left(\tilde{X}_{i}, \tilde{x}_{i}, \tilde{y}_{i}, t\right)$ is the electric 
field $\mathbf{E}_{i}\left(\mathbf{r}_{i}, X_{i}, t\right)$, determined in $\tilde{\mathbf{r}}_{i}, 
\tilde{X}_{i}$ coordinates. The integration of Eqs. (\ref{117}), (\ref{118}) with accounting for Eqs. 
(\ref{119})--(\ref{121}) and averaging over the fast time $t\gg \omega_{ci}^{-1}$ velocities 
$\bar{U}_{ix}\left(\tilde{X}_{i}, t\right)$ and  $\bar{U}_{iy}\left(\tilde{X}_{i}, t\right)$
gives for $\bar{U}_{ix}\left(\tilde{X}_{i}\right)$ and $\bar{U}_{iy}\left(\tilde{X}_{i}\right)$ solutions
\begin{eqnarray}
&\displaystyle
\bar{U}_{ix}\left(\tilde{X}_{i}\right)=\langle\langle \bar{U}_{ix}\left(\tilde{X}_{i}, t\right)\rangle\rangle
\nonumber
\\  
	&\displaystyle 
	=\frac{e_{i}^{2}}{m^{2}_{i}}\frac{1}{\omega_{ci}}\frac{1}{\left(2\pi\right)^{3}}\int d\mathbf{k}_{i}
	\frac{\partial}{\partial \tilde{X}_{i}}\left(\tilde{E}_{i0y}\left(\tilde{X}_{i}, \mathbf{k}_{i}\right)\right)
	\tilde{E}_{i0x}\left(\tilde{X}_{i}, \mathbf{k}_{i}\right)\frac{1}{\left(\omega^{2}_{ci}
	-\omega^{2}\left(\mathbf{k}_{i}\right)\right)},	
	\label{122}
	\\  
	&\displaystyle  
	\bar{U}_{iy}\left(\tilde{X}_{i}\right)=\langle\langle \bar{U}_{iy}\left(\tilde{X}_{i}, t\right)\rangle \rangle
	\nonumber
	\\  
	&\displaystyle 
	=\frac{e_{i}^{2}}{m^{2}_{i}}\frac{1}{\omega_{ci}}\frac{1}{\left(2\pi\right)^{3}}\int d\mathbf{k}_{i}
	\frac{\partial}{\partial \tilde{X}_{i}}\left(\tilde{E}_{i0x}\left(\tilde{X}_{i}, \mathbf{k}_{i}\right)\right)^{2}
	\frac{1}{\left(\omega^{2}_{ci}	-\omega^{2}\left(\mathbf{k}_{i}\right)\right)}.
\label{123} 
\end{eqnarray}
The  velocities $\bar{U}_{ex}\left(\tilde{X}_{e}\right)$  and $\bar{U}_{ey}\left(\tilde{X}_{e}\right)$  
of the electron convective flows are determined by Eqs.  (\ref{122}), (\ref{123}) with ion species index changed on the 
electron species index. It follows from Eqs. (\ref{122}),  (\ref{123}) that the electron convective velocities are in 
$\omega_{ce}/\omega_{ci}$ times less than the ion convective velocities $\bar{U}_{ix}\left(\tilde{X}_{i}\right)$  and 
$\bar{U}_{iy}\left(\tilde{X}_{i}\right)$ and, therefore, the convective motion of electrons may be neglected.

\bibliographystyle{plain}
\bibliography{refs}

\end{document}